\documentclass[11pt]{article}
\usepackage{graphicx}
\usepackage{fullpage}
\usepackage{titlesec}
\usepackage{amsmath}
\usepackage{amsthm}
\usepackage{color}
\usepackage{setspace}
\usepackage{courier}
\usepackage[toc, page]{appendix}
\usepackage{listings}
\usepackage{lmodern}
\usepackage{adjustbox}
\usepackage{amssymb}
\usepackage{underscore}
\usepackage{textcomp}
\usepackage[english]{babel}
\usepackage{url}
\usepackage{csquotes}

\newtheorem{statement}{Statement}

\newcommand{\statenumstate}{7,910-state }
\newcommand{\statenum}{7,910 }
\newcommand{\statenumcomma}{7,910, }
\newcommand{\bbstatenum}{$BB($7,910) }

\newcommand{\bbbitsperiod}{157,819.}

\newcommand{\zhaltstate}{\texttt{7862}}

\newcommand{\gbstatenum}{4,888 }
\newcommand{\gbstatenumstate}{4,888-state }
\newcommand{\bbgbstatenum}{$BB($4,888) }

\newcommand{\rmstatenum}{5,372 }
\newcommand{\rmstatenumstate}{5,372-state }

\newenvironment{nscenter}
 {\parskip=0pt\par\nopagebreak\centering}
 {\par\noindent\ignorespacesafterend}

\begin{document}

\title{A Relatively Small Turing Machine Whose Behavior Is Independent of Set Theory}
\author{
    Adam Yedidia\\
    \small\texttt{MIT}\\
    \small\texttt{adamy@mit.edu}
    \and
    Scott Aaronson\\
    \small\texttt{MIT}\\
    \small\texttt{aaronson@csail.mit.edu}
}
\maketitle

\begin{abstract}

Since the definition of the Busy Beaver function by Rad\'{o} in 1962, an interesting open question has been the smallest value of $n$ for which $BB(n)$ is independent of ZFC set theory. \ Is this $n$ approximately $10$, or closer to 1,000,000, or is it even larger? \ In this paper, we show that it is at most \statenum by presenting an explicit description of a \statenumstate Turing machine $Z$ with 1 tape and a 2-symbol alphabet that cannot be proved to run forever in ZFC (even though it presumably does), assuming ZFC is consistent. \ The machine is based on work of Harvey Friedman on independent statements involving order-invariant graphs. \ In doing so, we give the first known upper bound on the highest provable Busy Beaver number in ZFC. \ To create $Z$, we develop and use a higher-level language, Laconic, which is much more convenient than direct state manipulation. \ We also use Laconic to design two Turing machines, $G$ and $R$, that halt if and only if there are counterexamples to Goldbach's Conjecture and the Riemann Hypothesis, respectively.

\end{abstract}

\section{Introduction}

\subsection{Background and Motivation \label{sec:background}}

\emph{Zermelo-Fraenkel set theory with the axiom of choice}, more commonly known as ZFC, is an axiomatic system invented in the twentieth which has since been used as the foundation of most of modern mathematics. \ It encodes arithmetic by describing natural numbers as increasing sets of sets.

Like any axiomatic system capable of encoding arithmetic, ZFC is constrained by G\"{o}del's two incompleteness theorems. \ The first incompleteness theorem states that if ZFC is \emph{consistent} (it never proves both a statement and its opposite), then ZFC cannot also be \emph{complete} (able to prove every true statement). \ The second incompleteness theorem states that if ZFC is consistent, then ZFC cannot prove its own consistency. \ Because we have built modern mathematics on top of ZFC, we can reasonably be said to have assumed ZFC's consistency. \ This means that we must also believe that ZFC cannot prove its own consistency. \ This fact carries with it certain surprising conclusions.

In particular, consider a Turing machine $Z$ that enumerates, one after the other, each of the provable statements in ZFC. \ To describe how such a machine might be constructed, $Z$ could iterate over the axioms and inference rules of ZFC, applying each in every possible way to each conclusion or pair of conclusions that had been reached so far. \ We might ask $Z$ to halt if it ever reaches a contradiction; in other words, $Z$ will halt if and only if it finds a proof of $0 = 1$. \ Because this machine will enumerate \emph{every} provable statement in ZFC, it will run forever if and only if ZFC is consistent.

It follows that $Z$ is a Turing machine for which the question of its behavior (whether or not it halts when run indefinitely) is equivalent to the consistency of ZFC.\footnote{While we will talk about ZFC throughout this paper, rather than simple ZF set theory, this is simply a convention. \ For our purposes, the Axiom of Choice is irrelevant: the consistency of ZFC is equivalent to the consistency of simple ZF set theory,~\cite{godelcohen} and ZFC and ZF prove exactly the same arithmetical statements (which include, among other things, statements about whether Turing machines halt).~\cite{schoenfield}} \ Therefore, just as ZFC cannot prove its own consistency (assuming ZFC is consistent), ZFC also cannot prove that $Z$ will run forever. In other words, the statement, ``$Z$ will run forever'' is \emph{independent of} ZFC.

This is interesting because, while the undecidability of the halting problem tells us that there cannot exist an algorithmic method for determining whether an \emph{arbitrary} Turing machine loops or halts, $Z$ is an example of a \emph{specific} Turing machine whose behavior cannot be proven one way or the other using the foundation of modern mathematics. \ Mathematicians and computer scientists think of themselves as being able to determine how a given algorithm will behave if given enough time to stare at it; despite this intuition, $Z$ is a machine whose behavior we can never prove without assuming axioms more powerful than those generally assumed in modern mathematics.

\subsection{Turing Machines \label{sec:tm}}

There are many slightly different definitions of Turing machines. \ For example, some definitions allow the machine to have multiple tapes; others only allow it to have one; some allow an arbitrarily large alphabet, while others allow only two symbols, and so on. \ In most research regarding Turing machines, mathematicians don't concern themselves with which of these models to use, because any one can simulate the others (usually efficiently). \ However, because this work is concerned with upper-bounding the exact number of states required to perform certain tasks, it's important to define the model precisely. \ The model we choose here is traditional for the Busy Beaver function.

Formally, a $k$-state Turing machine is a 7-tuple $M = (Q, \Gamma, a, \Sigma, \delta, q_0, F)$, where: \\ \\
$Q$ is the set of $k$ \emph{states} $\{q_0, q_1, \dots, q_{k-2}, q_{k-1}\}$ \\
$\Gamma = \{a, b\}$ is the set of \emph{tape alphabet symbols} \\
\texttt{a} is the \emph{blank symbol} \\
$\Sigma = \empty$ is the set of \emph{input symbols} \\
$\delta = Q \times \Gamma \rightarrow (Q \cup F) \times \Gamma \times \{L, R\}$ is the \emph{transition function} \\
$q_0$ is the \emph{start state} \\
$F = \{\textrm{HALT}, \textrm{ERROR}\}$ is the set of \emph{halting states}. \\

A Turing machine's \emph{states} make up the Turing machine's easily-accessible, finite memory. \ The Turing machine's state is initialized to $q_0$.

The \emph{tape alphabet symbols} correspond to the symbols that can be written on the Turing machine's infinite tape.

In this work, all Turing machines are run on the all-\texttt{a} input.

The \emph{transition function} encodes the Turing machine's behavior. \ It takes two inputs: the current state of the Turing machine (an element of $Q \cup F$) and the symbol read off the tape (an element of $\Gamma$). \ It outputs three instructions: what state to enter (an element of $Q \cup F$), what symbol to write onto the tape (an element of $\Gamma$) and what direction to move the head in (an element of $\{L, R\}$). \ A transition function specifies the entire behavior of the Turing machine in all cases.

The \emph{start state} is the state that the Turing machine is in at initialization.

A \emph{halting transition} is a transition to a halting state, which causes the Turing machine to halt. \ While having three possible halting transitions is not necessary for our purposes, being able to differentiate between different types of halting (HALT and ERROR) is useful for testing.

\subsection{The Busy Beaver Function}

Consider the set of all Turing machines with $k$ states, for some positive integer $k$. \ We call a Turing machine $B$ a $k$\emph{-state Busy Beaver} if when run on the empty tape as input, $B$ halts, and also runs for at least as many steps before halting as all other halting $k$-state Turing machines.~\cite{busybeaver}

In other words, a Busy Beaver is a Turing machine that runs for at least as long as all other halting Turing machines with the same number of states. \ Another common definition for a Busy Beaver is a Turing machine that writes as many 1's on the tape as possible; because the number of 1's written is a somewhat arbitrary measure, it is not used in this work.

The \emph{Busy Beaver function}, written $BB(k)$, equals the number of steps it takes for a $k$-state Busy Beaver to halt. \ The Busy Beaver function has many striking properties. \ To begin with, it is not \emph{computable}; in other words, there does not exist an algorithm that takes $k$ as input and returns $BB(k)$, for arbitrary values of $k$. \ This follows directly from the undecidability of the halting problem. \ Suppose an algorithm existed to compute the Busy Beaver function; then given a $k$-state Turing machine $M$ as input, we could compute $BB(k)$ and run $M$ for $BB(k)$ steps. \ If, after $BB(k)$ steps, $M$ had not yet halted, we could safely conclude that $M$ would never halt. \ Thus, we could solve the halting problem, which we know is impossible.

By the same argument, $BB(k)$ must grow faster than any computable function. \ (To check this, assume that some computable function $f(k)$ grows faster than $BB(k)$, and substitute $f(k)$ for $BB(k)$ in the rest of the proof.) \ In particular, the Busy Beaver grows even faster than (for instance) the Ackermann function, a well-known fast-growing function.

Because finding the value of $BB(k)$ for a given $k$ requires so much work (one must fully explore the behavior of all $k$-state Turing machines), few explicit values of the Busy Beaver function are known. \ The known values are~\cite{bbfour,bbsmall}:

$$BB(1) = 1$$
$$BB(2) = 6$$
$$BB(3) = 21$$
$$BB(4) = 107$$

For $BB(5)$, $BB(6)$, and $BB(7)$ only lower bounds are known~\cite{bbvalues,bigbbvalues}:

$$BB(5) \ge \textrm{47,176,870}$$
$$BB(6) > 7.4 \times 10^{\textrm{36,534}}$$
$$BB(7) > 10^{10^{10^{10^{10^7}}}}$$

Additionally, $BB(22)$ is known to be larger than Graham's Number (a famous huge number from Ramsey theory, obtained by iterating the Ackermann function 64 times)~\cite{grahamsnumber}. \ Researchers have worked on pinning down the value of $BB(5)$ exactly, and some consider it to be possibly within reach.

Another way to discuss the Busy Beaver sequence is to say that modern mathematics has established a \emph{lower bound} of $4$ on the highest provable Busy Beaver value. \ In this paper, we prove the first known \emph{upper bound} on the highest provable Busy Beaver value in ZFC; that is, we give a value of $k$, namely \statenumcomma such that the value of $BB(k)$ cannot be proven in ZFC.

Intuitively, one might expect that while no algorithm may exist to compute $BB(k)$ for \emph{all} values of $k$, we could find the value of $BB(k)$ for any \emph{specific} $k$ using a procedure similar to the one we used to find the value of $BB(k)$ for $k \le 4$. \ The reason this is not so is closely tied to the existence of a machine like the G\"{o}delian machine $Z$, as described in Section~\ref{sec:background}. \ Suppose that $Z$ has $k$ states. Because $Z$'s behavior (whether it halts or loops) cannot be proven in ZFC, it follows that the value of $BB(k)$ also can't be proven in ZFC; if it could, then a proof would exist of $Z$'s behavior in ZFC. \ Such a proof would consist of a \emph{computation history} for $Z$, which is an explicit step-by-step description of $Z$'s behavior for a certain number of steps. \ If $Z$ halts, then a computation history leading up to $Z$'s halting would be the entire proof; if $Z$ loops, then a computation history that takes $BB(k)$ steps, combined with a proof of the value of $BB(k)$, would constitute a proof that $Z$ will run forever.

In this paper we construct a machine like $Z$, for which a proof that $Z$ runs forever would imply that ZFC was consistent. \ In doing so, we give an explicit upper bound on the highest Busy Beaver value provable in ZFC assuming the consistency of a slightly stronger set theory. \ Our machine, which we shall refer to as $Z$ hereafter, contains \statenum states. \ Therefore, we will never be able to prove the value of \bbstatenum without assuming more powerful axioms than those of ZFC. \ This upper bound is presumably very far from tight, but it is a first step.

Even to achieve a state count of \statenumcomma we will need three nontrivial ideas: Harvey Friedman's order-theoretic statements, \emph{on-tape processing}, and \emph{introspective encoding}. \ Without all three ideas, we found that the state count would be in the tens of thousands, hundreds of thousands, or even millions. \ We briefly introduce these ideas in the following subsection, and explore them in much greater detail in Section \ref{sec:compandproc}. \ The implementation of these ideas constitutes this paper's main technical contribution.

\subsection{Parsimony}

In most algorithmic study, efficiency is the primary concern. \ In designing $Z$, however, parsimony is the only thing that matters. \ One historical analogue is the practice of ``code-golfing'': a recreational pursuit adopted by some programmers in which the goal is to produce a piece of code in a given programming language, using as few characters as possible. \ Many examples of code-golfing can be found at~\cite{codegolf}. \ The goal of designing a Turing machine with as few states as possible to accomplish a certain task, without concern for the machine's efficiency or space usage, can be thought of as code-golfing with a particularly low-level programming language.

Part of the charm of Turing machines is that they give us a ``standard reference point'' for measuring complexity, unencumbered by the details of more sophisticated programming languages. \ Also, with Turing machines, there can be no suspicion that we engineered a programming formalism just for the purpose of code-golfing, or for making the concepts we want artificially simple to describe. \ This is why we prefer Turing machines as a tool for measuring complexity; not because they are particularly special, but simply because they are so primitive that their specifics will interfere minimally with what we mean by an algorithm being ``complicated.''

In this paper, we use three ideas for generating parsimonious Turing machines: Harvey Friedman's mathematical statements, \emph{on-tape processing}, and \emph{introspective} Turing machines. \ The last of these ideas was proposed, under a different name and with some variations, by Ben-Amram and Petersen in 2002 \cite{benamram}. \ These three ideas are explained in more detail in Subsections~\ref{sec:friedmanstate},~\ref{sec:ontape}, and~\ref{sec:introspect}, respectively, but we summarize them very briefly here.

The first idea is simply to use the research done by Friedman into finding simple-to-express statements that are equivalent to the consistency of various axiomatic systems. \ In particular, we use a statement discovered by Friedman to be equivalent to the consistency of a set theory stronger than ZFC (and whose consistency, therefore, would imply the consistency of ZFC).\footnote{Admittedly, it's not obvious that using Friedman's current statements \emph{does} decrease the state count of the Turing machines. \ It's possible that one could do as well or better by directly searching for contradictions in ZFC, and indeed, recent unpublished work by Stefan O'Rear has given some evidence for that~\cite{comments}. \ On the other hand, Friedman's statements can be translated into code without using the apparatus of first-order logic, which arguably gives us a conceptual simplification. \ In addition, statements like Friedman's seem like the most plausible path forward for \emph{further} reductions in the state count, beyond whatever lower limit one hits when one needs to encode the ZFC axioms explicitly.}~\cite{friedman}

The second idea, on-tape processing, is a way to encode high-level commands into a Turing machine parsimoniously. \ Instead of converting commands to groups of states directly, which incurs a multiplicative overhead based on how large these groups need to be, on-tape processing begins by writing the commands onto the tape, using as efficient an encoding as possible. \ Then, once the commands are on the tape, the commands are processed by a single group of states that understands how to interpret them.

The third idea, introspective Turing machines, is a way to write long strings onto the tape using as few states as possible. \ The idea is to encode information one of each state's transitions, instead of encoding information in each state's write field. \ This is advantageous because there are many choices for which state to point a transition to, but only two choices for which bit to write. \ Therefore, more information can be encoded in each state using this method.

\subsection{Implementation Overview}

To generate descriptions of Turing machines with nice mathematical properties entirely by hand is a daunting task. \ Rather than approach the problem directly, we created tools for generating parsimonious Turing machines while presenting an interface that is comfortably familiar to most programmers (and to us!).

\begin{figure}
\begin{center}
\includegraphics[scale=0.42]{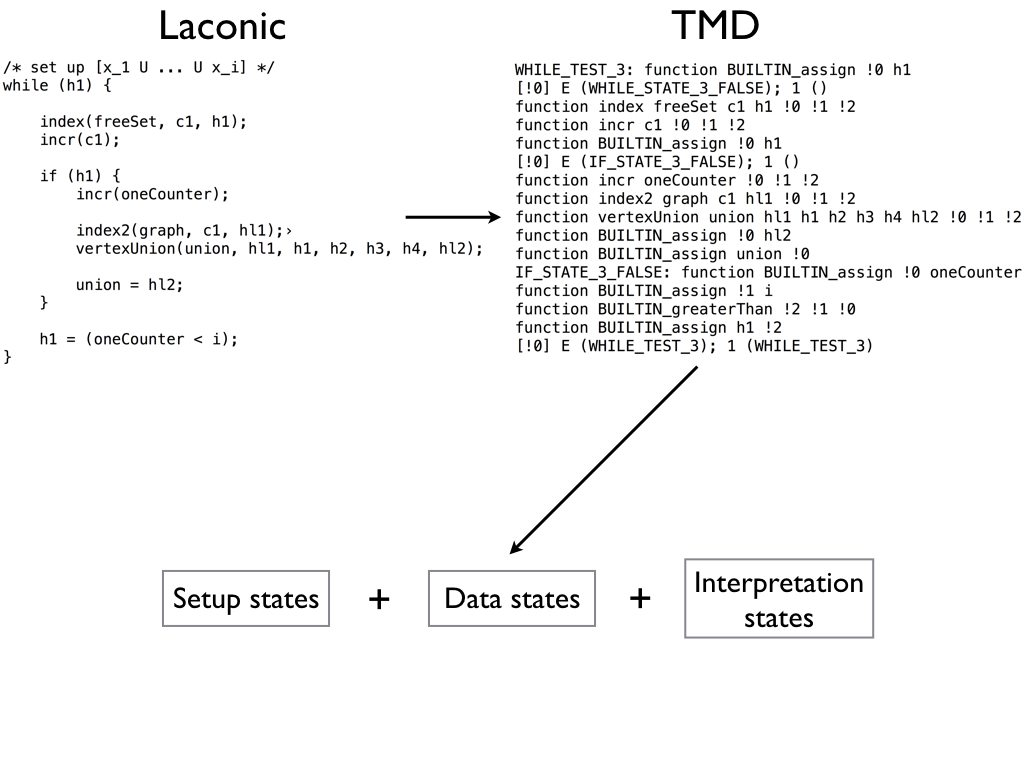}
\caption{A visual overview of the compilation process. \label{fig:compilation}}
\end{center}
\end{figure}

We created two tools. At the top level is the Laconic programming language, whose syntax and capabilities are similar to those of most programming languages, such as Java or Python. \ Beneath it we created a lower-level language called Turing Machine Descriptor (TMD). \ TMD is quite unlike most programming languages, and is better thought of as a convenient way to describe a multi-tape, 3-symbol Turing machine plus a function stack. \ The style of multi-tape Turing machine used in TMD is the commonly used ``one-tape-at-a-time'' abstraction: only one tape at a time can be interacted with, for reading, writing, and moving the head. Laconic compiles down to a TMD program, and TMD compiles down to a description of a single-tape, 2-symbol Turing machine. \ This process is illustrated in Figure~\ref{fig:compilation}.

We recommend that programmers hoping to use our tools to generate their own encodings of mathematical statements or algorithms as Turing machines use Laconic. \ Laconic's interface is perfect for somebody hoping to write in a ``traditional'' language. \ On the other hand, if the programmer wishes to improve upon Laconic's compilation process, writing code directly in TMD is likely to be the better option.

\section{Related Work}

Gregory Chaitin raised the problem of proving a version of our result in his book \emph{The Limits of Mathematics}.~\cite{chaitin} \ He wrote:

\begin{displayquote}
\noindent I would like to have somebody program out Zermelo-Fraenkel set theory in my version of LISP, which is pretty close to normal LISP as far as this task is concerned, just to see how many bits of complexity mathematicians normally assume \dots If you programmed ZF, you'd get a really sharp incompleteness result. \ It wouldn't say that you can get at most $H(ZF) + 15328$ bits of [Chaitin's halting probability] $\Omega$, it would say, perhaps, at most 96000 bits! \ We'd have a much more definite incompleteness theorem.
\end{displayquote}

We did not program ZF set theory in LISP, but we programmed it in an even simpler language---thereby answering Chaitin's call for an explicit number of bits to attach to the complexity of ZF set theory. (As many as required to fully describe our Turing machine---or more precisely, \bbbitsperiod)

This paper is not the first to attempt to quantify the complexity of arithmetical statements. \ Calude and Calude~\cite{calude} define a register machine of their own design, and provide quantifications of the complexity of Legendre's Conjecture, Fermat's Last Theorem, Goldbach's Conjecture, Dyson's Conjecture, the Riemann Hypothesis, and the Four Color Theorem.\footnote{Because Fermat's Last Theorem and the Four Color Theorem have been proved, their ``complexity'' is now known to be $1$---the minimum number of states in a Turing machine that runs forever.}
In addition, Koza~\cite{koza} and Pargellis~\cite{pargellis} each invent instruction sets that are particularly well-suited to representing self-reproducing programs simply, and show that starting from a ``primordial soup'' of such instructions distributed about a large memory, along with an increasing number of program threads, a rich ecosystem of increasingly efficient self-reproducing programs start to dominate the ``landscape.''

This paper differs from the previous work in two ways: firstly, it's the first to give explicit, relatively small machines whose behavior is provably independent of the standard axioms of modern mathematics. \ Secondly, to our knowledge, this paper is the first concrete study of parsimony to use Turing machines themselves as the model of computation---rather than (for example) a new programming language proposed by the authors, or a complex on-tape description of Turing machines! \ We consider it important to use the weakest and most common model of computation for complexity comparisons across different mathematical statements. \ This is because the more powerful and complex the model of computation used, the more of the complexity of the algorithm can be ``shunted'' onto the model of computation, and the greater the potential distortion created by the choice of model. \ As a reductio ad absurdum, we could imagine a programming language that included ``test the Riemann Hypothesis'' and ``test the consistency of ZFC'' as primitive operations. \ By using the ``weakest'' model of computation that's commonly known, we hope to avoid this pitfall and make it easier to interpret our results in a model-independent way.

Also related to the work of this paper is the famous search for the smallest universal Turing machine. \ Here a \emph{universal Turing machine} is a Turing machine that can simulate any other Turing machine, when a description of the latter is provided on its input tape. \ The smallest-known universal Turing machine has only $2$ states and a $3$-symbol alphabet, and was found by Alex Smith~\cite{universal} in 2007. \ From the perspective of this paper, the problem is that the known small universal Turing machines achieve their small size only at the cost of an extremely complicated description format for the input machine. \ I.e., most of the complexity gets ``shunted'' from the Turing machine itself to the input encoding format. \ By contrast, with small Turing machines to test $Con(ZFC)$, the Riemann Hypothesis, Goldbach's Conjecture, etc., and which run on an initially blank tape, there's no analogous trick for hiding the statement's complexity.

\section{A Turing Machine that Cannot Be Shown to Run Forever Using ZFC}

We present a \statenumstate Turing machine whose behavior is \emph{independent of ZFC}; it is not possible to prove that this machine halts or doesn't halt using the axioms of ZFC, assuming that a stronger set theory is consistent. \ It's therefore impossible to prove the value of \bbstatenum to be any given value without assuming axioms more powerful than ZFC, assuming that ZFC is consistent.

For an explicit listing of this machine, see Appendix~\ref{sec:explicitz}.

We call this machine $Z$. \ One way to build this machine would be to start with the axioms of ZFC and apply the inference rules of first-order logic repeatedly in each possible way so as to enumerate every statement ZFC could prove, and to halt if ever a contradiction was found. \ While this method is conceptually simple, to actually construct such a machine would lead to a huge number of states, because it would require writing a program to manipulate the axioms of ZFC and the inference rules of first-order logic, and then compiling that program all the way down to Turing machine states.

\subsection{Friedman's Mathematical Statement} \label{sec:friedmanstate}

Thankfully, a simpler method exists for creating $Z$. \ Friedman~\cite{friedman}
was able to derive a graph-theoretic statement whose truth implies the consistency of ZFC, and which will be false if ZFC is inconsistent.\footnotemark
\footnotetext{In fact, Friedman's statement is equivalent to the consistency of SRP (``stationary Ramsey property''), which is a system of axioms more powerful than ZFC. \ Because SRP is strictly more powerful than ZFC (it in fact consists of ZFC plus some additional axioms), the consistency of SRP implies the consistency of ZFC, and the inconsistency of ZFC implies the inconsistency of SRP.} \
Here is Friedman's statement (the notation will be explained in the rest of this section):

\begin{statement} \label{eq:friedman}
For all $k, n, r > 0$, every order invariant graph on $[\mathbb{Q}]^{\le k}$ has a free $\{x_1,\dots,x_r, \\
\textrm{ush}(x_1),...,\textrm{ush}(x_r)\}$ of complexity $\le (8knr)!$, each $\{x_1, \dots, x_{(8kni)!}\}$, for $i > 0$ and $(8kni!) \le r$, reducing $[x_1 \cup \dots \cup x_i \cup \{0,\dots,n\}]^{\le k}$. \cite{friedman}
\end{statement}

If $s$ is a set, the operation $(.)^{\le k}$ refers to the set of all subsets of $s$ with size at most $k$.

A graph on $[\mathbb{Q}]^{\le k}$ is an irreflexive symmetric relation on $[\mathbb{Q}]^{\le k}$. In other words, it can be thought of as a graph whose vertices are elements of $[\mathbb{Q}]^{\le k}$, and whose edges are undirected, connect pairs of vertices, and never connect vertices to themselves.

A \emph{free} set is a set such that no pair of elements in that set are connected by an edge.

A number of \emph{complexity} at most $c$ refers to a number that can be written as a fraction $a/b$, where $a$ and $b$ are both integers with absolute value less than or equal to $c$. \ A set has complexity at most $c$ if all the numbers it contains have complexity at most $c$.

An \emph{order invariant graph} is a graph containing a countably infinite number of nodes. \ In particular, it has one node for each finite set of rational numbers. \ The only numbers relevant to the statement are numbers of complexity $(8knr)!$ or smaller. \ In every description of nodes that follows, the term \emph{node} refers both to the object in the order invariant graph and to the set of numbers that it represents.

In an order invariant graph, two nodes $(a,b)$ have an edge between them if and only if each other pair of nodes $(c,d)$ that is \emph{order equivalent} with $(a,b)$ has an edge between them. \ Two pairs of nodes $(a, b)$ and $(c, d)$ are \emph{order equivalent} if $a$ and $c$ are the same size and $b$ and $d$ are the same size and if for all $1 \le i \le |a|$ and $1 \le j \le |b|$, the $i$-th element of $a$ is less than the $j$-th element of $b$ if and only if the $i$-th element of $c$ is less than the $j$-th element of $d$.

To give some trivial examples of order invariant graphs: the graph with no edges is order invariant, as is the complete graph. \ A less trivial example is a graph on $[\mathbb{Q}]^{\le 2}$, in which each node corresponds to a set of two rational numbers of a given complexity, and there is an edge between two nodes if and only if their corresponding sets $a$ and $b$ satisfy $|a| = |b| = 2$ and $a_1 < b_1 < a_2 < b_2$. \ (Because edges are undirected in order invariant graphs, such an edge will exist if \emph{either} assignment of the vertices to $a$ and $b$ satisfies the inequality above.)

The \emph{ush()} function takes as input a set and returns a copy of that set with all non-negative numbers in that set incremented by $1$.

For vertices $x$ and $y$, $x \le_{rlex} y$ if and only if $x = y$ or $x_{|x|-i} < y_{|y|-i}$ where $i$ is the least integer such that $x_{|x|-i} \not= y_{|y|-i}$.\footnote{Friedman recommended in private communication that we use the $\le_{rlex}$ comparator to compare vertices, instead of comparing their maximum elements as described in his manuscript.} \ (The $\le_{rlex}$ operation creates a lexicographic ordering over the vertices, weighting the last and largest elements of those vertices most heavily. Like with lexicographic orderings, if the two vertices are identical but one is longer, the shorter one comes first.)

Finally, a set of vertices $X$ \emph{reduces} a set of vertices $Y$ if and only if for all $y \in Y$, there exists $x \in X$ such that
either $x = y$ or $x \le_{rlex} y$ and an edge exists between $x$ and $y$.

\subsection{Implementation Methods}

To create $Z$, we needed to design a Turing machine that halts if Statement~\ref{eq:friedman} is false, and loops if Statement~\ref{eq:friedman} is true. \ Such a Turing Machine's behavior is necessarily independent of ZFC, because the truth or falsehood of Statement~\ref{eq:friedman} is independent of ZFC, assuming the consistency of SRP.~\cite{friedman} SRP is an extension of ZFC by certain relatively mild large cardinal hypotheses, and is widely regarded by set theorists as consistent. For more information about SRP, see~\cite{srp}.

To design such a Turing machine, we wrote a Laconic program which encodes Friedman's statement, then compiled the program down to a description of a single-tape, $2$-symbol Turing machine. \ What follows is an extremely brief description of the design of the Laconic program; for the documented Laconic code itself, along with a detailed explanation of the full compilation process, see~\cite{github}.

Our Laconic program begins by looping over all non-negative values for $k$, $n$, and $r$. \ For each trio $(k, n, r)$, our program generates a list $N$ of all numbers of complexity at most $(8knr)!$. \ These numbers represent the vertices in our putative order invariant graph. \ Because Laconic does not support floating-point numbers, the list is entirely composed of integers; it is a list of all numbers that can be written in the form $(((8knr)!)!)\frac{i}{j}$, where $i$ and $j$ are integers satisfying $-(8knr)! \le i \le (8knr)!$ and $1 \le j \le (8knr)!$. \ (Note that any number that can be expressed in this form is necessarily an integer, because of the large scaling factor in front.)

After we generate $N$, we generate the nodes in a potential order invariant graph by adding to $N$ all possible lists of $k$ or fewer numbers from $N$. \ We call this list of lists $V$.

We iterate over all binary lists of length $|V|^2$. \ Any such list $E$ represents a possible set of edges in the graph. \ To be more precise, we say that an edge exists between node $i$ and node $j$ (represented by $V_i$ and $V_j$ respectively) if and only if $E_{i|V| + j}$ is $1$.

For any graph $(V, E)$, we say that it is ``valid'' if the following three conditions hold:

\begin{enumerate}

\item No node has an edge to itself.
\item If an edge exists between node $i$ and node $j$, an edge also exists between node $j$ and node $i$.
\item The graph has a free $\{x_1,\dots,x_r, \textrm{ush}(x_1),...,\textrm{ush}(x_r)\}$, each  $\{x_1, \dots, x_{(8kni)!}\}$ reducing $[x_1 \cup \dots \cup x_i \cup \{0,\dots,n\}]^{\le k}$.

\end{enumerate}

For each list of nodes $V$, we loop over every possible binary list $E$, and if no pair $(V, E)$ yields a valid graph, we halt.

When verifying the validity of a graph, checking the first two conditions is trivial, but the third merits further explanation. \ In order to verify that a given graph $(V, E)$ has a free \\ $\{x_1,\dots,x_r, ush(x_1),...,ush(x_r)\}$, each  $\{x_1, \dots, x_{(8kni)!}\}$ reducing $[x_1 \cup \dots \cup x_i \cup \{0,\dots,n\}]^{\le k}$, we look at every possible subset of the nodes in $V$. \ For each subset, we verify that it has length $r$, that $\textrm{ush}(x_1),...,\textrm{ush}(x_r)$ all exist in $V$, and for each $i$ such that $(8kni)! \le r$, that $\{x_1, \dots, x_{(8kni)!}\}$ reduces $[x_1 \cup \dots \cup x_i \cup \{0,\dots,n\}]^{\le k}$. \ Once we have found such a subset, we know that the third conditon is satisfied.

\section{A Turing Machine that Encodes Goldbach's Conjecture} \label{sec:g}

We present a \gbstatenumstate Turing machine that \emph{encodes Goldbach's Conjecture}; in other words, to know whether this machine halts is to know whether Goldbach's Conjecture is true. \ It's therefore impossible to prove the value of \bbgbstatenum without simultaneously proving or disproving Goldbach's Conjecture.\footnote{Note that our tools were primarily meant to encode complex statements into Turing machines, such as Statement~\ref{eq:friedman}. \ Because Goldbach's Conjecture is so simple, it's feasible in that case to make dramatically smaller Turing machines through a more direct approach.  Indeed, after a preprint of this paper was circulated online, ``Jared S'' and ``code golf addict'' created Turing machines for Goldbach's Conjecture with $47$ and $31$ states respectively~\cite{comments}, though these machines have not yet been tested in detail.}

Recall that Goldbach's Conjecture is as follows:

\begin{statement}
\emph{Every even integer greater than 2 can be expressed as the sum of two primes.}
\label{goldbachstatement}
\end{statement}

Because Goldbach's Conjecture is so simple to state, the Laconic program encoding the statement is also quite simple. \ It can be found in Appendix~\ref{sec:applac}. \ A detailed explanation of the compilation process, documentation for the Laconic language, and an explicit description of this Turing machine are available at~\cite{github}.

\section{A Turing Machine that Encodes the Riemann Hypothesis}

We present a \rmstatenumstate Turing machine that \emph{encodes the Riemann Hypothesis}; in other words, to know whether this machine halts is to know whether the Riemann Hypothesis is true. \ An explicit description of this machine can be found at~\cite{github}

The Riemann Hypothesis is traditionally stated as follows:

\begin{statement}
\emph{The Riemann zeta function has its zeros only at the negative even integers and the complex numbers with real part 1/2.}
\label{goldbachstatement}
\end{statement}

\subsection{Equivalent Statement}

Instead of encoding the Riemann zeta function into a Laconic program, it is simpler to use the following statement, which was shown by Lagarias~\cite{riemann} to be equivalent to the Riemann Hypothesis:

\begin{statement} \label{eq:riemann}
For all integers $n \ge 1$,
$$\left(\left(\sum_{k \le \delta(n)} \frac{1}{k}\right) - \frac{n^2}{2}\right)^2 < 36n^3$$
\end{statement}

The function $\delta(n)$ used in Statement~\ref{eq:riemann} is defined as follows: \\

\begin{nscenter}
$\eta(j) = p$ if $j = p^k$, $p$ is prime, $k$ is a positive integer \\
$\eta(j) = 1$ otherwise
\end{nscenter}
$$\delta(x) = \prod_{n<x}\prod_{j \le n} \eta(j)$$

\subsection{Implementation Methods}

Statement \ref{eq:riemann} is equivalent to the following statement, which involves only positive integers\footnotemark:
\footnotetext{Although it is not immediately obvious, $\frac{\delta(n)!}{k}$ is necessarily an integer for all $k \le \delta(n)$, and $\frac{\delta(n)!}{2}$ is an integer for all $n > 1$.}

$$l(n) < r(n)$$ for all positive integers $n$, where

$$l(n) = a(n)^2 + b(n)^2$$
$$r(n) = 36n^3(\delta(n)!)^2 + 2a(n)b(n)$$.
$$a(n) = \sum_{k \le \delta(n)} \frac{\delta(n)!}{k}$$
$$b(n) = \frac{n^2 \delta(n)!}{2}$$

To check the Riemann Hypothesis, our program computes $a(n)$, $b(n)$, $l(n)$, and $r(n)$, in that order, for each possible value of $n$. \ If $l(n) \ge r(n)$, our program halts.

\section{Laconic}

Laconic is a programming language designed to be both user-friendly and easy to compile down to parsimonious Turing machine descriptions.

Laconic is a strongly-typed language that supports recursive functions. Laconic compiles to an intermediate language called TMD. \ TMD programs are spread across multiple files and grouped into directories. \ TMD directories are meant to represent sequences of commands that could be given to a multi-tape, $3$-symbol Turing machine, using the Turing machine abstraction that allows the machine to read and write from one head at a time.

For an example of a Laconic program, see Appendix~\ref{sec:applac}. \ For an illustration of the compilation process, see Figure~\ref{fig:compilation}.

\section{TMD}

TMD is a programming language designed to help the user describe the behavior of a multi-tape, $3$-symbol Turing machine with a function stack. Each tape is infinite in one direction and supports three symbols: \texttt{\_}, \texttt{1}, and \texttt{E}. \ The blank symbol is \texttt{\_}: that is, \texttt{\_} is the only symbol that can appear on the tape an infinite number of times. \ The tape must always have the form $\texttt{\_}?(1|E)^+\texttt{\_}^{\infty}$; in other words, each tape must always contain a string of \texttt{1}'s and \texttt{E}'s of size at least 1, possibly preceded by a \texttt{\_} symbol, and necessarily followed by an infinite number of copies of the \texttt{\_} symbol.

What is the purpose of having a language like TMD as an intermediary between Laconic and a description of a single-tape machine? \ The concept of tapes in a multi-tape Turing machine and the concept of variables in standard imperative programming languages map to one another very nicely. \ The idea of the Laconic-to-TMD compiler is to encode the value of each variable on one tape. \ Then, each Laconic command that manipulates the value of one or more variables compiles down to a TMD function call that manipulates the tapes that correspond to those variables appropriately.

As an example, consider the following Laconic command: \\ \\
\texttt{a=b*c;} \\

This Laconic command assigns the value of \texttt{b*c} to the variable \texttt{a}. \ It compiles down to the following TMD function call: \\ \\
\texttt{function BUILTIN\_multiply a b c} \\

This function call will result in \texttt{BUILTIN\_multiply} being run on the three tapes \texttt{a}, \texttt{b}, and \texttt{c}. \ This will cause the symbols on tape \texttt{a} to take on a representation of an integer whose value is equal to $bc$.

In turn, the TMD code compiles directly to a string of bits that are written onto the tape at the start of the Turing machine's execution.

A TMD directory consists of three types of files:

\begin{enumerate}
\item The \texttt{functions} file. \ This file contains a list of the names of all the functions used by the TMD program. \ The top function in the file is pushed onto the stack at initialization. \ When this top function returns, the Turing machine halts.
\item The \texttt{initvar} file. \ This file contains the non-blank symbols that start in each register (or tape) at initialization.
\item Any files used to describe TMD functions. \ These files all end in a \texttt{.tfn} extension and only have any relevance to the compiled program if they show up in the functions file.
\end{enumerate}

\section{Compilation and Processing}
\label{sec:compandproc}

There are two ways to think about the layout of the tape symbols: with a $4$-symbol alphabet ($\{\texttt{\_}, \texttt{1}, \texttt{H}, \texttt{E}\}$, blank symbol \texttt{\_}), and with a $2$-symbol alphabet ($\{\texttt{a}, \texttt{b}\}$, blank symbol \texttt{a}). \ The $2$-symbol alphabet version is the one that's ultimately used for the results in this paper, since we advertised a Turing machine that used only two symbols. \ However, in nearly all parts of the Turing machine, the $2$-symbol version of the machine is a direct translation of the $4$-symbol version, according to the following mapping: \\ \\
$\texttt{\_} \leftrightarrow \texttt{aa}$ \\
$\texttt{1} \leftrightarrow \texttt{ab}$ \\
$\texttt{H} \leftrightarrow \texttt{ba}$ \\
$\texttt{E} \leftrightarrow \texttt{bb}$ \\

The sections that follow sometimes refer to the \texttt{ERROR} state. \ Transitions to the \texttt{ERROR} state should never be taken under any circumstances, and are useful for debugging purposes.

\subsection{Concept} \label{sec:ontape}

A directory of TMD functions is converted at compilation time to a string of bits to be written onto the tape, along with other states designed to interpret these bits. \ The resulting Turing machine has three main components, or \emph{submachines}:

\begin{enumerate}
\item The \emph{initializer} sets up the basic structure of the variable registers and the function stack.
\item The \emph{printer} writes down the binary string that corresponds to the compiled TMD code.
\item The \emph{processor} interprets the compiled binary, modifying the variable registers and the function stack as necessary.
\end{enumerate}

The Turing machine's control flow proceeds from the initializer to the the printer to the interpreter. \ In other words, initializer states point only to initializer states or to printer states, printer states point only to printer states or to interpreter states, and interpreter states point only to interpreter states or the \texttt{HALT} state.

This division of labor, while seemingly straightforward, actually constitutes an important idea. \ The problem of the compiler is to convert a higher-level representation---a machine with many tapes, a larger alphabet, and a function stack---to the lower-level representation of a machine with a single tape, a $2$-symbol alphabet and no function stack. \ The immediately obvious solution, and the one taught in every computability theory class as a proof of the equivalence of different kinds of Turing machines, is to have every ``state'' in the higher-level machine compile down to many states in the lower-level machine. 

While simple, this approach is suboptimal in terms of the number of states. \ As is nearly always true when designing systems to be parsimonious, the clue that improvement is possible lies in the presence of repetition. \ Each state transition in the higher-level machine is converted to a group of lower-level states with the same basic structure. \ Why not instead explain how to perform this conversion exactly once, and then apply the conversion many times?

This idea is at the core of the division of labor described previously. \ We begin by writing a description of the higher-level machine onto the tape, and then ``run'' the higher-level machine by reading what is on the tape with a set of states that understands how to interpret the encoded higher-level machine. \ We refer to this idea as \emph{on-tape processing}.

In this paper, we use TMD as the representation of the higher-level machine.\footnotemark
\footnotetext{Note that instead of TMD, the on-tape processing scheme could be used for any language, assuming the designer provides both a processor and an encoding for that language. \ We chose TMD because it made the interpreter easy to write, but other minimalist languages, like Unlambda~\cite{unlambda}, Brainf*ck~\cite{brainfuck}, or Iota and Jot~\cite{iota}, might be good candidates for parsimonious designs, with the additional advantage of being already known to some programmers! \ Thanks to Luke Schaeffer for this point.}
The printer writes the TMD program onto the tape, and the processor executes it. As a result of using this scheme, we incur a constant \emph{additive} overhead---we have to include the processor in our final Turing machine---but we avoid the constant \emph{multiplicative} overhead required for the na\"ive scheme.

Is this additive overhead small enough to be worth it? \ We found that it is. \ Our implementation of the processor requires 3,860 states. \ (See Section~\ref{sec:cost} for a detailed breakdown of the state cost by submachine.) \ In contrast to this additive overhead of 3,860, the na\"ive approach incurs a large multiplicative overhead that depends in part on how many states must be used to represent each higher-level state transition, and in part on how efficient an encoding scheme can be devised for the on-tape approach. \ The following table compares the performance of on-tape processing to the performance of an implementation that used the na\"ive approach. \ The comparison is shown for three kinds of machines: a machine that halts if and only if Goldbach's Conjecture is false, a machine that halts if and only if the Riemann Hypothesis is false, and a machine whose behavior is independent of ZFC.

\begin{center}
    \begin{tabular}{||c c c||}
    \hline
    Program & States (Na\"ive) & States (On-Tape Processing) \\ [0.5ex]
    \hline
    Goldbach & 7,902 & \gbstatenum\\
    \hline
    Riemann & 36,146 & \rmstatenum\\
    \hline
    ZFC & 340,943 & \statenum\\
    \hline
    \end{tabular}
\end{center}

As can be seen from this table, on-tape interpretation results in huge gains, particularly in large and complex programs.

The subsections that follow describe each of the three submachines---the initializer, the printer, and the processor---in greater detail.

\subsection{The Initializer}

The initializer starts by writing a counter onto the tape which encodes how many registers there will be in the program. \ Using the value in that counter, it creates each register, with demarcation patterns between registers, and unique identifiers for each register. \ Each register's value begins with the pattern of non-blank symbols laid out in the \texttt{initvar} file. \ The initializer also creates the program counter, which starts at 0, and the function stack, which starts out with only a single function call to the top function in the \texttt{functions} file.

\begin{figure}
\begin{center}
\includegraphics[scale=0.42]{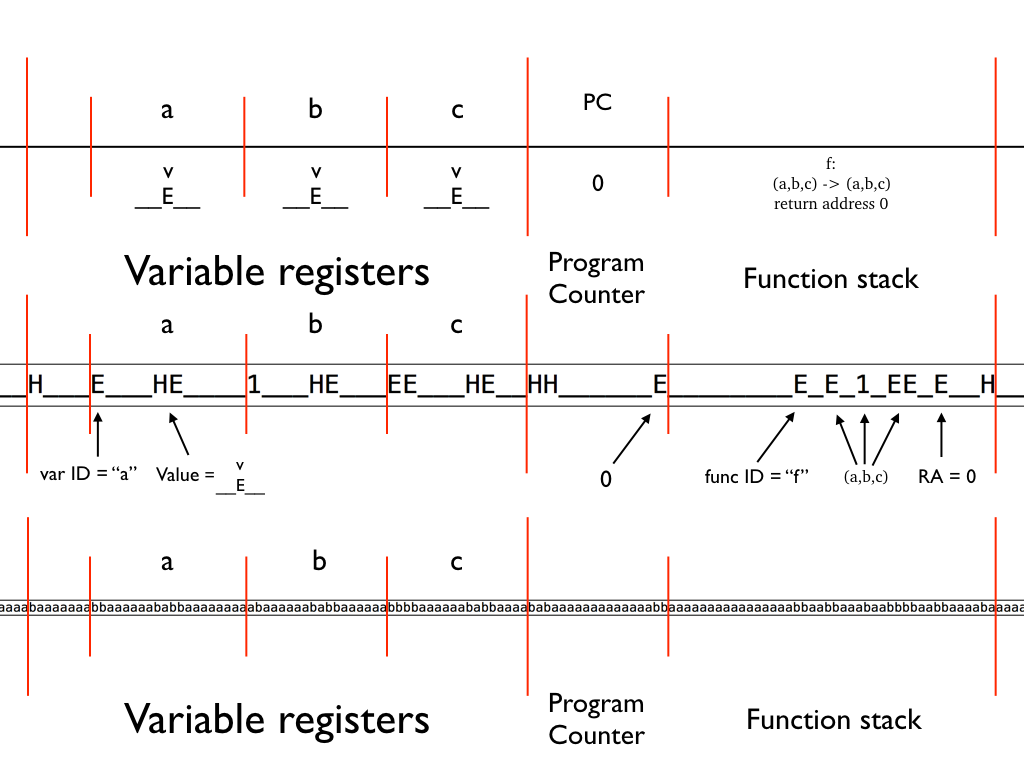}
\caption{The state of the Turing machine tape after the initializer completes. \ The TMD program being expressed in Turing machine form is described in full in Appendix~\ref{sec:apptmd}. \ The top bar is a high-level description of what each part of the Turing machine tape represents. \ The middle bar is an encoding of the tape in the standard $4$-symbol alphabet; the bottom bar is simply the translation of that tape into the $2$-symbol alphabet. \ For a more detailed explanation of how to interpret the tape patterns, see~\cite{github}. \label{fig:postinit}}
\end{center}
\end{figure}

Figure~\ref{fig:postinit} is a detailed diagram describing the tape's state when the initializer passes control to the printer.

\subsection{The Printer} \label{sec:introspect}

\subsubsection{Specification}

The printer writes down a long binary string which encodes the entirety of the TMD program onto the tape.

\begin{figure}
\begin{center}
\includegraphics[scale=0.42]{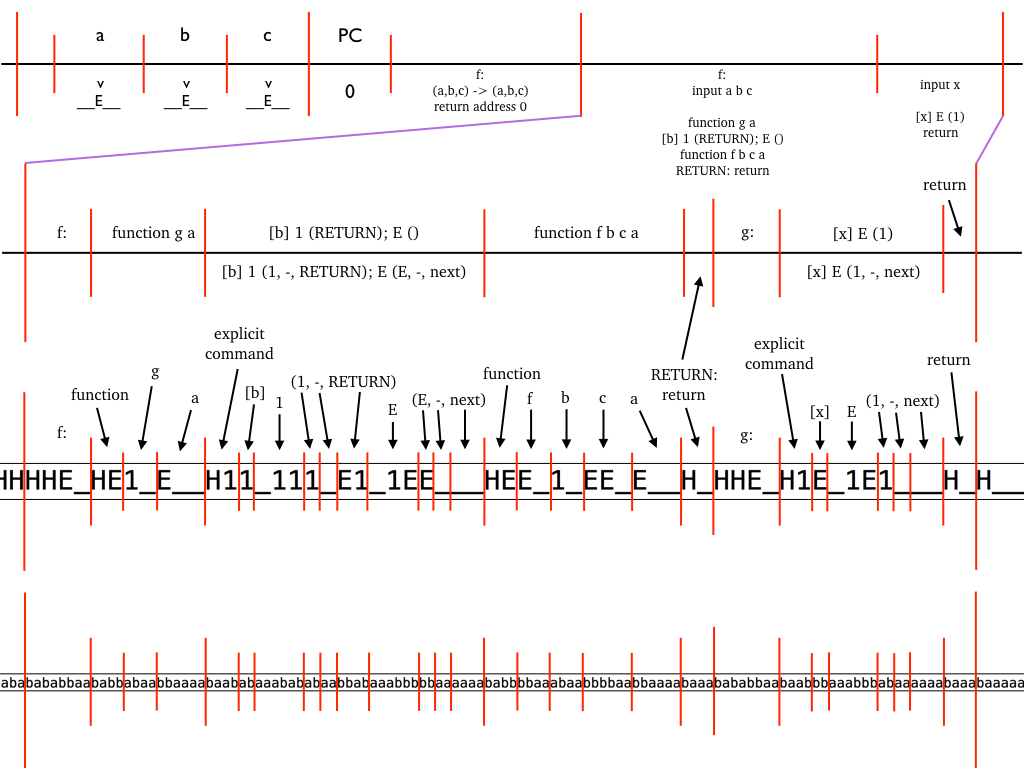}
\caption{The state of the Turing machine tape after the printer completes. \ The TMD program being expressed in Turing machine form is described in full in Appendix~\ref{sec:apptmd}. \ The top bar is a high-level description of the entire tape; unfortunately, at this point there are so many symbols on the tape that it is impossible to see everything at once. \ For a detailed view of the first two-thirds of the tape (registers, program counter, and stack), see Figure~\ref{fig:postinit}. \ The bottom three bars show a zoomed-in view of the program binary. \ From the top, the second bar gives a high-level description of what each part of the program binary means; the third bar gives the direct correspondence between $4$-symbol alphabet symbols on the tape and their meaning in TMD; the fourth and final bar gives the translation of the third bar into the $2$-symbol alphabet. \ For a more detailed explanation of the encoding of TMD into tape symbols, see~\cite{github}. \label{fig:postprog}}
\end{center}
\end{figure}

Figure~\ref{fig:postprog} shows the tape's state when the printer passes control to the processor.

\subsubsection{Introspection}

\begin{figure}
\begin{center}
\includegraphics[scale=0.28]{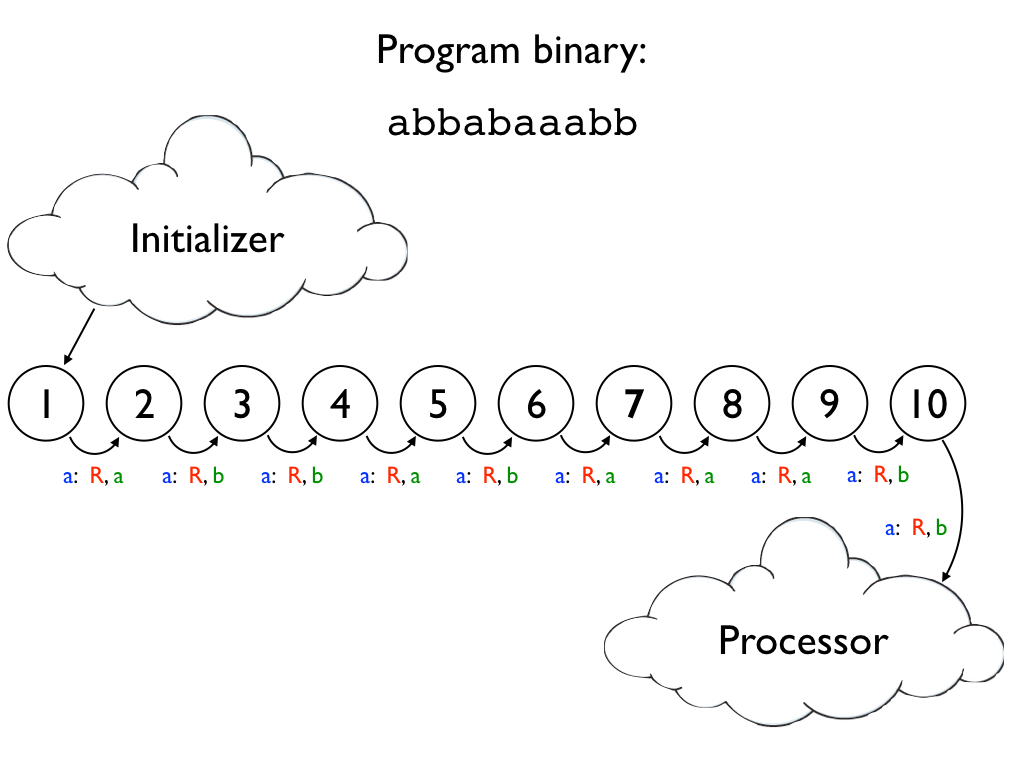}
\caption{A na\"ive implementation of the printer. \ In this example, the hypothetical program is ten bits long, and the printer uses ten states, one for each bit. \ In the diagram, the blue symbol is the symbol that is read on a transition, the red letter indicates the direction the head moves, and the green symbol indicates the symbol that it written. \ Note the lack of transitions on reading a \texttt{b}; this is because in this implementation, the printer will only ever read the blank symbol, which is \texttt{a}, since the head is always proceeding to untouched parts of the tape. \ It therefore makes no difference what behavior the Turing machine adopts upon reading a \texttt{b} in states 1-10 (and therefore \texttt{b} transitions are presumed to lead to the \texttt{ERROR} state) \label{fig:naiveprog}}
\includegraphics[scale=0.28]{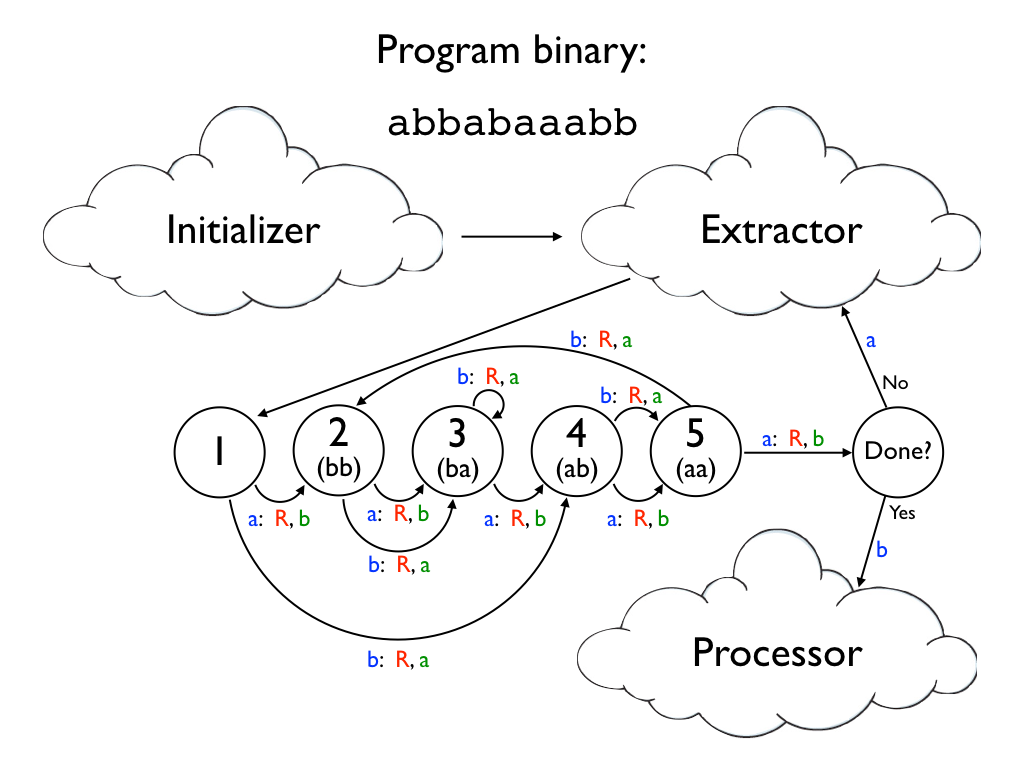}
\caption{An introspective implementation of the printer. \ In this example, the hypothetical program is $k=10$ bits long, and so the word size must be 2 (since $w=2$ is the largest $w$ such that $w2^w \le 10$). \ There are therefore $n_w = \left \lceil{\frac{k}{w}}\right \rceil = 5$ data states, each encoding two bits. \ The \texttt{b} transitions carry the information about the encoding; note that each one only points to one of the last four data states. \ The last four data states have in parentheses what word we mean to encode if we point to them. \label{fig:introspectprog}}
\end{center}
\end{figure}

Writing down a long binary string onto a Turing machine tape in a parsimonious fashion is not as straightforward as it might initially appear. \ The first idea that comes to mind is simply to use one state per symbol, with each state pointing to the next, as shown in Figure~\ref{fig:naiveprog}.

On closer examination, however, this approach is quite wasteful for all but the smallest binary files. \ Every \texttt{a} transition points to the next state in the sequence, and none of the \texttt{b} transitions are used at all! \ Indeed, the only information-bearing part of the state is the single bit contained in the choice of which symbol to write. \ But in theory, far more information than that could be encoded in each state. \ In a machine with $n$ states, each state could contain $2(\log_2(n) + 1)$ bits of information, because each of its two transitions could point to any of the $n$ states, and write either an \texttt{a} or a \texttt{b} onto the tape. \ Of course, this is only in theory; in practice, to extract the information contained in therefore Turing machine's states and translate it into bits on the tape is nontrivial.

We will use a scheme originally conceived by Ben-Amram and Petersen~\cite{benamram} and refined further and suggested to us by Luke Schaeffer. \ It does not achieve the optimal theoretical encoding described above, but is relatively simple to implement and understand, and is within a factor of $2$ of optimal for large binary strings. \ Schaeffer named Turing machines that use this idea \emph{introspective}.

Introspection works as follows. \ If the binary string contains $k$ bits, then let $w$ be the \emph{word size}. The word size $w$ takes the largest value it can such that $w2^w \le k$. \ We can split the binary string into $n_w = \left \lceil{\frac{k}{w}}\right \rceil$ \emph{words} of $w$ bits each (we can pad the last word with the blank symbol). \ In our scheme, each word in the bit-string is represented by a \emph{data state}. \ Each data state points to the state representing the next word in the sequence for its \texttt{a} transition, but which state the \texttt{b} transition points to encodes the next word. \ Every \texttt{b} transition points to one of the last $2^w$ data states, thereby encoding $w$ bits of information.

Of course, the encoding is useless until we specify how to extract the encoded bit-string from the data states. \ The extraction scheme works as follows. \ To query the $i^\textrm{th}$ data state for the bits it encodes, we run the data states on the string $\texttt{a}^{i-1}\texttt{b}\texttt{a}^{\infty}$ (a string of $i-1$ \texttt{a}'s followed by a \texttt{b} in the $i^\textrm{th}$ position). \ After running the data states on that string, what remains on the tape is the string $\texttt{b}^{i-1}\texttt{a}\texttt{b}^r\texttt{a}^{\infty}$, assuming that the $i^\textrm{th}$ data state pointed to the $r^\textrm{th}$-to-last data state. \ Thus, what we're left with is essentially a unary encoding of the ``value'' of the word in binary. \ Thus, the job of the extractor is to set up a binary counter which removes one \texttt{b} at a time and increments the counter appropriately. \ Then, afterward, the extractor reverts the tape back to the form $\texttt{a}^i\texttt{b}\texttt{a}^{\infty}$, shifts all symbols on the tape over by $w$ bits, and repeats the process. \ Finally, when the state beyond the last data state sees a \texttt{b} on the tape, we know that the process has completed, and we can pass control to the processor. \ Figure~\ref{fig:introspectprog} shows the whole procedure.

How much have we gained by using introspection for encoding the program binary, instead of the na\"ive approach? \ It depends on how large the program binary is. \ Using introspection incurs an $O(\log k)$ \emph{additive} overhead, because we have to include the extractor in our machine. \ (Our implementation of the extractor takes $10w + 17$ states. It's possible to build a constant-size extractor, but it's not worth it for our value of $w$) \ In return, we save a \emph{multiplicative} factor of $w$ (which scales with $\log k$) on the number of data states needed.

This is plainly not worth it for the $10$-bit example binary shown in Figs.~\ref{fig:naiveprog} and~\ref{fig:introspectprog}. \ For that binary, we require $69$ additional states for the extractor in order to save $5$ data states. \ For real programs, however, it is worth it, as can be seen from the following table.

\begin{center}
    \begin{tabular}{||c c c c c c c||}
    \hline
    Program & Binary Size & $w$ & $n_w$ & Extractor Size & States (Na\"ive) & States (Introspective) \\ [0.5ex]
    \hline\hline
    Example TMD & 116 & 4 & 29 & 57 & 116 & 86 \\
    \hline
    Goldbach & 4,964 & 9 & 552 & 107 & 4,964 & 659 \\
    \hline
    Riemann & 9,532 & 10 & 1,024 & 117 & 9,532 & 1,141 \\
    \hline
    ZFC & 38,864 & 11 & 3,534 & 127 & 38,864 & 3,661 \\
    \hline
    \end{tabular}
\end{center}

One minor detail concerns the numbers presented for the Riemann program. \ Ordinarily, with a binary of size 9,532, we would opt to split the program into 1,060 words of 9 bits each plus a 107-state extractor, since 9 is the greatest $w$ such that $w2^w <$ 9,532. \ But because 9,532 is so close to the ``magic number'' 10,240, it's actually more parsimonious to pad the program with copies of the blank symbol until it's 10,240 bits long, and split it into 1,024 words of $10$ bits each plus a $117$-state extractor.

\subsection{The Processor}

The processor's job is to interpret the code written onto the tape and modify the variable registers and function stack accordingly. \ The processor does this by the following sequence of steps:  \\ \\
START:
\begin{enumerate}
\item Find the function call at the top of the stack. Mark the function $f$ in the code whose ID matches that of the top function call.
\item Read the current program counter. Mark the line of code $l$ in $f$ whose line number matches the program counter.
\item Read $l$. Depending on what type of command $l$ is, carry out one of the following three lists of tasks.
\end{enumerate}

\noindent IF $l$ IS AN EXPLICIT TAPE COMMAND:
\begin{enumerate}
\item Read the variable name off $l$. Index the variable name into the list of variables in the top function on the stack. This list of variables corresponds to the mapping between the function's local variables and the register names.
\item Match the indexed variable to its corresponding register $r$. Mark $r$. Read the symbol $s_r$ to the right of the head marker in that register.
\item Travel back to $l$, remembering the value of $s_r$ using states. Find and mark the reaction $x$ corresponding to the symbol. See what symbol $s_w$ should be written in response to reading $s_r$.
\item Travel back to $r$, remembering the value of $s_w$ using states. Replace $s_r$ with $s_w$.
\item Travel back to $x$. See which direction $d$ the head should move in response to reading $s_r$.
\item Travel back to $r$, remembering the value of $d$ using states. Move the head marker accordingly.
\item Travel back to $x$. See if a jump is specified. If a jump is specified, copy the jump address onto the program counter. Otherwise, increment the program counter by 1.
\item Go back to START.
\end{enumerate}

\noindent IF $l$ IS A FUNCTION CALL:
\begin{enumerate}
\item Write the function's name to the top of the stack.
\item For each variable in the function call, index the variable name into the list of variables in the top function on the stack. This list of variables corresponds to the mapping between the function's local variables and the register names. Push the corresponding register names in the order that they correspond to the variables in the function call.
\item Copy the current program counter to the return address of the newborn function call at the top of the stack.
\item Replace the current program counter with 0 (meaning ``read the first line of code'').
\item Go back to START.
\end{enumerate}

\noindent IF $l$ IS A RETURN STATEMENT:
\begin{enumerate}
\item Replace the current program counter with $f$'s return address.
\item Increment the program counter by 1.
\item Erase the call to $f$ from the top of the stack.
\item Check if the stack is now empty. If so, halt.
\item Go back to START.
\end{enumerate}

\subsection{Cost Analysis} \label{sec:cost}

It's worthwhile to analyze the relative contributions of the initializer, the printer, and the processor to the machine's final state count. \ The following table lists the number of states in each submachine for each of the four different TMD programs under discussion.

\begin{center}
    \begin{tabular}{||c c c c c||}
    \hline
    Program & Initializer & Printer & Processor & Total \\ [0.5ex]
    \hline\hline
    Example TMD & 349 & 86 & 3,860 & 4,295 \\
    \hline
    Goldbach & 369 & 659 & 3,860 & \gbstatenum \\
    \hline
    Riemann & 371 & 1,141 & 3,860 & \rmstatenum \\
    \hline
    ZFC & 389 & 3,661 & 3,860 & \statenum \\
    \hline
    \end{tabular}
\end{center}

As can be seen from this table, the processor makes the largest contribution to all four programs. \ Improving the processor, therefore, is probably the best approach for improving upon the bounds we present. \ Equally clear, however, is that for programs more complicated than the ones presented here, the cost of the printer will grow almost linearly but the cost of the processor will stay the same. \ The cost of the initializer grows very slightly with the complexity of programs because of the need to initialize additional registers.

Improving the printer, and with it the TMD and Laconic languages, is probably the best approach for reducing state count for very large and complex programs.

\section{Future Work}

This paper still leaves a three orders-of-magnitude gap between the smallest $n$, namely \statenumcomma for which $BB(n)$ is known to be independent of ZF set theory, and the largest $n$, namely $4$, for which $BB(n)$ is known to be determinable. \ We regard it as a fascinating problem to pin down the truth here: for example, is it conceivable that $BB(10)$ or even $BB(6)$ might be independent of ZF? \ If so, that would arguably force a qualitative change in our understanding of the G\"{o}del incompleteness phenomenon---showing that incompleteness from strong set theories rears its head for much simpler arithmetical questions than had previously been known.

A more immediate question is how much further $Z$'s state count can be reduced. \ We are optimistic about the possibility of further reductions. \ For example, one could adapt the processor-printer model to use a better language than TMD. \ Ideally, one wants a language whose processor contains fewer states than TMD's, and whose typical programs are \emph{also} shorter than TMD programs. \ A few ideas have been proposed for this~\cite{comments}, many of them related in some way to lambda calculus.

Other future work might involve further use of our Laconic language to upper-bound the `complexities' of mathematical statements and algorithms, in as standardized and model-independent a way as possible. \ Perhaps Laconic could be used to measure the complexity of other well-known conjectures, or even to compare different algorithms for solving the same problem to each other (e.g., to try to quantify the notion that an insertion sort is simpler than a merge sort)!

\section{Acknowledgements}

We thank Prof.\ Harvey Friedman for having done the crucial theoretical work that made this project feasible. \ Prof.\ Friedman was endlessly available over email, and provided us with detailed clarifications when we needed them.

We thank Luke Schaeffer for his early help, as well as his help designing introspective Turing machines.

We thank Alex Arkhipov for introducing us to the term ``code golfing.''

We thank the commenters on Scott Aaronson's blog~\cite{comments} for their ideas and suggestions.

Supported by an Alan T.\ Waterman Award from the National Science Foundation, under grant no.\ 1249349.

\begin{appendices}

\section{Example Laconic Program: Goldbach's Conjecture} \label{sec:applac}

The following is an example Laconic program, which compiles down to the Turing machine $G$ mentioned in Section~\ref{sec:g} (which halts if and only Goldbach's Conjecture is false).

{ \scriptsize \tt
\begin{lstlisting}
func zero(x) {
    x = 0;
    return;
}

func one(x) {
    x = 1;
    return;
}

func incr(x) {
    x = x + 1;
    return;
}

/* Computes x modulo y */
func modulus(x, y, out) {
    out = x;
    
    while (out >= y) {
        out = out - y;
    }
    
    return;
}

func assignXtoYminusX(x, y) {
    x = y - x;
    return;
}

/* Figures out if x is prime, and puts the output in y */
/* Does not modify x, modifies y */
func isPrime(x, h, y) {
    if (x == 1) {
        zero(y);
        return;
    }
    
    y = 2;
    
    while (x > y) {
        modulus(x, y, h);
    
        if (h == 0) {
            zero(y);
            return;
        }
        incr(y);
    }
    
    return;
}

int evenNumber;
int primeCounter;
int isThisOnePrime;
int foundSum;
int h;

evenNumber = 2;
one(foundSum);

while (foundSum) {
    zero(foundSum);
    evenNumber = evenNumber + 2;
    one(primeCounter);
    
    while (primeCounter < evenNumber) {    
        isPrime(primeCounter, h, isThisOnePrime);
        
        if (isThisOnePrime) {
            assignXtoYminusX(primeCounter, evenNumber);
            isPrime(primeCounter, h, isThisOnePrime);
            assignXtoYminusX(primeCounter, evenNumber);
            
            if (isThisOnePrime) {
                print evenNumber;
                print primeCounter;
                
                one(foundSum);
            }
        }
        
        incr(primeCounter);  
    }         
}

halt;
\end{lstlisting}
}

For detailed documentation of the Laconic programming language, see~\cite{github}. \ To find this file specifically, navigate to \texttt{parsimony/src/laconic/laconic\_files/goldbach.lac} at~\cite{github}.

\section{Example TMD Program} \label{sec:apptmd}

The following is an example TMD directory, which compiles down to a binary string to be written on a Turing machine tape. \ It's the example used in illustrations throughout this paper, most notably in the example compilation shown in Figs.~\ref{fig:postinit} and~\ref{fig:postprog}. \ The program calls itself recursively three times until the starting symbol on each tape, \texttt{E}, is replaced with a \texttt{1}, at which point the program halts.

This TMD directory is called \texttt{example\_tmd\_dir}, and contains four files: \texttt{f.tmd}, \texttt{g.tmd}, \texttt{initvar}, and \texttt{functions}. \\ \\

\texttt{f.tmd}:
{ \scriptsize \tt
\begin{lstlisting}
input a b c

// Recursively writes a 1 on every tape.

function g a 
[b] 1 (RETURN); E ()
function f b c a
RETURN: return
\end{lstlisting}
}
\texttt{g.tmd}:
{ \scriptsize \tt
\begin{lstlisting}
input x

// Writes a 1 on the input tape.

[x] E (1)
return
\end{lstlisting}
}

\texttt{functions}:
{ \scriptsize \tt
\begin{lstlisting}
f
g
\end{lstlisting}
}

\texttt{initvar}:
{ \scriptsize \tt
\begin{lstlisting}
E
\end{lstlisting}
}

For detailed documentation of the TMD programming language, see~\cite{github}. \ To find this directory specifically, navigate to \texttt{parsimony/src/tmd/tmd\_dirs/example\_tmd\_dir/} at~\cite{github}.

\section{Explicit Listing of $Z$} \label{sec:explicitz}

\begin{figure}[h]
\begin{center}
\includegraphics[scale=0.4]{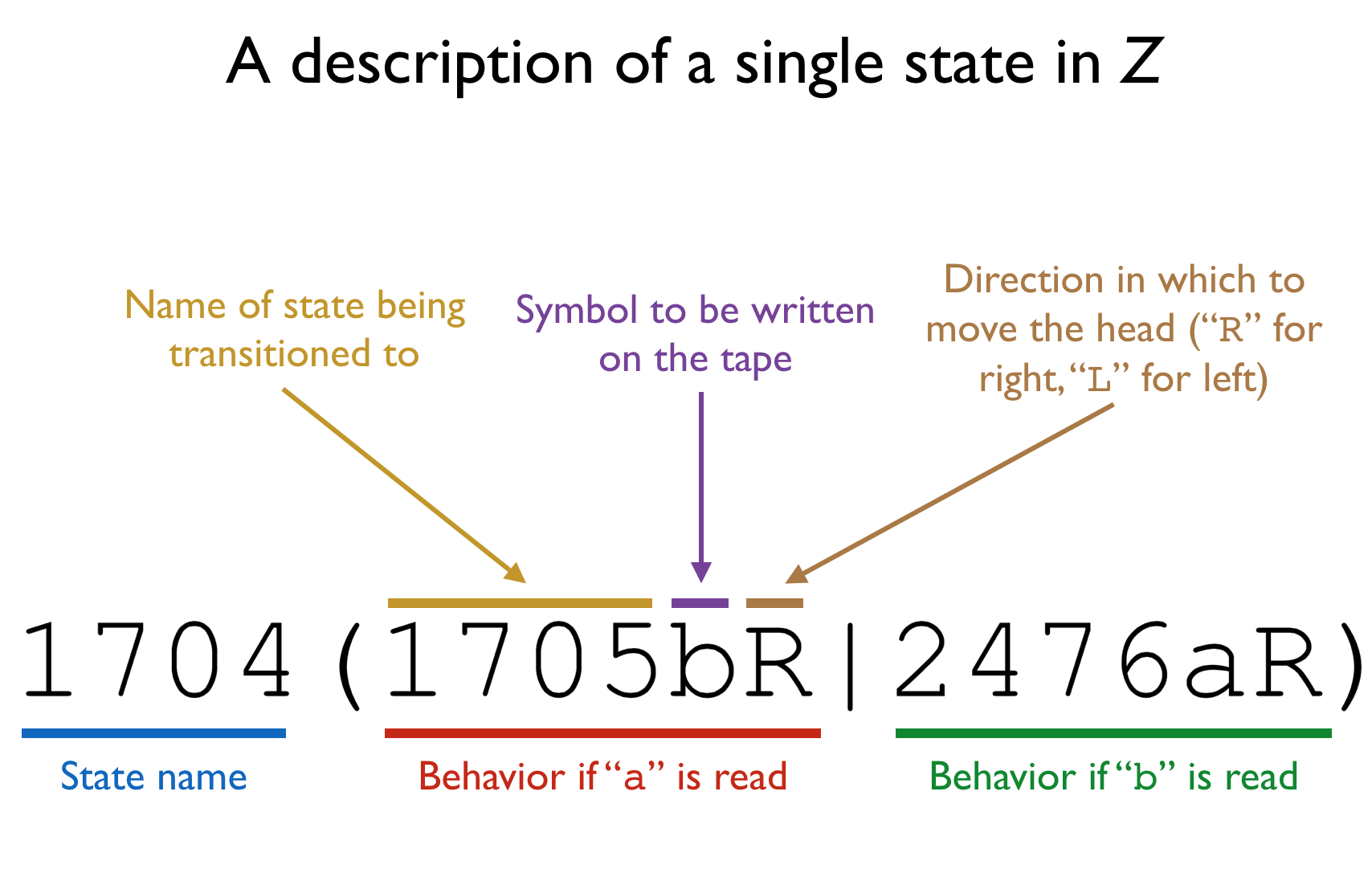}
\caption{This figure explains how to read a description of a single state. \ Note that ``\texttt{ERROR-}'' or ``\texttt{HALT--}'' denote transitions to the \texttt{ERROR} or \texttt{HALT} states, respectively (no further information is provided because what symbol is written and which direction the head moves are at that point irrelevant). \label{fig:syntax}}
\end{center}
\end{figure}

We present below an explicit listing of $Z$. \ For a more easily readable version of $Z$, complete with descriptive state names, see~\cite{github}.

We ran this Turing machine for 10,000,000,000 steps (more than half a day on our simulators) and within that time it did not halt. \ We note, however, that $Z$ was designed for parsimony rather than efficiency, and that this ``experiment'' is of little consequence! \ We similarly ran a Turing machine designed to test the conjecture that all perfect squares are less than $5$, and it ran for 2,895,083,899 steps (a couple hours on our simulator) before it found the counterexample $9$ and halted.

Figure~\ref{fig:syntax} explains how to interpret the description shown below. \ In addition, note the following:

\begin{enumerate}

\item The tape has a $2$-symbol alphabet, with tape symbols $\{\texttt{a}, \texttt{b}\}$ and blank symbol \texttt{a} (in other words, \texttt{a} is the only symbol that can appear an infinite times on the tape).
\item The start state of $Z$ is \texttt{0000}.
\item $Z$ will never transition to the \texttt{ERROR} state. \ Any transition to the \texttt{ERROR} state could be replaced by a transition to any other state (including \texttt{HALT}) and the Turing machine's behavior would remain identical.
\item $Z$ contains only one transition to the \texttt{HALT} state, out of state \zhaltstate.

\end{enumerate}

{\fontsize{4.5}{4}\selectfont\tt\noindent0000(0001bR|ERROR-) 0001(0004bR|ERROR-) 0002(0003bR|ERROR-) 0003(0012aR|0012bR) 0004(0005aR|ERROR-) 0005(0006bR|ERROR-) 0006(0007bR|ERROR-) 0007(0008bR|ERROR-) 0008(0009bR|ERROR-) 0009(0010bR|ERROR-) 0010(0011bR|ERROR-) 0011(0002bR|ERROR-) 0012(0013aR|ERROR-) 0013(0014aR|0014bR) 0014(0015aR|ERROR-) 0015(0057aR|0057bR) 0016(0017bR|ERROR-) 0017(0018bR|ERROR-) 0018(0019aR|ERROR-) 0019(0020aR|0020bR) 0020(0021aR|ERROR-) 0021(0022aR|0022bR) 0022(0023aR|ERROR-) 0023(0024aR|0024bR) 0024(0025aR|ERROR-) 0025(0067aR|0067bR) 0026(0027aR|0032bR) 0027(0028aL|0030bL) 0028(0029aL|0029bL) 0029(0026aL|0026bL) 0030(0031aL|0031bL) 0031(0026aL|0026bL) 0032(0033aL|0035bL) 0033(0034aL|0034bL) 0034(0037aL|0037bL) 0035(0036aL|0036bL) 0036(0026aL|0026bL) 0037(0038aR|0041bR) 0038(0044aR|0039bL) 0039(0040bR|0040bR) 0040(0049aL|0049bL) 0041(ERROR-|0042bL) 0042(0043aL|0043aL) 0043(0037aL|0037bL) 0044(0045aR|0048bR) 0045(ERROR-|0046aL) 0046(0047aR|0047aR) 0047(0049aL|0049bL) 0048(0071aR|ERROR-) 0049(0050aR|0051bR) 0050(0049aR|0049bR) 0051(0052aR|0049bR) 0052(0053aR|0054bR) 0053(0052aR|0052bR) 0054(0055aL|0052bR) 0055(0056aR|0056aR) 0056(0012aR|0012bR) 0057(0058aR|ERROR-) 0058(0059aR|0059bR) 0059(0060aR|ERROR-) 0060(0061aR|0061bR) 0061(0062aR|ERROR-) 0062(0063aR|0063bR) 0063(0064aR|ERROR-) 0064(0065aR|0065bR) 0065(0066aR|ERROR-) 0066(0016aR|0016bR) 0067(0068bR|ERROR-) 0068(0069bR|ERROR-) 0069(0070bL|ERROR-) 0070(0026aL|0026bL) 0071(0072aR|0075bR) 0072(0071aR|0073bL) 0073(0074aR|0074bR) 0074(0078aL|0078bL) 0075(ERROR-|0076bL) 0076(0077aR|0077bR) 0077(0078aL|0078bL) 0078(0079aR|0082bR) 0079(0080aL|0078bR) 0080(0081bR|0081bR) 0081(0083aR|0083bR) 0082(0078aR|0078bR) 0083(0084aR|0085bR) 0084(0083aR|0083bR) 0085(0086aL|0083bR) 0086(0087aL|0087bL) 0087(0088aL|0088bL) 0088(0089aR|0094bR) 0089(0090aL|0092bL) 0090(0091aR|0091bR) 0091(0099aL|0099bL) 0092(0093aL|0093bL) 0093(0088aL|0088bL) 0094(0095aL|0097bL) 0095(0096aL|0096bL) 0096(0088aL|0088bL) 0097(0098aL|0098bL) 0098(0088aL|0088bL) 0099(0100aR|0105bR) 0100(0101aL|0103bL) 0101(0102aL|0102bL) 0102(0099aL|0099bL) 0103(0104aR|0104bR) 0104(0110aL|0110bL) 0105(0106aL|0108bL) 0106(0107aR|0107aR) 0107(0139aL|0139bL) 0108(0109aR|0109bR) 0109(0110aL|0110bL) 0110(0111aR|0116bR) 0111(0112bL|0114bL) 0112(0113bL|0113bL) 0113(0130aL|0130bL) 0114(0115bL|0115bL) 0115(0110aL|0110bL) 0116(ERROR-|0117bL) 0117(0118aR|0118aR) 0118(0119aL|0119bL) 0119(0120aR|0125bR) 0120(0121aL|0123bL) 0121(0122aR|0122bR) 0122(0130aL|0130bL) 0123(0124aL|0124bL) 0124(0119aL|0119bL) 0125(0126aL|0128bL) 0126(0127aL|0127bL) 0127(0119aL|0119bL) 0128(0129aL|0129bL) 0129(0119aL|0119bL) 0130(0131aR|0136bR) 0131(0132aL|0134bL) 0132(0133aL|0133bL) 0133(0130aL|0130bL) 0134(0135aR|0135bR) 0135(0088aL|0088bL) 0136(ERROR-|0137bL) 0137(0138aR|0138bR) 0138(0088aL|0088bL) 0139(0140aR|0143bR) 0140(0139aR|0141bL) 0141(0142aR|0142bR) 0142(0146aL|0146bL) 0143(ERROR-|0144bL) 0144(0145aR|0145bR) 0145(0146aL|0146bL) 0146(0147aR|0150bR) 0147(0148aL|0146bR) 0148(0149aR|0149bR) 0149(0071aL|0071bL) 0150(0151aL|0146bR) 0151(0152aR|0152aR) 0152(0153aR|0153bR) 0153(0154aR|ERROR-) 0154(0155aR|0155bR) 0155(0156bL|ERROR-) 0156(0157aL|0157bL) 0157(0158aR|0163bR) 0158(0159aL|0161bL) 0159(0160aL|0160bL) 0160(0157aL|0157bL) 0161(0162aR|0162bR) 0162(0166aL|0166bL) 0163(0197aR|0164bL) 0164(0165aR|0165bR) 0165(0166aL|0166bL) 0166(0167aR|0172bR) 0167(0168aL|0170bL) 0168(0169bL|0169bL) 0169(0177aL|0177bL) 0170(0171aL|0171bL) 0171(0166aL|0166bL) 0172(0173aL|0175bL) 0173(0174aL|0174bL) 0174(0166aL|0166bL) 0175(0176aL|0176bL) 0176(0166aL|0166bL) 0177(0178aR|0183bR) 0178(0179aL|0181bL) 0179(0180aL|0180bL) 0180(0177aL|0177bL) 0181(0182aR|0182bR) 0182(0186aL|0186bL) 0183(ERROR-|0184bL) 0184(0185aR|0185bR) 0185(0186aL|0186bL) 0186(0187aR|0192bR) 0187(0188aL|0190bL) 0188(0189aR|0189bR) 0189(0157aL|0157bL) 0190(0191aL|0191bL) 0191(0186aL|0186bL) 0192(0193aL|0195bL) 0193(0194aL|0194bL) 0194(0186aL|0186bL) 0195(0196aL|0196bL) 0196(0186aL|0186bL) 0197(0198aR|0199bR) 0198(0197aR|0197bR) 0199(0200aR|0197bR) 0200(0201aR|0204bR) 0201(0202aL|0197bR) 0202(0203bR|0203bR) 0203(0205aR|0205bR) 0204(0200aR|0197bR) 0205(0206aR|ERROR-) 0206(0207aR|0207bR) 0207(0208aR|ERROR-) 0208(0209aR|0209bR) 0209(0210aR|ERROR-) 0210(0219aR|0219bR) 0211(0212bR|ERROR-) 0212(0213bR|ERROR-) 0213(0214aR|ERROR-) 0214(0233aR|0233bR) 0215(0216bR|ERROR-) 0216(0217bR|ERROR-) 0217(0218bL|ERROR-) 0218(0263aL|0263bL) 0219(0220aR|ERROR-) 0220(0221aR|0221bR) 0221(0222aR|ERROR-) 0222(0223aR|0223bR) 0223(0224aR|ERROR-) 0224(0225aR|0225bR) 0225(0226aR|ERROR-) 0226(0227aR|0227bR) 0227(0228aR|ERROR-) 0228(0229aR|0229bR) 0229(0230aR|ERROR-) 0230(0231aR|0231bR) 0231(0232aR|ERROR-) 0232(0211aR|0211bR) 0233(0234bR|ERROR-) 0234(0237bR|ERROR-) 0235(0236bR|ERROR-) 0236(0245aR|0245bR) 0237(0238aR|ERROR-) 0238(0239bR|ERROR-) 0239(0240bR|ERROR-) 0240(0241bR|ERROR-) 0241(0242bR|ERROR-) 0242(0243bR|ERROR-) 0243(0244aR|ERROR-) 0244(0235bR|ERROR-) 0245(0246aR|ERROR-) 0246(0247aR|0247bR) 0247(0248aR|ERROR-) 0248(0249aR|0249bR) 0249(0250aR|ERROR-) 0250(0251aR|0251bR) 0251(0252aR|ERROR-) 0252(0253aR|0253bR) 0253(0254aR|ERROR-) 0254(0255aR|0255bR) 0255(0256aR|ERROR-) 0256(0257aR|0257bR) 0257(0258aR|ERROR-) 0258(0259aR|0259bR) 0259(0260aR|ERROR-) 0260(0261aR|0261bR) 0261(0262aR|ERROR-) 0262(0215aR|0215bR) 0263(0264aR|0269bR) 0264(0265aL|0267bL) 0265(0266aL|0266bL) 0266(0263aL|0263bL) 0267(0268aL|0268bL) 0268(0263aL|0263bL) 0269(0270aL|0272bL) 0270(0271aL|0271bL) 0271(0274aL|0274bL) 0272(0273aL|0273bL) 0273(0263aL|0263bL) 0274(0275aR|0278bR) 0275(0281aR|0276bL) 0276(0277bR|0277bR) 0277(0288aL|0288bL) 0278(ERROR-|0279bL) 0279(0280aL|0280aL) 0280(0274aL|0274bL) 0281(0282aR|0285bR) 0282(ERROR-|0283aL) 0283(0284aR|0284aR) 0284(0288aL|0288bL) 0285(0286aL|ERROR-) 0286(0287aR|0287aR) 0287(0300aR|0300bR) 0288(0289aR|0290bR) 0289(0288aR|0288bR) 0290(0291aR|0288bR) 0291(0292aR|0293bR) 0292(0291aR|0291bR) 0293(0294aL|0291bR) 0294(0295aR|0295aR) 0295(0296aR|0296bR) 0296(0297bR|ERROR-) 0297(0298bR|ERROR-) 0298(0299bL|ERROR-) 0299(0263aL|0263bL) 0300(0301aR|0302bR) 0301(0300aR|0300bR) 0302(0300aR|0303bR) 0303(0304aR|0305bR) 0304(0386aR|0386bR) 0305(0386aR|0386bR) 0306(0307bR|ERROR-) 0307(0308aR|0308bR) 0308(0309aR|0310bR) 0309(0308aR|0308bR) 0310(0311aL|0308bR) 0311(0312aL|0312bL) 0312(0313aL|0313bL) 0313(0314aR|0317bR) 0314(0338aR|0315bL) 0315(0316bL|0316bL) 0316(0313aL|0313bL) 0317(0338aR|0318bL) 0318(0319aR|0319aR) 0319(0360aL|0360bL) 0320(0321aR|0324bR) 0321(0320aR|0322aL) 0322(0323aR|0323aR) 0323(0329aR|0329bR) 0324(0325aL|0327aL) 0325(0326aR|0326aR) 0326(0347aR|0347bR) 0327(0328aR|0328aR) 0328(0338aR|0338bR) 0329(0330aR|0333bR) 0330(0331bL|0329bR) 0331(0332aR|0332aR) 0332(0320aR|0320bR) 0333(0334bL|0336bL) 0334(0335aR|0335aR) 0335(0347aR|0347bR) 0336(0337aR|0337aR) 0337(0338aR|0338bR) 0338(0339aR|0344bR) 0339(0340bL|0342bL) 0340(0341bR|0341bR) 0341(0320aR|0320bR) 0342(0343bR|0343bR) 0343(0329aR|0329bR) 0344(0345bL|0338bR) 0345(0346bR|0346bR) 0346(0347aR|0347bR) 0347(0348bL|ERROR-) 0348(0349aL|0349bL) 0349(0350aR|0355bR) 0350(0351aL|0353bL) 0351(0352aL|0352bL) 0352(0349aL|0349bL) 0353(0354aR|0354bR) 0354(0313aL|0313bL) 0355(0356aL|0358bL) 0356(0357aR|0357aR) 0357(0371aR|0371bR) 0358(0359aL|0359bL) 0359(0349aL|0349bL) 0360(0361aR|0366bR) 0361(0362aL|0364bL) 0362(0363aL|0363bL) 0363(0313aL|0313bL) 0364(0365aL|0365bL) 0365(0360aL|0360bL) 0366(0367aL|0369bL) 0367(0368aR|0368aR) 0368(0371aR|0371bR) 0369(0370aL|0370bL) 0370(0360aL|0360bL) 0371(0372aR|0375bR) 0372(0373aL|0371bR) 0373(0374bR|0374bR) 0374(0308aR|0308bR) 0375(0376aL|0371bR) 0376(0377aR|0377aR) 0377(0378aR|0378bR) 0378(0379bR|ERROR-) 0379(0380bR|ERROR-) 0380(0381aR|ERROR-) 0381(0382aR|0382bR) 0382(0383aR|ERROR-) 0383(0384aR|0384bR) 0384(0385bR|ERROR-) 0385(0389aR|0389bR) 0386(0387aR|0388bR) 0387(0306aR|0306bR) 0388(0306aR|0306bR) 0389(0390aR|ERROR-) 0390(0391bL|ERROR-) 0391(0395aR|ERROR-) 0392(0393aL|0392aL) 0393(0393aL|0394bR) 0394(4050aR|ERROR-) 0395(0396bR|2559aR) 0396(0397bR|3077aR) 0397(0398bR|2269aR) 0398(0399bR|2637aR) 0399(0400bR|3293aR) 0400(0401bR|2588aR) 0401(0402bR|3610aR) 0402(0403bR|2065aR) 0403(0404bR|2265aR) 0404(0405bR|1939aR) 0405(0406bR|3411aR) 0406(0407bR|2579aR) 0407(0408bR|2234aR) 0408(0409bR|2070aR) 0409(0410bR|2234aR) 0410(0411bR|2067aR) 0411(0412bR|2259aR) 0412(0413bR|2587aR) 0413(0414bR|3002aR) 0414(0415bR|3737aR) 0415(0416bR|1978aR) 0416(0417bR|2569aR) 0417(0418bR|3185aR) 0418(0419bR|2440aR) 0419(0420bR|2654aR) 0420(0421bR|2617aR) 0421(0422bR|3757aR) 0422(0423bR|2820aR) 0423(0424bR|3838aR) 0424(0425bR|2649aR) 0425(0426bR|3293aR) 0426(0427bR|2637aR) 0427(0428bR|3293aR) 0428(0429bR|2585aR) 0429(0430bR|2259aR) 0430(0431bR|2587aR) 0431(0432bR|3002aR) 0432(0433bR|3737aR) 0433(0434bR|1978aR) 0434(0435bR|2569aR) 0435(0436bR|3185aR) 0436(0437bR|2440aR) 0437(0438bR|2654aR) 0438(0439bR|2617aR) 0439(0440bR|3753aR) 0440(0441bR|2820aR) 0441(0442bR|3838aR) 0442(0443bR|2651aR) 0443(0444bR|3293aR) 0444(0445bR|2637aR) 0445(0446bR|3293aR) 0446(0447bR|2585aR) 0447(0448bR|3283aR) 0448(0449bR|2587aR) 0449(0450bR|3002aR) 0450(0451bR|3737aR) 0451(0452bR|1978aR) 0452(0453bR|2569aR) 0453(0454bR|3185aR) 0454(0455bR|2440aR) 0455(0456bR|2654aR) 0456(0457bR|2617aR) 0457(0458bR|3741aR) 0458(0459bR|2820aR) 0459(0460bR|3834aR) 0460(0461bR|2012aR) 0461(0462bR|3183aR) 0462(0463bR|1892aR) 0463(0464bR|1902aR) 0464(0465bR|3595aR) 0465(0466bR|3698aR) 0466(0467bR|2083aR) 0467(0468bR|3678aR) 0468(0469bR|3420aR) 0469(0470bR|2224aR) 0470(0471bR|2441aR) 0471(0472bR|2250aR) 0472(0473bR|2651aR) 0473(0474bR|2269aR) 0474(0475bR|2587aR) 0475(0476bR|3002aR) 0476(0477bR|2582aR) 0477(0478bR|2138aR) 0478(0479bR|2788aR) 0479(0480bR|3278aR) 0480(0481bR|2657aR) 0481(0482bR|3294aR) 0482(0483bR|2075aR) 0483(0484bR|3283aR) 0484(0485bR|2059aR) 0485(0486bR|3434aR) 0486(0487bR|2859aR) 0487(0488bR|3434aR) 0488(0489bR|2696aR) 0489(0490bR|2458aR) 0490(0491bR|2945aR) 0491(0492bR|3434aR) 0492(0493bR|2865aR) 0493(0494bR|2208aR) 0494(0495bR|1939aR) 0495(0496bR|1910aR) 0496(0497bR|2646aR) 0497(0498bR|1966aR) 0498(0499bR|2115aR) 0499(0500bR|3762aR) 0500(0501bR|2789aR) 0501(0502bR|3006aR) 0502(0503bR|2657aR) 0503(0504bR|3294aR) 0504(0505bR|2075aR) 0505(0506bR|3283aR) 0506(0507bR|2587aR) 0507(0508bR|2238aR) 0508(0509bR|2859aR) 0509(0510bR|1978aR) 0510(0511bR|2441aR) 0511(0512bR|2999aR) 0512(0513bR|3427aR) 0513(0514bR|2330aR) 0514(0515bR|3449aR) 0515(0516bR|3760aR) 0516(0517bR|3745aR) 0517(0518bR|1967aR) 0518(0519bR|1916aR) 0519(0520bR|3254aR) 0520(0521bR|2695aR) 0521(0522bR|1969aR) 0522(0523bR|1929aR) 0523(0524bR|2993aR) 0524(0525bR|2533aR) 0525(0526bR|2224aR) 0526(0527bR|3205aR) 0527(0528bR|2160aR) 0528(0529bR|2070aR) 0529(0530bR|2234aR) 0530(0531bR|2449aR) 0531(0532bR|3394aR) 0532(0533bR|2569aR) 0533(0534bR|3249aR) 0534(0535bR|2437aR) 0535(0536bR|2208aR) 0536(0537bR|3207aR) 0537(0538bR|3294aR) 0538(0539bR|2637aR) 0539(0540bR|2910aR) 0540(0541bR|3595aR) 0541(0542bR|3690aR) 0542(0543bR|2939aR) 0543(0544bR|3437aR) 0544(0545bR|2659aR) 0545(0546bR|3394aR) 0546(0547bR|2657aR) 0547(0548bR|3360aR) 0548(0549bR|3193aR) 0549(0550bR|3437aR) 0550(0551bR|2657aR) 0551(0552bR|3397aR) 0552(0553bR|2044aR) 0553(0554bR|2933aR) 0554(0555bR|2407aR) 0555(0556bR|3287aR) 0556(0557bR|3033aR) 0557(0558bR|1977aR) 0558(0559bR|2107aR) 0559(0560bR|2334aR) 0560(0561bR|3428aR) 0561(0562bR|2911aR) 0562(0563bR|2565aR) 0563(0564bR|3190aR) 0564(0565bR|2695aR) 0565(0566bR|2282aR) 0566(0567bR|3077aR) 0567(0568bR|1978aR) 0568(0569bR|2708aR) 0569(0570bR|2992aR) 0570(0571bR|2405aR) 0571(0572bR|2144aR) 0572(0573bR|2533aR) 0573(0574bR|1888aR) 0574(0575bR|2565aR) 0575(0576bR|2159aR) 0576(0577bR|2044aR) 0577(0578bR|1966aR) 0578(0579bR|3033aR) 0579(0580bR|3838aR) 0580(0581bR|2939aR) 0581(0582bR|3758aR) 0582(0583bR|3769aR) 0583(0584bR|2997aR) 0584(0585bR|2083aR) 0585(0586bR|3278aR) 0586(0587bR|2893aR) 0587(0588bR|2921aR) 0588(0589bR|2649aR) 0589(0590bR|3690aR) 0590(0591bR|3463aR) 0591(0592bR|3001aR) 0592(0593bR|2625aR) 0593(0594bR|3289aR) 0594(0595bR|2585aR) 0595(0596bR|3278aR) 0596(0597bR|3089aR) 0597(0598bR|3487aR) 0598(0599bR|2569aR) 0599(0600bR|2270aR) 0600(0601bR|3143aR) 0601(0602bR|1973aR) 0602(0603bR|1931aR) 0603(0604bR|3262aR) 0604(0605bR|2081aR) 0605(0606bR|3305aR) 0606(0607bR|2659aR) 0607(0608bR|3434aR) 0608(0609bR|3431aR) 0609(0610bR|3434aR) 0610(0611bR|2659aR) 0611(0612bR|2270aR) 0612(0613bR|2593aR) 0613(0614bR|1981aR) 0614(0615bR|2070aR) 0615(0616bR|2234aR) 0616(0617bR|2068aR) 0617(0618bR|3561aR) 0618(0619bR|2659aR) 0619(0620bR|2269aR) 0620(0621bR|2639aR) 0621(0622bR|1962aR) 0622(0623bR|3429aR) 0623(0624bR|3248aR) 0624(0625bR|2584aR) 0625(0626bR|2882aR) 0626(0627bR|1916aR) 0627(0628bR|2225aR) 0628(0629bR|2440aR) 0629(0630bR|2397aR) 0630(0631bR|2619aR) 0631(0632bR|2286aR) 0632(0633bR|3595aR) 0633(0634bR|3488aR) 0634(0635bR|2105aR) 0635(0636bR|3293aR) 0636(0637bR|2587aR) 0637(0638bR|3002aR) 0638(0639bR|2582aR) 0639(0640bR|2234aR) 0640(0641bR|2065aR) 0641(0642bR|2259aR) 0642(0643bR|2619aR) 0643(0644bR|2269aR) 0644(0645bR|2639aR) 0645(0646bR|1950aR) 0646(0647bR|3429aR) 0647(0648bR|2144aR) 0648(0649bR|2569aR) 0649(0650bR|2935aR) 0650(0651bR|2913aR) 0651(0652bR|3374aR) 0652(0653bR|3463aR) 0653(0654bR|3359aR) 0654(0655bR|2443aR) 0655(0656bR|3674aR) 0656(0657bR|2939aR) 0657(0658bR|3437aR) 0658(0659bR|2659aR) 0659(0660bR|3394aR) 0660(0661bR|2657aR) 0661(0662bR|3353aR) 0662(0663bR|2859aR) 0663(0664bR|3759aR) 0664(0665bR|2020aR) 0665(0666bR|3610aR) 0666(0667bR|2065aR) 0667(0668bR|2286aR) 0668(0669bR|2820aR) 0669(0670bR|2224aR) 0670(0671bR|3164aR) 0671(0672bR|1902aR) 0672(0673bR|2939aR) 0673(0674bR|3437aR) 0674(0675bR|2659aR) 0675(0676bR|3397aR) 0676(0677bR|2020aR) 0677(0678bR|2929aR) 0678(0679bR|1939aR) 0679(0680bR|3289aR) 0680(0681bR|2857aR) 0681(0682bR|2265aR) 0682(0683bR|2587aR) 0683(0684bR|2234aR) 0684(0685bR|2441aR) 0685(0686bR|3255aR) 0686(0687bR|3425aR) 0687(0688bR|3354aR) 0688(0689bR|3079aR) 0689(0690bR|3262aR) 0690(0691bR|2857aR) 0691(0692bR|2346aR) 0692(0693bR|2657aR) 0693(0694bR|3294aR) 0694(0695bR|2075aR) 0695(0696bR|3284aR) 0696(0697bR|1915aR) 0697(0698bR|3680aR) 0698(0699bR|2115aR) 0699(0700bR|3703aR) 0700(0701bR|2917aR) 0701(0702bR|2269aR) 0702(0703bR|2587aR) 0703(0704bR|3002aR) 0704(0705bR|2582aR) 0705(0706bR|2922aR) 0706(0707bR|2780aR) 0707(0708bR|3374aR) 0708(0709bR|3664aR) 0709(0710bR|2547aR) 0710(0711bR|1893aR) 0711(0712bR|3854aR) 0712(0713bR|3580aR) 0713(0714bR|2906aR) 0714(0715bR|2839aR) 0715(0716bR|3028aR) 0716(0717bR|2119aR) 0717(0718bR|2250aR) 0718(0719bR|2070aR) 0719(0720bR|3023aR) 0720(0721bR|2065aR) 0721(0722bR|1975aR) 0722(0723bR|3476aR) 0723(0724bR|3795aR) 0724(0725bR|1892aR) 0725(0726bR|2359aR) 0726(0727bR|3756aR) 0727(0728bR|2486aR) 0728(0729bR|3149aR) 0729(0730bR|2202aR) 0730(0731bR|2966aR) 0731(0732bR|2159aR) 0732(0733bR|2629aR) 0733(0734bR|2370aR) 0734(0735bR|2699aR) 0735(0736bR|3859aR) 0736(0737bR|3113aR) 0737(0738bR|2000aR) 0738(0739bR|3205aR) 0739(0740bR|3796aR) 0740(0741bR|3238aR) 0741(0742bR|1887aR) 0742(0743bR|2867aR) 0743(0744bR|3674aR) 0744(0745bR|3589aR) 0745(0746bR|3284aR) 0746(0747bR|2632aR) 0747(0748bR|3374aR) 0748(0749bR|3804aR) 0749(0750bR|3849aR) 0750(0751bR|2777aR) 0751(0752bR|3760aR) 0752(0753bR|2070aR) 0753(0754bR|2934aR) 0754(0755bR|3859aR) 0755(0756bR|3401aR) 0756(0757bR|2893aR) 0757(0758bR|1885aR) 0758(0759bR|2697aR) 0759(0760bR|3796aR) 0760(0761bR|2115aR) 0761(0762bR|3696aR) 0762(0763bR|1937aR) 0763(0764bR|3393aR) 0764(0765bR|2572aR) 0765(0766bR|3616aR) 0766(0767bR|2577aR) 0767(0768bR|3397aR) 0768(0769bR|2105aR) 0769(0770bR|3440aR) 0770(0771bR|2584aR) 0771(0772bR|2393aR) 0772(0773bR|2444aR) 0773(0774bR|3296aR) 0774(0775bR|3161aR) 0775(0776bR|3377aR) 0776(0777bR|2870aR) 0777(0778bR|2497aR) 0778(0779bR|2524aR) 0779(0780bR|3394aR) 0780(0781bR|2708aR) 0781(0782bR|3651aR) 0782(0783bR|2721aR) 0783(0784bR|2240aR) 0784(0785bR|3303aR) 0785(0786bR|3253aR) 0786(0787bR|2070aR) 0787(0788bR|3024aR) 0788(0789bR|3047aR) 0789(0790bR|3189aR) 0790(0791bR|1942aR) 0791(0792bR|3022aR) 0792(0793bR|3601aR) 0793(0794bR|3767aR) 0794(0795bR|3475aR) 0795(0796bR|2377aR) 0796(0797bR|2896aR) 0797(0798bR|2547aR) 0798(0799bR|2596aR) 0799(0800bR|2370aR) 0800(0801bR|2791aR) 0801(0802bR|3838aR) 0802(0803bR|1939aR) 0803(0804bR|3703aR) 0804(0805bR|3047aR) 0805(0806bR|2351aR) 0806(0807bR|2637aR) 0807(0808bR|3296aR) 0808(0809bR|2128aR) 0809(0810bR|2547aR) 0810(0811bR|2596aR) 0811(0812bR|2370aR) 0812(0813bR|2791aR) 0813(0814bR|3838aR) 0814(0815bR|2067aR) 0815(0816bR|3509aR) 0816(0817bR|2859aR) 0817(0818bR|3446aR) 0818(0819bR|2870aR) 0819(0820bR|2498aR) 0820(0821bR|1925aR) 0821(0822bR|3834aR) 0822(0823bR|1932aR) 0823(0824bR|3545aR) 0824(0825bR|2572aR) 0825(0826bR|3651aR) 0826(0827bR|2724aR) 0827(0828bR|2359aR) 0828(0829bR|2947aR) 0829(0830bR|2295aR) 0830(0831bR|3756aR) 0831(0832bR|2394aR) 0832(0833bR|3151aR) 0833(0834bR|3374aR) 0834(0835bR|2916aR) 0835(0836bR|2912aR) 0836(0837bR|3335aR) 0837(0838bR|3893aR) 0838(0839bR|2131aR) 0839(0840bR|3412aR) 0840(0841bR|1939aR) 0841(0842bR|3306aR) 0842(0843bR|3420aR) 0843(0844bR|3610aR) 0844(0845bR|3668aR) 0845(0846bR|2936aR) 0846(0847bR|2584aR) 0847(0848bR|2474aR) 0848(0849bR|3149aR) 0849(0850bR|3231aR) 0850(0851bR|2838aR) 0851(0852bR|3017aR) 0852(0853bR|2820aR) 0853(0854bR|3545aR) 0854(0855bR|2407aR) 0855(0856bR|3447aR) 0856(0857bR|3603aR) 0857(0858bR|3401aR) 0858(0859bR|2820aR) 0859(0860bR|3565aR) 0860(0861bR|2710aR) 0861(0862bR|2218aR) 0862(0863bR|3144aR) 0863(0864bR|2677aR) 0864(0865bR|2083aR) 0865(0866bR|3394aR) 0866(0867bR|2533aR) 0867(0868bR|3447aR) 0868(0869bR|3473aR) 0869(0870bR|2286aR) 0870(0871bR|2865aR) 0871(0872bR|2350aR) 0872(0873bR|2919aR) 0873(0874bR|1911aR) 0874(0875bR|3431aR) 0875(0876bR|3893aR) 0876(0877bR|2131aR) 0877(0878bR|3411aR) 0878(0879bR|1889aR) 0879(0880bR|2353aR) 0880(0881bR|2867aR) 0881(0882bR|3760aR) 0882(0883bR|2065aR) 0883(0884bR|2387aR) 0884(0885bR|2533aR) 0885(0886bR|3850aR) 0886(0887bR|3431aR) 0887(0888bR|3283aR) 0888(0889bR|1881aR) 0889(0890bR|3535aR) 0890(0891bR|2128aR) 0891(0892bR|2548aR) 0892(0893bR|2105aR) 0893(0894bR|2373aR) 0894(0895bR|2828aR) 0895(0896bR|3859aR) 0896(0897bR|3113aR) 0897(0898bR|1984aR) 0898(0899bR|3161aR) 0899(0900bR|2998aR) 0900(0901bR|2966aR) 0901(0902bR|3017aR) 0902(0903bR|2792aR) 0903(0904bR|2413aR) 0904(0905bR|1932aR) 0905(0906bR|2004aR) 0906(0907bR|2627aR) 0907(0908bR|1973aR) 0908(0909bR|2134aR) 0909(0910bR|1881aR) 0910(0911bR|2708aR) 0911(0912bR|2003aR) 0912(0913bR|1921aR) 0913(0914bR|3318aR) 0914(0915bR|2966aR) 0915(0916bR|3583aR) 0916(0917bR|3045aR) 0917(0918bR|3446aR) 0918(0919bR|3739aR) 0919(0920bR|3186aR) 0920(0921bR|3790aR) 0921(0922bR|2297aR) 0922(0923bR|2053aR) 0923(0924bR|3833aR) 0924(0925bR|2055aR) 0925(0926bR|3859aR) 0926(0927bR|3115aR) 0927(0928bR|3506aR) 0928(0929bR|3748aR) 0929(0930bR|3024aR) 0930(0931bR|3193aR) 0931(0932bR|1998aR) 0932(0933bR|3089aR) 0933(0934bR|3838aR) 0934(0935bR|2963aR) 0935(0936bR|3859aR) 0936(0937bR|3113aR) 0937(0938bR|2290aR) 0938(0939bR|2791aR) 0939(0940bR|3393aR) 0940(0941bR|2060aR) 0941(0942bR|3295aR) 0942(0943bR|2620aR) 0943(0944bR|3284aR) 0944(0945bR|3238aR) 0945(0946bR|3018aR) 0946(0947bR|1940aR) 0947(0948bR|2388aR) 0948(0949bR|3238aR) 0949(0950bR|1966aR) 0950(0951bR|3141aR) 0951(0952bR|2369aR) 0952(0953bR|2556aR) 0953(0954bR|2375aR) 0954(0955bR|3213aR) 0955(0956bR|3375aR) 0956(0957bR|2859aR) 0957(0958bR|3424aR) 0958(0959bR|2859aR) 0959(0960bR|3439aR) 0960(0961bR|2125aR) 0961(0962bR|3250aR) 0962(0963bR|2073aR) 0963(0964bR|3440aR) 0964(0965bR|3205aR) 0965(0966bR|1982aR) 0966(0967bR|3745aR) 0967(0968bR|3294aR) 0968(0969bR|3045aR) 0969(0970bR|2003aR) 0970(0971bR|2532aR) 0971(0972bR|1910aR) 0972(0973bR|2660aR) 0973(0974bR|3568aR) 0974(0975bR|2073aR) 0975(0976bR|3859aR) 0976(0977bR|3113aR) 0977(0978bR|1906aR) 0978(0979bR|3612aR) 0979(0980bR|3616aR) 0980(0981bR|2583aR) 0981(0982bR|1982aR) 0982(0983bR|3737aR) 0983(0984bR|1983aR) 0984(0985bR|2120aR) 0985(0986bR|2818aR) 0986(0987bR|3161aR) 0987(0988bR|3540aR) 0988(0989bR|2627aR) 0989(0990bR|3509aR) 0990(0991bR|2134aR) 0991(0992bR|2934aR) 0992(0993bR|2641aR) 0993(0994bR|2336aR) 0994(0995bR|3171aR) 0995(0996bR|3540aR) 0996(0997bR|3238aR) 0997(0998bR|3022aR) 0998(0999bR|1937aR) 0999(1000bR|1910aR) 1000(1001bR|3862aR) 1001(1002bR|3583aR) 1002(1003bR|2949aR) 1003(1004bR|3447aR) 1004(1005bR|2951aR) 1005(1006bR|3447aR) 1006(1007bR|3756aR) 1007(1008bR|2409aR) 1008(1009bR|2599aR) 1009(1010bR|3509aR) 1010(1011bR|2824aR) 1011(1012bR|2399aR) 1012(1013bR|2084aR) 1013(1014bR|3022aR) 1014(1015bR|3790aR) 1015(1016bR|2297aR) 1016(1017bR|2088aR) 1017(1018bR|2741aR) 1018(1019bR|1942aR) 1019(1020bR|3280aR) 1020(1021bR|3591aR) 1021(1022bR|1910aR) 1022(1023bR|2646aR) 1023(1024bR|1946aR) 1024(1025bR|3151aR) 1025(1026bR|3373aR) 1026(1027bR|2449aR) 1027(1028bR|2359aR) 1028(1029bR|2907aR) 1029(1030bR|3536aR) 1030(1031bR|3373aR) 1031(1032bR|3545aR) 1032(1033bR|2710aR) 1033(1034bR|3273aR) 1034(1035bR|2127aR) 1035(1036bR|1998aR) 1036(1037bR|2105aR) 1037(1038bR|3767aR) 1038(1039bR|3603aR) 1039(1040bR|3017aR) 1040(1041bR|2127aR) 1041(1042bR|3373aR) 1042(1043bR|2449aR) 1043(1044bR|1888aR) 1044(1045bR|3335aR) 1045(1046bR|3166aR) 1046(1047bR|2777aR) 1047(1048bR|3838aR) 1048(1049bR|2088aR) 1049(1050bR|2742aR) 1050(1051bR|2595aR) 1051(1052bR|3680aR) 1052(1053bR|3772aR) 1053(1054bR|3022aR) 1054(1055bR|2127aR) 1055(1056bR|3373aR) 1056(1057bR|2780aR) 1057(1058bR|3191aR) 1058(1059bR|2937aR) 1059(1060bR|3540aR) 1060(1061bR|2105aR) 1061(1062bR|3504aR) 1062(1063bR|3161aR) 1063(1064bR|1984aR) 1064(1065bR|2664aR) 1065(1066bR|2818aR) 1066(1067bR|3335aR) 1067(1068bR|3189aR) 1068(1069bR|2076aR) 1069(1070bR|3616aR) 1070(1071bR|2020aR) 1071(1072bR|3314aR) 1072(1073bR|3745aR) 1073(1074bR|2998aR) 1074(1075bR|3790aR) 1075(1076bR|2314aR) 1076(1077bR|3548aR) 1077(1078bR|2994aR) 1078(1079bR|2865aR) 1079(1080bR|3754aR) 1080(1081bR|2785aR) 1081(1082bR|3850aR) 1082(1083bR|3623aR) 1083(1084bR|3397aR) 1084(1085bR|2817aR) 1085(1086bR|2250aR) 1086(1087bR|3656aR) 1087(1088bR|3397aR) 1088(1089bR|2812aR) 1089(1090bR|3833aR) 1090(1091bR|2053aR) 1091(1092bR|3859aR) 1092(1093bR|3113aR) 1093(1094bR|1970aR) 1094(1095bR|3661aR) 1095(1096bR|1902aR) 1096(1097bR|2966aR) 1097(1098bR|3583aR) 1098(1099bR|2937aR) 1099(1100bR|2000aR) 1100(1101bR|3205aR) 1101(1102bR|3796aR) 1102(1103bR|3238aR) 1103(1104bR|1887aR) 1104(1105bR|2867aR) 1105(1106bR|3760aR) 1106(1107bR|3431aR) 1107(1108bR|2373aR) 1108(1109bR|2115aR) 1109(1110bR|3866aR) 1110(1111bR|3478aR) 1111(1112bR|1967aR) 1112(1113bR|2870aR) 1113(1114bR|2501aR) 1114(1115bR|2819aR) 1115(1116bR|2330aR) 1116(1117bR|2120aR) 1117(1118bR|2375aR) 1118(1119bR|3213aR) 1119(1120bR|3295aR) 1120(1121bR|2859aR) 1121(1122bR|3510aR) 1122(1123bR|3737aR) 1123(1124bR|3248aR) 1124(1125bR|2076aR) 1125(1126bR|2259aR) 1126(1127bR|2599aR) 1127(1128bR|2375aR) 1128(1129bR|3213aR) 1129(1130bR|3278aR) 1130(1131bR|2952aR) 1131(1132bR|2398aR) 1132(1133bR|2076aR) 1133(1134bR|2335aR) 1134(1135bR|1929aR) 1135(1136bR|3306aR) 1136(1137bR|3080aR) 1137(1138bR|2481aR) 1138(1139bR|2620aR) 1139(1140bR|3610aR) 1140(1141bR|2780aR) 1141(1142bR|2208aR) 1142(1143bR|2584aR) 1143(1144bR|2481aR) 1144(1145bR|2600aR) 1145(1146bR|2482aR) 1146(1147bR|2076aR) 1147(1148bR|3565aR) 1148(1149bR|2595aR) 1149(1150bR|3439aR) 1150(1151bR|2454aR) 1151(1152bR|2238aR) 1152(1153bR|3080aR) 1153(1154bR|2481aR) 1154(1155bR|2596aR) 1155(1156bR|3567aR) 1156(1157bR|2451aR) 1157(1158bR|3837aR) 1158(1159bR|2595aR) 1159(1160bR|3439aR) 1160(1161bR|2454aR) 1161(1162bR|2250aR) 1162(1163bR|3080aR) 1163(1164bR|2416aR) 1164(1165bR|2587aR) 1165(1166bR|3838aR) 1166(1167bR|2951aR) 1167(1168bR|2370aR) 1168(1169bR|2684aR) 1169(1170bR|3393aR) 1170(1171bR|2565aR) 1171(1172bR|3258aR) 1172(1173bR|2684aR) 1173(1174bR|3614aR) 1174(1175bR|2780aR) 1175(1176bR|2144aR) 1176(1177bR|2584aR) 1177(1178bR|2458aR) 1178(1179bR|2620aR) 1179(1180bR|3258aR) 1180(1181bR|2857aR) 1181(1182bR|3247aR) 1182(1183bR|1929aR) 1183(1184bR|3530aR) 1184(1185bR|2949aR) 1185(1186bR|3006aR) 1186(1187bR|2865aR) 1187(1188bR|2249aR) 1188(1189bR|2663aR) 1189(1190bR|3258aR) 1190(1191bR|3790aR) 1191(1192bR|2298aR) 1192(1193bR|2437aR) 1193(1194bR|3509aR) 1194(1195bR|2857aR) 1195(1196bR|2240aR) 1196(1197bR|2663aR) 1197(1198bR|2351aR) 1198(1199bR|1942aR) 1199(1200bR|3583aR) 1200(1201bR|2949aR) 1201(1202bR|3511aR) 1202(1203bR|3041aR) 1203(1204bR|2294aR) 1204(1205bR|2955aR) 1205(1206bR|3702aR) 1206(1207bR|2870aR) 1207(1208bR|2497aR) 1208(1209bR|2444aR) 1209(1210bR|3312aR) 1210(1211bR|2116aR) 1211(1212bR|1911aR) 1212(1213bR|3043aR) 1213(1214bR|3442aR) 1214(1215bR|2833aR) 1215(1216bR|3761aR) 1216(1217bR|2128aR) 1217(1218bR|2547aR) 1218(1219bR|2588aR) 1219(1220bR|2370aR) 1220(1221bR|2695aR) 1221(1222bR|3833aR) 1222(1223bR|2524aR) 1223(1224bR|3002aR) 1224(1225bR|3094aR) 1225(1226bR|2240aR) 1226(1227bR|2128aR) 1227(1228bR|2547aR) 1228(1229bR|2060aR) 1229(1230bR|2259aR) 1230(1231bR|1925aR) 1231(1232bR|3770aR) 1232(1233bR|3089aR) 1233(1234bR|3767aR) 1234(1235bR|3756aR) 1235(1236bR|2414aR) 1236(1237bR|2068aR) 1237(1238bR|3567aR) 1238(1239bR|2838aR) 1239(1240bR|3583aR) 1240(1241bR|2908aR) 1241(1242bR|2369aR) 1242(1243bR|2572aR) 1243(1244bR|3651aR) 1244(1245bR|2715aR) 1245(1246bR|3514aR) 1246(1247bR|3737aR) 1247(1248bR|1974aR) 1248(1249bR|2966aR) 1249(1250bR|3583aR) 1250(1251bR|3475aR) 1251(1252bR|3381aR) 1252(1253bR|2084aR) 1253(1254bR|3545aR) 1254(1255bR|2710aR) 1255(1256bR|3833aR) 1256(1257bR|2088aR) 1257(1258bR|2414aR) 1258(1259bR|3152aR) 1259(1260bR|2547aR) 1260(1261bR|2011aR) 1261(1262bR|3518aR) 1262(1263bR|3475aR) 1263(1264bR|3424aR) 1264(1265bR|3163aR) 1265(1266bR|3377aR) 1266(1267bR|2695aR) 1267(1268bR|2351aR) 1268(1269bR|1942aR) 1269(1270bR|3583aR) 1270(1271bR|2908aR) 1271(1272bR|2374aR) 1272(1273bR|2777aR) 1273(1274bR|3849aR) 1274(1275bR|2817aR) 1275(1276bR|3680aR) 1276(1277bR|2639aR) 1277(1278bR|3373aR) 1278(1279bR|2065aR) 1279(1280bR|2004aR) 1280(1281bR|2627aR) 1281(1282bR|1973aR) 1282(1283bR|2056aR) 1283(1284bR|2741aR) 1284(1285bR|1939aR) 1285(1286bR|2926aR) 1286(1287bR|3780aR) 1287(1288bR|2933aR) 1288(1289bR|2627aR) 1289(1290bR|2360aR) 1290(1291bR|2440aR) 1291(1292bR|2474aR) 1292(1293bR|3151aR) 1293(1294bR|2330aR) 1294(1295bR|3606aR) 1295(1296bR|1946aR) 1296(1297bR|3151aR) 1297(1298bR|2330aR) 1298(1299bR|3606aR) 1299(1300bR|3018aR) 1300(1301bR|3661aR) 1301(1302bR|1882aR) 1302(1303bR|3141aR) 1303(1304bR|2375aR) 1304(1305bR|3213aR) 1305(1306bR|2206aR) 1306(1307bR|3612aR) 1307(1308bR|2259aR) 1308(1309bR|2596aR) 1309(1310bR|3394aR) 1310(1311bR|2789aR) 1311(1312bR|3837aR) 1312(1313bR|2823aR) 1313(1314bR|2288aR) 1314(1315bR|3332aR) 1315(1316bR|2933aR) 1316(1317bR|2627aR) 1317(1318bR|3906aR) 1318(1319bR|2693aR) 1319(1320bR|2369aR) 1320(1321bR|1937aR) 1321(1322bR|3833aR) 1322(1323bR|2524aR) 1323(1324bR|3262aR) 1324(1325bR|3478aR) 1325(1326bR|2234aR) 1326(1327bR|3606aR) 1327(1328bR|1910aR) 1328(1329bR|3089aR) 1329(1330bR|3767aR) 1330(1331bR|3463aR) 1331(1332bR|2000aR) 1332(1333bR|2580aR) 1333(1334bR|3651aR) 1334(1335bR|2723aR) 1335(1336bR|1974aR) 1336(1337bR|3748aR) 1337(1338bR|2359aR) 1338(1339bR|3756aR) 1339(1340bR|2415aR) 1340(1341bR|2664aR) 1341(1342bR|2480aR) 1342(1343bR|2664aR) 1343(1344bR|2818aR) 1344(1345bR|3205aR) 1345(1346bR|3319aR) 1346(1347bR|3089aR) 1347(1348bR|3795aR) 1348(1349bR|2521aR) 1349(1350bR|3765aR) 1350(1351bR|2859aR) 1351(1352bR|2336aR) 1352(1353bR|2596aR) 1353(1354bR|3567aR) 1354(1355bR|2108aR) 1355(1356bR|3651aR) 1356(1357bR|2748aR) 1357(1358bR|2928aR) 1358(1359bR|3303aR) 1359(1360bR|3248aR) 1360(1361bR|2855aR) 1361(1362bR|1911aR) 1362(1363bR|3756aR) 1363(1364bR|2458aR) 1364(1365bR|2629aR) 1365(1366bR|2288aR) 1366(1367bR|3205aR) 1367(1368bR|3383aR) 1368(1369bR|3089aR) 1369(1370bR|3539aR) 1370(1371bR|2451aR) 1371(1372bR|3698aR) 1372(1373bR|2867aR) 1373(1374bR|3354aR) 1374(1375bR|3463aR) 1375(1376bR|3287aR) 1376(1377bR|2919aR) 1377(1378bR|3509aR) 1378(1379bR|2824aR) 1379(1380bR|2482aR) 1380(1381bR|2664aR) 1381(1382bR|2485aR) 1382(1383bR|2838aR) 1383(1384bR|2224aR) 1384(1385bR|2596aR) 1385(1386bR|3027aR) 1386(1387bR|2588aR) 1387(1388bR|3284aR) 1388(1389bR|2065aR) 1389(1390bR|2358aR) 1390(1391bR|2632aR) 1391(1392bR|2818aR) 1392(1393bR|3205aR) 1393(1394bR|3319aR) 1394(1395bR|3065aR) 1395(1396bR|3698aR) 1396(1397bR|2859aR) 1397(1398bR|3510aR) 1398(1399bR|3747aR) 1399(1400bR|1998aR) 1400(1401bR|3094aR) 1401(1402bR|1950aR) 1402(1403bR|3089aR) 1403(1404bR|3767aR) 1404(1405bR|3081aR) 1405(1406bR|3511aR) 1406(1407bR|3756aR) 1407(1408bR|2394aR) 1408(1409bR|3151aR) 1409(1410bR|3373aR) 1410(1411bR|2617aR) 1411(1412bR|2374aR) 1412(1413bR|2817aR) 1413(1414bR|3353aR) 1414(1415bR|2839aR) 1415(1416bR|3283aR) 1416(1417bR|2012aR) 1417(1418bR|2373aR) 1418(1419bR|2780aR) 1419(1420bR|3393aR) 1420(1421bR|1925aR) 1421(1422bR|3796aR) 1422(1423bR|3238aR) 1423(1424bR|1961aR) 1424(1425bR|2710aR) 1425(1426bR|1967aR) 1426(1427bR|2870aR) 1427(1428bR|2497aR) 1428(1429bR|2444aR) 1429(1430bR|1994aR) 1430(1431bR|2917aR) 1431(1432bR|2237aR) 1432(1433bR|2857aR) 1433(1434bR|3023aR) 1434(1435bR|2628aR) 1435(1436bR|1905aR) 1436(1437bR|2593aR) 1437(1438bR|3859aR) 1438(1439bR|3113aR) 1439(1440bR|3375aR) 1440(1441bR|1942aR) 1441(1442bR|2158aR) 1442(1443bR|2708aR) 1443(1444bR|1977aR) 1444(1445bR|2660aR) 1445(1446bR|2167aR) 1446(1447bR|3081aR) 1447(1448bR|2272aR) 1448(1449bR|3289aR) 1449(1450bR|3680aR) 1450(1451bR|2627aR) 1451(1452bR|3912aR) 1452(1453bR|3163aR) 1453(1454bR|3442aR) 1454(1455bR|3081aR) 1455(1456bR|2270aR) 1456(1457bR|3048aR) 1457(1458bR|2481aR) 1458(1459bR|2084aR) 1459(1460bR|3609aR) 1460(1461bR|2780aR) 1461(1462bR|1952aR) 1462(1463bR|2584aR) 1463(1464bR|2461aR) 1464(1465bR|1932aR) 1465(1466bR|3295aR) 1466(1467bR|2620aR) 1467(1468bR|1905aR) 1468(1469bR|2593aR) 1469(1470bR|3838aR) 1470(1471bR|2917aR) 1471(1472bR|3394aR) 1472(1473bR|2660aR) 1473(1474bR|3027aR) 1474(1475bR|2020aR) 1475(1476bR|2161aR) 1476(1477bR|2075aR) 1477(1478bR|3850aR) 1478(1479bR|2105aR) 1479(1480bR|3418aR) 1480(1481bR|3592aR) 1481(1482bR|3667aR) 1482(1483bR|1924aR) 1483(1484bR|1983aR) 1484(1485bR|2585aR) 1485(1486bR|3439aR) 1486(1487bR|2454aR) 1487(1488bR|2912aR) 1488(1489bR|1929aR) 1489(1490bR|2350aR) 1490(1491bR|2920aR) 1491(1492bR|2480aR) 1492(1493bR|2585aR) 1493(1494bR|3859aR) 1494(1495bR|3113aR) 1495(1496bR|2223aR) 1496(1497bR|2085aR) 1497(1498bR|3358aR) 1498(1499bR|3559aR) 1499(1500bR|3283aR) 1500(1501bR|1883aR) 1501(1502bR|3680aR) 1502(1503bR|2070aR) 1503(1504bR|2250aR) 1504(1505bR|2696aR) 1505(1506bR|2718aR) 1506(1507bR|3429aR) 1507(1508bR|3278aR) 1508(1509bR|2893aR) 1509(1510bR|1951aR) 1510(1511bR|2661aR) 1511(1512bR|2250aR) 1512(1513bR|2812aR) 1513(1514bR|3255aR) 1514(1515bR|2937aR) 1515(1516bR|2234aR) 1516(1517bR|2827aR) 1517(1518bR|3702aR) 1518(1519bR|2692aR) 1519(1520bR|2161aR) 1520(1521bR|2107aR) 1521(1522bR|3744aR) 1522(1523bR|2582aR) 1523(1524bR|1946aR) 1524(1525bR|2780aR) 1525(1526bR|1909aR) 1526(1527bR|2625aR) 1527(1528bR|3760aR) 1528(1529bR|3163aR) 1529(1530bR|2161aR) 1530(1531bR|2699aR) 1531(1532bR|3696aR) 1532(1533bR|2451aR) 1533(1534bR|3398aR) 1534(1535bR|2065aR) 1535(1536bR|2272aR) 1536(1537bR|2569aR) 1537(1538bR|3833aR) 1538(1539bR|2052aR) 1539(1540bR|2930aR) 1540(1541bR|2791aR) 1541(1542bR|3358aR) 1542(1543bR|3559aR) 1543(1544bR|3283aR) 1544(1545bR|2619aR) 1545(1546bR|3767aR) 1546(1547bR|3091aR) 1547(1548bR|2359aR) 1548(1549bR|3065aR) 1549(1550bR|2350aR) 1550(1551bR|3560aR) 1551(1552bR|2414aR) 1552(1553bR|2812aR) 1553(1554bR|2998aR) 1554(1555bR|2637aR) 1555(1556bR|3310aR) 1556(1557bR|3560aR) 1557(1558bR|2818aR) 1558(1559bR|3163aR) 1559(1560bR|3514aR) 1560(1561bR|3465aR) 1561(1562bR|2270aR) 1562(1563bR|3048aR) 1563(1564bR|2398aR) 1564(1565bR|3620aR) 1565(1566bR|3257aR) 1566(1567bR|2660aR) 1567(1568bR|2167aR) 1568(1569bR|3756aR) 1569(1570bR|2398aR) 1570(1571bR|3141aR) 1571(1572bR|3375aR) 1572(1573bR|1929aR) 1573(1574bR|3393aR) 1574(1575bR|2060aR) 1575(1576bR|1994aR) 1576(1577bR|3593aR) 1577(1578bR|2270aR) 1578(1579bR|3790aR) 1579(1580bR|2302aR) 1580(1581bR|2956aR) 1581(1582bR|3567aR) 1582(1583bR|2632aR) 1583(1584bR|2457aR) 1584(1585bR|2789aR) 1585(1586bR|2288aR) 1586(1587bR|3205aR) 1587(1588bR|3703aR) 1588(1589bR|3756aR) 1589(1590bR|2416aR) 1590(1591bR|2451aR) 1591(1592bR|3394aR) 1592(1593bR|2052aR) 1593(1594bR|2271aR) 1594(1595bR|1939aR) 1595(1596bR|3289aR) 1596(1597bR|2651aR) 1597(1598bR|3837aR) 1598(1599bR|2567aR) 1599(1600bR|3505aR) 1600(1601bR|2627aR) 1601(1602bR|3418aR) 1602(1603bR|2916aR) 1603(1604bR|3568aR) 1604(1605bR|2569aR) 1605(1606bR|2370aR) 1606(1607bR|2649aR) 1607(1608bR|3440aR) 1608(1609bR|3193aR) 1609(1610bR|2266aR) 1610(1611bR|2820aR) 1611(1612bR|1888aR) 1612(1613bR|3289aR) 1613(1614bR|3354aR) 1614(1615bR|3047aR) 1615(1616bR|3287aR) 1616(1617bR|3881aR) 1617(1618bR|2976aR) 1618(1619bR|2085aR) 1619(1620bR|1998aR) 1620(1621bR|2649aR) 1621(1622bR|3422aR) 1622(1623bR|3748aR) 1623(1624bR|3249aR) 1624(1625bR|2127aR) 1625(1626bR|3253aR) 1626(1627bR|2115aR) 1627(1628bR|2350aR) 1628(1629bR|3881aR) 1629(1630bR|3376aR) 1630(1631bR|2582aR) 1631(1632bR|1904aR) 1632(1633bR|2629aR) 1633(1634bR|1998aR) 1634(1635bR|2649aR) 1635(1636bR|3422aR) 1636(1637bR|3748aR) 1637(1638bR|3249aR) 1638(1639bR|2127aR) 1639(1640bR|3253aR) 1640(1641bR|2113aR) 1641(1642bR|2350aR) 1642(1643bR|3883aR) 1643(1644bR|2975aR) 1644(1645bR|2833aR) 1645(1646bR|3761aR) 1646(1647bR|2075aR) 1647(1648bR|3759aR) 1648(1649bR|1893aR) 1649(1650bR|1975aR) 1650(1651bR|3067aR) 1651(1652bR|2271aR) 1652(1653bR|2620aR) 1653(1654bR|3261aR) 1654(1655bR|2692aR) 1655(1656bR|3249aR) 1656(1657bR|1929aR) 1657(1658bR|2259aR) 1658(1659bR|1913aR) 1659(1660bR|3290aR) 1660(1661bR|2949aR) 1661(1662bR|3006aR) 1662(1663bR|2780aR) 1663(1664bR|3249aR) 1664(1665bR|1929aR) 1665(1666bR|2234aR) 1666(1667bR|2125aR) 1667(1668bR|1946aR) 1668(1669bR|2684aR) 1669(1670bR|2225aR) 1670(1671bR|2075aR) 1671(1672bR|3696aR) 1672(1673bR|2569aR) 1673(1674bR|1977aR) 1674(1675bR|2587aR) 1675(1676bR|2370aR) 1676(1677bR|2820aR) 1677(1678bR|2935aR) 1678(1679bR|3081aR) 1679(1680bR|1994aR) 1680(1681bR|2859aR) 1681(1682bR|1905aR) 1682(1683bR|2067aR) 1683(1684bR|3289aR) 1684(1685bR|2865aR) 1685(1686bR|1973aR) 1686(1687bR|2107aR) 1687(1688bR|2350aR) 1688(1689bR|3896aR) 1689(1690bR|2398aR) 1690(1691bR|2823aR) 1691(1692bR|3424aR) 1692(1693bR|2569aR) 1693(1694bR|1977aR) 1694(1695bR|2639aR) 1695(1696bR|3253aR) 1696(1697bR|2115aR) 1697(1698bR|3760aR) 1698(1699bR|2582aR) 1699(1700bR|1902aR) 1700(1701bR|3468aR) 1701(1702bR|1910aR) 1702(1703bR|2585aR) 1703(1704bR|2266aR) 1704(1705bR|2867aR) 1705(1706bR|3273aR) 1706(1707bR|2692aR) 1707(1708bR|3250aR) 1708(1709bR|2637aR) 1709(1710bR|3277aR) 1710(1711bR|2820aR) 1711(1712bR|3651aR) 1712(1713bR|2713aR) 1713(1714bR|3008aR) 1714(1715bR|2695aR) 1715(1716bR|3290aR) 1716(1717bR|2952aR) 1717(1718bR|2397aR) 1718(1719bR|2697aR) 1719(1720bR|3702aR) 1720(1721bR|2652aR) 1721(1722bR|1975aR) 1722(1723bR|3756aR) 1723(1724bR|2742aR) 1724(1725bR|2449aR) 1725(1726bR|2288aR) 1726(1727bR|3292aR) 1727(1728bR|3226aR) 1728(1729bR|3432aR) 1729(1730bR|3550aR) 1730(1731bR|2710aR) 1731(1732bR|3583aR) 1732(1733bR|2939aR) 1733(1734bR|3354aR) 1734(1735bR|3464aR) 1735(1736bR|2398aR) 1736(1737bR|2108aR) 1737(1738bR|2266aR) 1738(1739bR|2961aR) 1739(1740bR|3370aR) 1740(1741bR|3592aR) 1741(1742bR|2485aR) 1742(1743bR|2585aR) 1743(1744bR|3853aR) 1744(1745bR|2617aR) 1745(1746bR|3440aR) 1746(1747bR|2584aR) 1747(1748bR|2415aR) 1748(1749bR|1937aR) 1749(1750bR|3393aR) 1750(1751bR|2564aR) 1751(1752bR|1994aR) 1752(1753bR|3097aR) 1753(1754bR|3696aR) 1754(1755bR|2451aR) 1755(1756bR|3398aR) 1756(1757bR|2065aR) 1757(1758bR|2272aR) 1758(1759bR|2569aR) 1759(1760bR|3838aR) 1760(1761bR|3461aR) 1761(1762bR|2003aR) 1762(1763bR|2407aR) 1763(1764bR|3018aR) 1764(1765bR|2867aR) 1765(1766bR|3017aR) 1766(1767bR|2691aR) 1767(1768bR|3760aR) 1768(1769bR|3739aR) 1769(1770bR|1951aR) 1770(1771bR|2661aR) 1771(1772bR|1978aR) 1772(1773bR|2695aR) 1773(1774bR|2253aR) 1774(1775bR|2820aR) 1775(1776bR|3546aR) 1776(1777bR|2827aR) 1777(1778bR|3424aR) 1778(1779bR|2067aR) 1779(1780bR|2350aR) 1780(1781bR|3748aR) 1781(1782bR|2993aR) 1782(1783bR|2125aR) 1783(1784bR|2934aR) 1784(1785bR|2107aR) 1785(1786bR|3854aR) 1786(1787bR|2105aR) 1787(1788bR|3434aR) 1788(1789bR|3592aR) 1789(1790bR|3546aR) 1790(1791bR|1937aR) 1791(1792bR|2288aR) 1792(1793bR|2067aR) 1793(1794bR|2350aR) 1794(1795bR|3745aR) 1795(1796bR|2166aR) 1796(1797bR|2788aR) 1797(1798bR|3546aR) 1798(1799bR|2057aR) 1799(1800bR|3761aR) 1800(1801bR|2593aR) 1801(1802bR|3503aR) 1802(1803bR|1931aR) 1803(1804bR|3696aR) 1804(1805bR|2451aR) 1805(1806bR|3394aR) 1806(1807bR|2788aR) 1807(1808bR|2259aR) 1808(1809bR|2588aR) 1809(1810bR|2999aR) 1810(1811bR|3556aR) 1811(1812bR|2929aR) 1812(1813bR|2627aR) 1813(1814bR|3440aR) 1814(1815bR|3373aR) 1815(1816bR|3616aR) 1816(1817bR|2535aR) 1817(1818bR|1982aR) 1818(1819bR|3780aR) 1819(1820bR|3226aR) 1820(1821bR|3432aR) 1821(1822bR|3616aR) 1822(1823bR|2567aR) 1823(1824bR|1975aR) 1824(1825bR|3756aR) 1825(1826bR|2677aR) 1826(1827bR|1939aR) 1827(1828bR|3296aR) 1828(1829bR|3303aR) 1829(1830bR|3894aR) 1830(1831bR|2646aR) 1831(1832bR|3273aR) 1832(1833bR|2115aR) 1833(1834bR|3703aR) 1834(1835bR|3604aR) 1835(1836bR|3790aR) 1836(1837bR|2896aR) 1837(1838bR|2547aR) 1838(1839bR|2428aR) 1839(1840bR|2354aR) 1840(1841bR|2857aR) 1841(1842bR|1903aR) 1842(1843bR|2084aR) 1843(1844bR|3651aR) 1844(1845bR|2721aR) 1845(1846bR|3023aR) 1846(1847bR|2620aR) 1847(1848bR|3546aR) 1848(1849bR|2828aR) 1849(1850bR|1994aR) 1850(1851bR|3593aR) 1851(1852bR|2370aR) 1852(1853bR|2789aR) 1853(1854bR|2374aR) 1854(1855bR|2105aR) 1855(1856bR|3696aR) 1856(1857bR|2584aR) 1857(1858bR|2410aR) 1858(1859bR|3141aR) 1859(1860bR|2336aR) 1860(1861bR|2585aR) 1861(1862bR|3838aR) 1862(1863bR|3467aR) 1863(1864bR|3838aR) 1864(1865bR|3593aR) 1865(1866bR|3837aR) 1866(1867bR|2585aR) 1867(1868bR|3702aR) 1868(1869bR|2859aR) 1869(1870bR|3505aR) 1870(1871bR|2125aR) 1871(1872bR|1968aR) 1872(1873bR|2697aR) 1873(1874bR|3702aR) 1874(1875bR|2652aR) 1875(1876bR|3626aR) 1876(1877bR|3429aR) 1877(1878bR|2350aR) 1878(1879bR|2966aR) 1879(1880bR|3583aR) 1880(1881bR|3083aR) 1881(1882bR|3539aR) 1882(1883bR|2620aR) 1883(1884bR|2370aR) 1884(1885bR|2407aR) 1885(1886bR|3360aR) 1886(1887bR|3207aR) 1887(1888bR|2359aR) 1888(1889bR|2945aR) 1889(1890bR|2266aR) 1890(1891bR|3745aR) 1891(1892bR|3294aR) 1892(1893bR|3465aR) 1893(1894bR|2369aR) 1894(1895bR|2563aR) 1895(1896bR|3702aR) 1896(1897bR|2660aR) 1897(1898bR|3565aR) 1898(1899bR|2628aR) 1899(1900bR|2250aR) 1900(1901bR|3748aR) 1901(1902bR|1998aR) 1902(1903bR|3790aR) 1903(1904bR|2302aR) 1904(1905bR|3596aR) 1905(1906bR|3567aR) 1906(1907bR|1940aR) 1907(1908bR|3562aR) 1908(1909bR|2949aR) 1909(1910bR|1998aR) 1910(1911bR|3748aR) 1911(1912bR|3258aR) 1912(1913bR|3748aR) 1913(1914bR|1906aR) 1914(1915bR|2857aR) 1915(1916bR|1906aR) 1916(1917bR|2116aR) 1917(1918bR|1911aR) 1918(1919bR|2907aR) 1919(1920bR|2351aR) 1920(1921bR|2059aR) 1921(1922bR|2370aR) 1922(1923bR|2791aR) 1923(1924bR|3399aR) 1924(1925bR|3213aR) 1925(1926bR|1983aR) 1926(1927bR|2620aR) 1927(1928bR|3283aR) 1928(1929bR|2425aR) 1929(1930bR|3535aR) 1930(1931bR|2084aR) 1931(1932bR|3022aR) 1932(1933bR|3790aR) 1933(1934bR|2301aR) 1934(1935bR|2441aR) 1935(1936bR|3505aR) 1936(1937bR|2637aR) 1937(1938bR|1886aR) 1938(1939bR|2917aR) 1939(1940bR|2259aR) 1940(1941bR|2555aR) 1941(1942bR|3442aR) 1942(1943bR|2708aR) 1943(1944bR|3022aR) 1944(1945bR|2949aR) 1945(1946bR|3006aR) 1946(1947bR|2859aR) 1947(1948bR|3440aR) 1948(1949bR|2125aR) 1949(1950bR|3306aR) 1950(1951bR|3606aR) 1951(1952bR|2138aR) 1952(1953bR|3077aR) 1953(1954bR|2237aR) 1954(1955bR|2865aR) 1955(1956bR|3273aR) 1956(1957bR|2657aR) 1957(1958bR|2350aR) 1958(1959bR|3883aR) 1959(1960bR|3440aR) 1960(1961bR|2637aR) 1961(1962bR|2930aR) 1962(1963bR|2075aR) 1963(1964bR|3850aR) 1964(1965bR|2105aR) 1965(1966bR|3744aR) 1966(1967bR|2584aR) 1967(1968bR|2409aR) 1968(1969bR|2085aR) 1969(1970bR|2266aR) 1970(1971bR|3465aR) 1971(1972bR|3306aR) 1972(1973bR|3080aR) 1973(1974bR|2654aR) 1974(1975bR|3429aR) 1975(1976bR|1952aR) 1976(1977bR|2569aR) 1977(1978bR|2167aR) 1978(1979bR|3081aR) 1979(1980bR|3360aR) 1980(1981bR|3193aR) 1981(1982bR|3487aR) 1982(1983bR|2070aR) 1983(1984bR|2976aR) 1984(1985bR|1929aR) 1985(1986bR|3278aR) 1986(1987bR|2893aR) 1987(1988bR|2138aR) 1988(1989bR|3620aR) 1989(1990bR|1910aR) 1990(1991bR|2660aR) 1991(1992bR|2162aR) 1992(1993bR|2639aR) 1993(1994bR|1973aR) 1994(1995bR|2073aR) 1995(1996bR|3022aR) 1996(1997bR|2651aR) 1997(1998bR|2370aR) 1998(1999bR|2428aR) 1999(2000bR|2353aR) 2000(2001bR|2695aR) 2001(2002bR|2351aR) 2002(2003bR|2059aR) 2003(2004bR|2286aR) 2004(2005bR|3748aR) 2005(2006bR|1973aR) 2006(2007bR|2859aR) 2007(2008bR|3488aR) 2008(2009bR|2637aR) 2009(2010bR|3006aR) 2010(2011bR|2684aR) 2011(2012bR|1975aR) 2012(2013bR|3460aR) 2013(2014bR|2933aR) 2014(2015bR|2451aR) 2015(2016bR|3411aR) 2016(2017bR|2436aR) 2017(2018bR|1909aR) 2018(2019bR|2661aR) 2019(2020bR|2237aR) 2020(2021bR|2692aR) 2021(2022bR|3610aR) 2022(2023bR|2780aR) 2023(2024bR|1898aR) 2024(2025bR|3589aR) 2025(2026bR|1952aR) 2026(2027bR|3205aR) 2027(2028bR|2250aR) 2028(2029bR|3748aR) 2029(2030bR|2166aR) 2030(2031bR|2859aR) 2031(2032bR|2286aR) 2032(2033bR|2949aR) 2033(2034bR|3028aR) 2034(2035bR|2055aR) 2035(2036bR|1977aR) 2036(2037bR|1939aR) 2037(2038bR|3411aR) 2038(2039bR|2436aR) 2039(2040bR|1910aR) 2040(2041bR|2693aR) 2041(2042bR|2237aR) 2042(2043bR|2692aR) 2043(2044bR|3610aR) 2044(2045bR|2780aR) 2045(2046bR|1898aR) 2046(2047bR|3589aR) 2047(2048bR|1952aR) 2048(2049bR|3195aR) 2049(2050bR|3290aR) 2050(2051bR|3465aR) 2051(2052bR|3439aR) 2052(2053bR|2443aR) 2053(2054bR|3393aR) 2054(2055bR|2084aR) 2055(2056bR|1970aR) 2056(2057bR|2083aR) 2057(2058bR|3838aR) 2058(2059bR|2437aR) 2059(2060bR|2287aR) 2060(2061bR|2833aR) 2061(2062bR|3761aR) 2062(2063bR|2593aR) 2063(2064bR|3504aR) 2064(2065bR|3291aR) 2065(2066bR|3679aR) 2066(2067bR|2627aR) 2067(2068bR|3511aR) 2068(2069bR|3756aR) 2069(2070bR|2482aR) 2070(2071bR|3661aR) 2071(2072bR|3162aR) 2072(2073bR|2697aR) 2073(2074bR|3859aR) 2074(2075bR|3123aR) 2075(2076bR|3759aR) 2076(2077bR|2105aR) 2077(2078bR|3906aR) 2078(2079bR|2565aR) 2079(2080bR|3849aR) 2080(2081bR|2820aR) 2081(2082bR|3017aR) 2082(2083bR|2893aR) 2083(2084bR|3247aR) 2084(2085bR|2870aR) 2085(2086bR|2497aR) 2086(2087bR|2436aR) 2087(2088bR|3022aR) 2088(2089bR|3737aR) 2089(2090bR|3249aR) 2090(2091bR|2699aR) 2091(2092bR|3761aR) 2092(2093bR|2075aR) 2093(2094bR|3838aR) 2094(2095bR|3595aR) 2095(2096bR|3854aR) 2096(2097bR|2780aR) 2097(2098bR|1888aR) 2098(2099bR|2584aR) 2099(2100bR|2410aR) 2100(2101bR|3612aR) 2101(2102bR|2254aR) 2102(2103bR|2917aR) 2103(2104bR|2259aR) 2104(2105bR|2628aR) 2105(2106bR|2260aR) 2106(2107bR|2577aR) 2107(2108bR|2334aR) 2108(2109bR|3592aR) 2109(2110bR|3546aR) 2110(2111bR|2085aR) 2111(2112bR|2288aR) 2112(2113bR|2585aR) 2113(2114bR|3440aR) 2114(2115bR|3163aR) 2115(2116bR|1994aR) 2116(2117bR|3091aR) 2117(2118bR|3423aR) 2118(2119bR|2070aR) 2119(2120bR|1905aR) 2120(2121bR|2821aR) 2121(2122bR|1978aR) 2122(2123bR|2867aR) 2123(2124bR|3354aR) 2124(2125bR|3559aR) 2125(2126bR|3277aR) 2126(2127bR|2857aR) 2127(2128bR|2272aR) 2128(2129bR|2585aR) 2129(2130bR|3440aR) 2130(2131bR|3163aR) 2131(2132bR|3446aR) 2132(2133bR|3091aR) 2133(2134bR|3423aR) 2134(2135bR|2070aR) 2135(2136bR|3253aR) 2136(2137bR|2076aR) 2137(2138bR|3249aR) 2138(2139bR|2870aR) 2139(2140bR|2498aR) 2140(2141bR|2708aR) 2141(2142bR|3550aR) 2142(2143bR|2021aR) 2143(2144bR|3447aR) 2144(2145bR|3756aR) 2145(2146bR|2394aR) 2146(2147bR|3152aR) 2147(2148bR|2547aR) 2148(2149bR|1923aR) 2149(2150bR|3446aR) 2150(2151bR|3083aR) 2151(2152bR|2286aR) 2152(2153bR|3432aR) 2153(2154bR|2482aR) 2154(2155bR|2084aR) 2155(2156bR|3613aR) 2156(2157bR|2780aR) 2157(2158bR|2934aR) 2158(2159bR|2646aR) 2159(2160bR|1898aR) 2160(2161bR|3612aR) 2161(2162bR|1906aR) 2162(2163bR|2083aR) 2163(2164bR|3680aR) 2164(2165bR|3201aR) 2165(2166bR|3263aR) 2166(2167bR|1931aR) 2167(2168bR|3702aR) 2168(2169bR|2660aR) 2169(2170bR|2162aR) 2170(2171bR|2617aR) 2171(2172bR|3838aR) 2172(2173bR|3045aR) 2173(2174bR|2260aR) 2174(2175bR|2023aR) 2175(2176bR|1981aR) 2176(2177bR|2820aR) 2177(2178bR|3863aR) 2178(2179bR|2913aR) 2179(2180bR|3002aR) 2180(2181bR|3620aR) 2181(2182bR|2162aR) 2182(2183bR|2617aR) 2183(2184bR|3850aR) 2184(2185bR|2105aR) 2185(2186bR|3790aR) 2186(2187bR|2838aR) 2187(2188bR|1969aR) 2188(2189bR|2593aR) 2189(2190bR|3859aR) 2190(2191bR|3113aR) 2191(2192bR|2207aR) 2192(2193bR|2085aR) 2193(2194bR|3360aR) 2194(2195bR|2585aR) 2195(2196bR|3440aR) 2196(2197bR|3207aR) 2197(2198bR|3312aR) 2198(2199bR|3332aR) 2199(2200bR|2933aR) 2200(2201bR|2627aR) 2201(2202bR|3905aR) 2202(2203bR|2452aR) 2203(2204bR|1982aR) 2204(2205bR|3593aR) 2205(2206bR|2270aR) 2206(2207bR|3748aR) 2207(2208bR|3262aR) 2208(2209bR|3779aR) 2209(2210bR|3679aR) 2210(2211bR|2451aR) 2211(2212bR|3411aR) 2212(2213bR|2009aR) 2213(2214bR|3514aR) 2214(2215bR|3467aR) 2215(2216bR|3761aR) 2216(2217bR|2075aR) 2217(2218bR|3854aR) 2218(2219bR|2780aR) 2219(2220bR|2230aR) 2220(2221bR|2595aR) 2221(2222bR|3833aR) 2222(2223bR|2663aR) 2223(2224bR|3290aR) 2224(2225bR|2952aR) 2225(2226bR|2398aR) 2226(2227bR|2956aR) 2227(2228bR|2254aR) 2228(2229bR|2917aR) 2229(2230bR|2260aR) 2230(2231bR|2577aR) 2231(2232bR|2358aR) 2232(2233bR|2632aR) 2233(2234bR|2818aR) 2234(2235bR|3195aR) 2235(2236bR|3186aR) 2236(2237bR|3612aR) 2237(2238bR|3651aR) 2238(2239bR|2721aR) 2239(2240bR|3247aR) 2240(2241bR|1928aR) 2241(2242bR|2481aR) 2242(2243bR|2444aR) 2243(2244bR|3566aR) 2244(2245bR|2659aR) 2245(2246bR|3422aR) 2246(2247bR|2940aR) 2247(2248bR|3567aR) 2248(2249bR|1927aR) 2249(2250bR|3838aR) 2250(2251bR|2587aR) 2251(2252bR|2270aR) 2252(2253bR|2651aR) 2253(2254bR|3838aR) 2254(2255bR|3461aR) 2255(2256bR|2167aR) 2256(2257bR|3081aR) 2257(2258bR|2250aR) 2258(2259bR|2859aR) 2259(2260bR|3434aR) 2260(2261bR|2966aR) 2261(2262bR|2223aR) 2262(2263bR|2053aR) 2263(2264bR|1969aR) 2264(2265bR|2454aR) 2265(2266bR|2253aR) 2266(2267bR|2659aR) 2267(2268bR|3837aR) 2268(2269bR|2535aR) 2269(2270bR|3022aR) 2270(2271bR|2917aR) 2271(2272bR|1969aR) 2272(2273bR|2441aR) 2273(2274bR|3283aR) 2274(2275bR|2585aR) 2275(2276bR|3290aR) 2276(2277bR|3745aR) 2277(2278bR|2207aR) 2278(2279bR|2021aR) 2279(2280bR|1975aR) 2280(2281bR|3083aR) 2281(2282bR|3434aR) 2282(2283bR|3745aR) 2283(2284bR|3185aR) 2284(2285bR|2116aR) 2285(2286bR|3185aR) 2286(2287bR|2057aR) 2287(2288bR|3002aR) 2288(2289bR|2637aR) 2289(2290bR|1886aR) 2290(2291bR|3045aR) 2291(2292bR|1975aR) 2292(2293bR|3081aR) 2293(2294bR|2238aR) 2294(2295bR|3748aR) 2295(2296bR|1951aR) 2296(2297bR|1942aR) 2297(2298bR|2161aR) 2298(2299bR|2569aR) 2299(2300bR|3001aR) 2300(2301bR|2639aR) 2301(2302bR|1968aR) 2302(2303bR|1929aR) 2303(2304bR|2998aR) 2304(2305bR|2646aR) 2305(2306bR|2229aR) 2306(2307bR|2651aR) 2307(2308bR|3838aR) 2308(2309bR|2916aR) 2309(2310bR|3027aR) 2310(2311bR|2587aR) 2311(2312bR|2266aR) 2312(2313bR|3745aR) 2313(2314bR|3438aR) 2314(2315bR|2940aR) 2315(2316bR|1904aR) 2316(2317bR|3291aR) 2317(2318bR|3354aR) 2318(2319bR|2915aR) 2319(2320bR|3760aR) 2320(2321bR|3748aR) 2321(2322bR|1903aR) 2322(2323bR|2637aR) 2323(2324bR|3293aR) 2324(2325bR|2812aR) 2325(2326bR|3566aR) 2326(2327bR|2659aR) 2327(2328bR|3422aR) 2328(2329bR|2940aR) 2329(2330bR|3567aR) 2330(2331bR|1927aR) 2331(2332bR|2370aR) 2332(2333bR|2393aR) 2333(2334bR|3434aR) 2334(2335bR|2948aR) 2335(2336bR|1904aR) 2336(2337bR|3207aR) 2337(2338bR|3293aR) 2338(2339bR|2859aR) 2339(2340bR|2226aR) 2340(2341bR|2116aR) 2341(2342bR|3001aR) 2342(2343bR|2587aR) 2343(2344bR|2270aR) 2344(2345bR|2865aR) 2345(2346bR|1973aR) 2346(2347bR|2083aR) 2347(2348bR|3374aR) 2348(2349bR|3076aR) 2349(2350bR|3546aR) 2350(2351bR|3081aR) 2351(2352bR|2234aR) 2352(2353bR|2075aR) 2353(2354bR|3397aR) 2354(2355bR|1927aR) 2355(2356bR|1978aR) 2356(2357bR|2115aR) 2357(2358bR|3434aR) 2358(2359bR|3737aR) 2359(2360bR|1967aR) 2360(2361bR|2444aR) 2361(2362bR|2229aR) 2362(2363bR|2116aR) 2363(2364bR|1905aR) 2364(2365bR|2627aR) 2365(2366bR|3509aR) 2366(2367bR|2649aR) 2367(2368bR|3422aR) 2368(2369bR|2940aR) 2369(2370bR|1968aR) 2370(2371bR|3205aR) 2371(2372bR|1969aR) 2372(2373bR|1942aR) 2373(2374bR|2201aR) 2374(2375bR|2657aR) 2375(2376bR|3422aR) 2376(2377bR|3748aR) 2377(2378bR|3249aR) 2378(2379bR|2454aR) 2379(2380bR|3583aR) 2380(2381bR|2940aR) 2381(2382bR|2354aR) 2382(2383bR|2821aR) 2383(2384bR|1978aR) 2384(2385bR|2857aR) 2385(2386bR|3007aR) 2386(2387bR|2108aR) 2387(2388bR|3257aR) 2388(2389bR|2660aR) 2389(2390bR|3651aR) 2390(2391bR|2723aR) 2391(2392bR|3488aR) 2392(2393bR|1928aR) 2393(2394bR|2480aR) 2394(2395bR|1937aR) 2395(2396bR|2259aR) 2396(2397bR|2617aR) 2397(2398bR|3354aR) 2398(2399bR|3777aR) 2399(2400bR|2249aR) 2400(2401bR|2659aR) 2401(2402bR|3022aR) 2402(2403bR|2893aR) 2403(2404bR|3354aR) 2404(2405bR|2660aR) 2405(2406bR|3568aR) 2406(2407bR|2021aR) 2407(2408bR|3027aR) 2408(2409bR|2567aR) 2409(2410bR|2229aR) 2410(2411bR|2065aR) 2411(2412bR|3027aR) 2412(2413bR|2619aR) 2413(2414bR|3694aR) 2414(2415bR|3745aR) 2415(2416bR|3189aR) 2416(2417bR|2617aR) 2417(2418bR|3434aR) 2418(2419bR|3428aR) 2419(2420bR|2976aR) 2420(2421bR|2056aR) 2421(2422bR|2485aR) 2422(2423bR|2057aR) 2423(2424bR|3027aR) 2424(2425bR|2619aR) 2425(2426bR|2266aR) 2426(2427bR|3745aR) 2427(2428bR|3694aR) 2428(2429bR|3428aR) 2429(2430bR|2976aR) 2430(2431bR|3323aR) 2431(2432bR|3354aR) 2432(2433bR|2916aR) 2433(2434bR|3256aR) 2434(2435bR|3204aR) 2435(2436bR|2912aR) 2436(2437bR|1928aR) 2437(2438bR|2881aR) 2438(2439bR|1921aR) 2439(2440bR|3503aR) 2440(2441bR|1925aR) 2441(2442bR|2229aR) 2442(2443bR|2070aR) 2443(2444bR|2249aR) 2444(2445bR|2619aR) 2445(2446bR|3853aR) 2446(2447bR|2065aR) 2447(2448bR|2266aR) 2448(2449bR|2115aR) 2449(2450bR|3905aR) 2450(2451bR|2396aR) 2451(2452bR|2286aR) 2452(2453bR|2948aR) 2453(2454bR|2933aR) 2454(2455bR|2065aR) 2455(2456bR|3018aR) 2456(2457bR|2637aR) 2457(2458bR|3354aR) 2458(2459bR|2780aR) 2459(2460bR|3625aR) 2460(2461bR|2617aR) 2461(2462bR|3674aR) 2462(2463bR|3592aR) 2463(2464bR|3546aR) 2464(2465bR|2916aR) 2465(2466bR|2928aR) 2466(2467bR|2023aR) 2467(2468bR|2231aR) 2468(2469bR|2916aR) 2469(2470bR|2226aR) 2470(2471bR|3431aR) 2471(2472bR|1973aR) 2472(2473bR|2449aR) 2473(2474bR|3283aR) 2474(2475bR|2617aR) 2475(2476bR|3290aR) 2476(2477bR|2859aR) 2477(2478bR|3690aR) 2478(2479bR|2940aR) 2479(2480bR|3550aR) 2480(2481bR|2619aR) 2481(2482bR|3354aR) 2482(2483bR|2785aR) 2483(2484bR|3838aR) 2484(2485bR|3428aR) 2485(2486bR|2935aR) 2486(2487bR|3036aR) 2487(2488bR|2976aR) 2488(2489bR|1928aR) 2489(2490bR|2457aR) 2490(2491bR|2779aR) 2491(2492bR|3694aR) 2492(2493bR|3452aR) 2493(2494bR|3625aR) 2494(2495bR|2617aR) 2495(2496bR|3418aR) 2496(2497bR|2820aR) 2497(2498bR|3834aR) 2498(2499bR|2788aR) 2499(2500bR|2928aR) 2500(2501bR|3321aR) 2501(2502bR|3354aR) 2502(2503bR|2940aR) 2503(2504bR|3249aR) 2504(2505bR|2454aR) 2505(2506bR|2249aR) 2506(2507bR|2617aR) 2507(2508bR|3853aR) 2508(2509bR|2065aR) 2509(2510bR|2269aR) 2510(2511bR|2579aR) 2511(2512bR|3411aR) 2512(2513bR|2617aR) 2513(2514bR|3354aR) 2514(2515bR|3777aR) 2515(2516bR|2249aR) 2516(2517bR|2681aR) 2517(2518bR|3374aR) 2518(2519bR|3883aR) 2519(2520bR|2225aR) 2520(2521bR|2577aR) 2521(2522bR|3310aR) 2522(2523bR|3428aR) 2523(2524bR|2976aR) 2524(2525bR|2056aR) 2525(2526bR|2398aR) 2526(2527bR|3077aR) 2527(2528bR|3680aR) 2528(2529bR|1927aR) 2529(2530bR|2165aR) 2530(2531bR|2582aR) 2531(2532bR|3166aR) 2532(2533bR|3429aR) 2533(2534bR|2144aR) 2534(2535bR|2569aR) 2535(2536bR|2935aR) 2536(2537bR|2907aR) 2537(2538bR|3006aR) 2538(2539bR|2651aR) 2539(2540bR|3482aR) 2540(2541bR|3428aR) 2541(2542bR|3566aR) 2542(2543bR|3431aR) 2543(2544bR|1975aR) 2544(2545bR|3089aR) 2545(2546bR|2249aR) 2546(2547bR|2867aR) 2547(2548bR|1973aR) 2548(2549bR|2113aR) 2549(2550bR|3278aR) 2550(2551bR|2893aR) 2551(2552bR|2162aR) 2552(2553bR|3079aR) 2553(2554bR|1993aR) 2554(2555bR|2619aR) 2555(2556bR|2334aR) 2556(2557bR|2859aR) 2557(2558bR|3678aR) 2558(2559bR|3432aR) 2559(2560bR|2714aR) 2560(2561bR|2780aR) 2561(2562bR|2974aR) 2562(2563bR|3592aR) 2563(2564bR|3546aR) 2564(2565bR|2916aR) 2565(2566bR|2928aR) 2566(2567bR|2023aR) 2567(2568bR|2231aR) 2568(2569bR|2913aR) 2569(2570bR|3774aR) 2570(2571bR|2683aR) 2571(2572bR|3678aR) 2572(2573bR|3452aR) 2573(2574bR|2992aR) 2574(2575bR|3163aR) 2575(2576bR|2226aR) 2576(2577bR|1931aR) 2577(2578bR|2249aR) 2578(2579bR|2788aR) 2579(2580bR|3190aR) 2580(2581bR|2595aR) 2581(2582bR|2330aR) 2582(2583bR|2785aR) 2583(2584bR|3694aR) 2584(2585bR|3748aR) 2585(2586bR|2927aR) 2586(2587bR|2536aR) 2587(2588bR|2482aR) 2588(2589bR|1937aR) 2589(2590bR|2259aR) 2590(2591bR|1931aR) 2591(2592bR|3354aR) 2592(2593bR|2785aR) 2593(2594bR|3694aR) 2594(2595bR|3748aR) 2595(2596bR|2927aR) 2596(2597bR|1928aR) 2597(2598bR|2485aR) 2598(2599bR|2443aR) 2599(2600bR|2370aR) 2600(2601bR|2107aR) 2601(2602bR|3357aR) 2602(2603bR|2779aR) 2603(2604bR|3838aR) 2604(2605bR|3431aR) 2605(2606bR|2231aR) 2606(2607bR|3547aR) 2607(2608bR|3679aR) 2608(2609bR|2436aR) 2609(2610bR|3250aR) 2610(2611bR|2056aR) 2611(2612bR|2413aR) 2612(2613bR|2595aR) 2613(2614bR|3486aR) 2614(2615bR|3428aR) 2615(2616bR|3568aR) 2616(2617bR|1925aR) 2617(2618bR|3283aR) 2618(2619bR|2619aR) 2619(2620bR|2352aR) 2620(2621bR|3161aR) 2621(2622bR|2334aR) 2622(2623bR|2785aR) 2623(2624bR|3678aR) 2624(2625bR|3777aR) 2625(2626bR|3273aR) 2626(2627bR|2649aR) 2627(2628bR|3278aR) 2628(2629bR|2893aR) 2629(2630bR|2142aR) 2630(2631bR|3044aR) 2631(2632bR|2229aR) 2632(2633bR|2065aR) 2633(2634bR|3018aR) 2634(2635bR|2637aR) 2635(2636bR|1982aR) 2636(2637bR|2779aR) 2637(2638bR|3690aR) 2638(2639bR|3460aR) 2639(2640bR|3625aR) 2640(2641bR|2617aR) 2641(2642bR|3753aR) 2642(2643bR|2820aR) 2643(2644bR|3226aR) 2644(2645bR|3737aR) 2645(2646bR|2991aR) 2646(2647bR|2437aR) 2647(2648bR|3189aR) 2648(2649bR|2107aR) 2649(2650bR|3744aR) 2650(2651bR|2571aR) 2651(2652bR|1993aR) 2652(2653bR|2619aR) 2653(2654bR|2334aR) 2654(2655bR|2857aR) 2655(2656bR|1902aR) 2656(2657bR|3036aR) 2657(2658bR|3005aR) 2658(2659bR|2595aR) 2659(2660bR|3310aR) 2660(2661bR|3431aR) 2661(2662bR|2253aR) 2662(2663bR|2820aR) 2663(2664bR|1887aR) 2664(2665bR|2535aR) 2665(2666bR|1973aR) 2666(2667bR|2449aR) 2667(2668bR|3283aR) 2668(2669bR|2553aR) 2669(2670bR|3774aR) 2670(2671bR|2083aR) 2671(2672bR|3422aR) 2672(2673bR|3431aR) 2673(2674bR|2253aR) 2674(2675bR|2820aR) 2675(2676bR|2143aR) 2676(2677bR|2437aR) 2677(2678bR|3253aR) 2678(2679bR|2065aR) 2679(2680bR|3018aR) 2680(2681bR|2637aR) 2681(2682bR|2922aR) 2682(2683bR|2659aR) 2683(2684bR|3502aR) 2684(2685bR|3612aR) 2685(2686bR|2237aR) 2686(2687bR|2595aR) 2687(2688bR|2286aR) 2688(2689bR|2916aR) 2689(2690bR|2933aR) 2690(2691bR|2625aR) 2691(2692bR|3759aR) 2692(2693bR|1895aR) 2693(2694bR|1909aR) 2694(2695bR|2065aR) 2695(2696bR|3018aR) 2696(2697bR|2625aR) 2697(2698bR|2346aR) 2698(2699bR|2859aR) 2699(2700bR|3678aR) 2700(2701bR|3420aR) 2701(2702bR|3561aR) 2702(2703bR|2785aR) 2703(2704bR|3678aR) 2704(2705bR|3777aR) 2705(2706bR|3017aR) 2706(2707bR|2657aR) 2707(2708bR|3278aR) 2708(2709bR|2896aR) 2709(2710bR|3566aR) 2710(2711bR|3431aR) 2711(2712bR|1975aR) 2712(2713bR|2917aR) 2713(2714bR|3017aR) 2714(2715bR|2619aR) 2715(2716bR|2334aR) 2716(2717bR|2867aR) 2717(2718bR|1973aR) 2718(2719bR|2076aR) 2719(2720bR|3249aR) 2720(2721bR|2870aR) 2721(2722bR|2502aR) 2722(2723bR|2817aR) 2723(2724bR|3673aR) 2724(2725bR|2120aR) 2725(2726bR|2999aR) 2726(2727bR|3475aR) 2727(2728bR|2933aR) 2728(2729bR|2407aR) 2729(2730bR|3680aR) 2730(2731bR|3303aR) 2731(2732bR|2912aR) 2732(2733bR|1884aR) 2733(2734bR|3290aR) 2734(2735bR|3772aR) 2735(2736bR|3256aR) 2736(2737bR|1943aR) 2737(2738bR|3895aR) 2738(2739bR|2907aR) 2739(2740bR|3536aR) 2740(2741bR|3373aR) 2741(2742bR|3561aR) 2742(2743bR|2600aR) 2743(2744bR|2477aR) 2744(2745bR|2708aR) 2745(2746bR|3562aR) 2746(2747bR|2060aR) 2747(2748bR|3651aR) 2748(2749bR|2715aR) 2749(2750bR|3506aR) 2750(2751bR|2125aR) 2751(2752bR|3306aR) 2752(2753bR|3094aR) 2753(2754bR|2238aR) 2754(2755bR|2838aR) 2755(2756bR|2165aR) 2756(2757bR|2083aR) 2757(2758bR|3488aR) 2758(2759bR|3321aR) 2759(2760bR|3680aR) 2760(2761bR|2023aR) 2761(2762bR|3287aR) 2762(2763bR|2951aR) 2763(2764bR|2282aR) 2764(2765bR|3748aR) 2765(2766bR|2165aR) 2766(2767bR|2859aR) 2767(2768bR|3504aR) 2768(2769bR|2859aR) 2769(2770bR|2330aR) 2770(2771bR|3077aR) 2771(2772bR|3028aR) 2772(2773bR|2407aR) 2773(2774bR|1993aR) 2774(2775bR|2115aR) 2775(2776bR|3905aR) 2776(2777bR|2427aR) 2777(2778bR|3518aR) 2778(2779bR|3475aR) 2779(2780bR|3423aR) 2780(2781bR|2443aR) 2781(2782bR|3354aR) 2782(2783bR|3748aR) 2783(2784bR|2166aR) 2784(2785bR|2859aR) 2785(2786bR|3503aR) 2786(2787bR|1942aR) 2787(2788bR|1885aR) 2788(2789bR|2780aR) 2789(2790bR|2162aR) 2790(2791bR|2639aR) 2791(2792bR|2933aR) 2792(2793bR|2083aR) 2793(2794bR|2350aR) 2794(2795bR|3881aR) 2795(2796bR|3185aR) 2796(2797bR|2827aR) 2797(2798bR|3697aR) 2798(2799bR|2593aR) 2799(2800bR|3504aR) 2800(2801bR|1937aR) 2801(2802bR|3394aR) 2802(2803bR|2684aR) 2803(2804bR|1975aR) 2804(2805bR|3452aR) 2805(2806bR|2930aR) 2806(2807bR|2451aR) 2807(2808bR|3411aR) 2808(2809bR|2057aR) 2809(2810bR|3306aR) 2810(2811bR|3798aR) 2811(2812bR|1909aR) 2812(2813bR|2684aR) 2813(2814bR|2229aR) 2814(2815bR|2127aR) 2815(2816bR|2165aR) 2816(2817bR|2073aR) 2817(2818bR|3759aR) 2818(2819bR|1942aR) 2819(2820bR|1906aR) 2820(2821bR|2684aR) 2821(2822bR|2229aR) 2822(2823bR|2127aR) 2823(2824bR|2165aR) 2824(2825bR|2119aR) 2825(2826bR|3280aR) 2826(2827bR|3204aR) 2827(2828bR|1970aR) 2828(2829bR|2128aR) 2829(2830bR|2547aR) 2830(2831bR|2404aR) 2831(2832bR|2370aR) 2832(2833bR|2444aR) 2833(2834bR|3632aR) 2834(2835bR|2529aR) 2835(2836bR|3674aR) 2836(2837bR|3665aR) 2837(2838bR|2003aR) 2838(2839bR|2533aR) 2839(2840bR|3834aR) 2840(2841bR|1932aR) 2841(2842bR|3616aR) 2842(2843bR|1929aR) 2843(2844bR|3833aR) 2844(2845bR|2060aR) 2845(2846bR|3399aR) 2846(2847bR|3213aR) 2847(2848bR|1962aR) 2848(2849bR|3084aR) 2849(2850bR|1994aR) 2850(2851bR|3593aR) 2851(2852bR|2270aR) 2852(2853bR|3748aR) 2853(2854bR|3258aR) 2854(2855bR|3779aR) 2855(2856bR|3679aR) 2856(2857bR|1895aR) 2857(2858bR|3287aR) 2858(2859bR|2940aR) 2859(2860bR|2287aR) 2860(2861bR|2116aR) 2861(2862bR|1905aR) 2862(2863bR|2637aR) 2863(2864bR|3375aR) 2864(2865bR|2125aR) 2865(2866bR|1905aR) 2866(2867bR|2125aR) 2867(2868bR|1887aR) 2868(2869bR|2067aR) 2869(2870bR|3424aR) 2870(2871bR|3291aR) 2871(2872bR|3680aR) 2872(2873bR|2451aR) 2873(2874bR|3411aR) 2874(2875bR|2628aR) 2875(2876bR|2003aR) 2876(2877bR|2585aR) 2877(2878bR|3795aR) 2878(2879bR|2437aR) 2879(2880bR|2254aR) 2880(2881bR|2920aR) 2881(2882bR|2481aR) 2882(2883bR|2660aR) 2883(2884bR|3550aR) 2884(2885bR|2577aR) 2885(2886bR|3506aR) 2886(2887bR|2963aR) 2887(2888bR|3423aR) 2888(2889bR|2070aR) 2889(2890bR|3253aR) 2890(2891bR|2119aR) 2891(2892bR|3280aR) 2892(2893bR|3373aR) 2893(2894bR|3567aR) 2894(2895bR|2088aR) 2895(2896bR|2482aR) 2896(2897bR|2838aR) 2897(2898bR|1885aR) 2898(2899bR|2789aR) 2899(2900bR|2288aR) 2900(2901bR|3205aR) 2901(2902bR|3703aR) 2902(2903bR|3756aR) 2903(2904bR|2394aR) 2904(2905bR|3152aR) 2905(2906bR|2547aR) 2906(2907bR|2017aR) 2907(2908bR|3775aR) 2908(2909bR|2585aR) 2909(2910bR|3439aR) 2910(2911bR|2454aR) 2911(2912bR|2233aR) 2912(2913bR|2696aR) 2913(2914bR|2650aR) 2914(2915bR|3431aR) 2915(2916bR|1910aR) 2916(2917bR|2646aR) 2917(2918bR|1951aR) 2918(2919bR|2827aR) 2919(2920bR|3423aR) 2920(2921bR|2059aR) 2921(2922bR|2369aR) 2922(2923bR|2428aR) 2923(2924bR|1984aR) 2924(2925bR|2073aR) 2925(2926bR|3439aR) 2926(2927bR|2454aR) 2927(2928bR|2233aR) 2928(2929bR|2792aR) 2929(2930bR|2650aR) 2930(2931bR|3429aR) 2931(2932bR|3190aR) 2932(2933bR|2646aR) 2933(2934bR|1951aR) 2934(2935bR|2663aR) 2935(2936bR|2266aR) 2936(2937bR|2949aR) 2937(2938bR|3027aR) 2938(2939bR|2585aR) 2939(2940bR|3795aR) 2940(2941bR|2587aR) 2941(2942bR|3447aR) 2942(2943bR|2939aR) 2943(2944bR|3487aR) 2944(2945bR|1929aR) 2945(2946bR|3397aR) 2946(2947bR|2449aR) 2947(2948bR|2281aR) 2948(2949bR|2820aR) 2949(2950bR|3838aR) 2950(2951bR|2917aR) 2951(2952bR|3394aR) 2952(2953bR|2660aR) 2953(2954bR|3027aR) 2954(2955bR|2052aR) 2955(2956bR|2161aR) 2956(2957bR|2075aR) 2957(2958bR|3850aR) 2958(2959bR|2105aR) 2959(2960bR|3482aR) 2960(2961bR|3592aR) 2961(2962bR|3566aR) 2962(2963bR|3593aR) 2963(2964bR|2376aR) 2964(2965bR|3163aR) 2965(2966bR|3295aR) 2966(2967bR|2789aR) 2967(2968bR|1978aR) 2968(2969bR|2684aR) 2969(2970bR|3609aR) 2970(2971bR|2780aR) 2971(2972bR|3278aR) 2972(2973bR|2824aR) 2973(2974bR|2398aR) 2974(2975bR|2828aR) 2975(2976bR|2233aR) 2976(2977bR|2660aR) 2977(2978bR|2167aR) 2978(2979bR|3420aR) 2979(2980bR|2934aR) 2980(2981bR|3591aR) 2981(2982bR|3834aR) 2982(2983bR|2821aR) 2983(2984bR|2004aR) 2984(2985bR|3238aR) 2985(2986bR|2222aR) 2986(2987bR|2966aR) 2987(2988bR|2159aR) 2988(2989bR|2825aR) 2989(2990bR|3838aR) 2990(2991bR|2569aR) 2991(2992bR|3859aR) 2992(2993bR|3115aR) 2993(2994bR|2161aR) 2994(2995bR|2595aR) 2995(2996bR|3353aR) 2996(2997bR|2859aR) 2997(2998bR|3674aR) 2998(2999bR|2952aR) 2999(3000bR|2485aR) 3000(3001bR|2057aR) 3001(3002bR|3837aR) 3002(3003bR|2785aR) 3003(3004bR|3678aR) 3004(3005bR|3420aR) 3005(3006bR|3626aR) 3006(3007bR|2105aR) 3007(3008bR|3437aR) 3008(3009bR|2627aR) 3009(3010bR|3905aR) 3010(3011bR|2593aR) 3011(3012bR|3674aR) 3012(3013bR|3748aR) 3013(3014bR|2911aR) 3014(3015bR|2454aR) 3015(3016bR|2249aR) 3016(3017bR|2588aR) 3017(3018bR|3562aR) 3018(3019bR|3452aR) 3019(3020bR|2928aR) 3020(3021bR|1896aR) 3021(3022bR|2717aR) 3022(3023bR|2780aR) 3023(3024bR|1898aR) 3024(3025bR|2820aR) 3025(3026bR|3834aR) 3026(3027bR|2692aR) 3027(3028bR|2912aR) 3028(3029bR|3207aR) 3029(3030bR|1906aR) 3030(3031bR|2637aR) 3031(3032bR|3354aR) 3032(3033bR|2664aR) 3033(3034bR|2464aR) 3034(3035bR|2023aR) 3035(3036bR|1973aR) 3036(3037bR|1942aR) 3037(3038bR|3178aR) 3038(3039bR|3431aR) 3039(3040bR|3184aR) 3040(3041bR|2584aR) 3041(3042bR|2416aR) 3042(3043bR|1937aR) 3043(3044bR|2003aR) 3044(3045bR|2617aR) 3045(3046bR|2330aR) 3046(3047bR|3748aR) 3047(3048bR|2927aR) 3048(3049bR|2125aR) 3049(3050bR|3018aR) 3050(3051bR|2105aR) 3051(3052bR|3353aR) 3052(3053bR|2867aR) 3053(3054bR|2165aR) 3054(3055bR|2113aR) 3055(3056bR|3278aR) 3056(3057bR|2893aR) 3057(3058bR|3353aR) 3058(3059bR|2660aR) 3059(3060bR|3568aR) 3060(3061bR|1925aR) 3061(3062bR|3027aR) 3062(3063bR|2020aR) 3063(3064bR|2976aR) 3064(3065bR|1895aR) 3065(3066bR|1975aR) 3066(3067bR|3089aR) 3067(3068bR|1982aR) 3068(3069bR|2859aR) 3069(3070bR|3678aR) 3070(3071bR|3432aR) 3071(3072bR|2410aR) 3072(3073bR|2787aR) 3073(3074bR|3674aR) 3074(3075bR|3428aR) 3075(3076bR|3545aR) 3076(3077bR|2449aR) 3077(3078bR|1994aR) 3078(3079bR|2107aR) 3079(3080bR|3398aR) 3080(3081bR|1895aR) 3081(3082bR|1997aR) 3082(3083bR|2067aR) 3083(3084bR|3411aR) 3084(3085bR|1881aR) 3085(3086bR|3313aR) 3086(3087bR|2593aR) 3087(3088bR|3509aR) 3088(3089bR|2777aR) 3089(3090bR|3678aR) 3090(3091bR|3452aR) 3091(3092bR|3545aR) 3092(3093bR|1925aR) 3093(3094bR|3504aR) 3094(3095bR|1939aR) 3095(3096bR|3424aR) 3096(3097bR|1895aR) 3097(3098bR|1973aR) 3098(3099bR|2454aR) 3099(3100bR|1881aR) 3100(3101bR|2084aR) 3101(3102bR|3022aR) 3102(3103bR|2919aR) 3103(3104bR|2253aR) 3104(3105bR|2820aR) 3105(3106bR|1887aR) 3106(3107bR|1925aR) 3107(3108bR|2161aR) 3108(3109bR|2571aR) 3109(3110bR|1981aR) 3110(3111bR|2617aR) 3111(3112bR|2330aR) 3112(3113bR|2785aR) 3113(3114bR|3694aR) 3114(3115bR|3737aR) 3115(3116bR|1887aR) 3116(3117bR|2620aR) 3117(3118bR|1910aR) 3118(3119bR|2788aR) 3119(3120bR|3190aR) 3120(3121bR|2585aR) 3121(3122bR|2266aR) 3122(3123bR|2657aR) 3123(3124bR|3438aR) 3124(3125bR|3036aR) 3125(3126bR|2208aR) 3126(3127bR|1895aR) 3127(3128bR|1973aR) 3128(3129bR|2449aR) 3129(3130bR|3283aR) 3130(3131bR|2617aR) 3131(3132bR|2282aR) 3132(3133bR|2859aR) 3133(3134bR|3678aR) 3134(3135bR|3068aR) 3135(3136bR|3561aR) 3136(3137bR|2107aR) 3137(3138bR|2329aR) 3138(3139bR|2779aR) 3139(3140bR|3838aR) 3140(3141bR|3044aR) 3141(3142bR|2976aR) 3142(3143bR|3171aR) 3143(3144bR|2353aR) 3144(3145bR|2660aR) 3145(3146bR|2165aR) 3146(3147bR|2617aR) 3147(3148bR|2330aR) 3148(3149bR|2785aR) 3149(3150bR|3694aR) 3150(3151bR|3739aR) 3151(3152bR|3023aR) 3152(3153bR|2083aR) 3153(3154bR|3680aR) 3154(3155bR|2449aR) 3155(3156bR|1993aR) 3156(3157bR|2619aR) 3157(3158bR|2334aR) 3158(3159bR|2859aR) 3159(3160bR|3674aR) 3160(3161bR|2952aR) 3161(3162bR|2485aR) 3162(3163bR|2065aR) 3163(3164bR|2369aR) 3164(3165bR|2439aR) 3165(3166bR|2165aR) 3166(3167bR|1937aR) 3167(3168bR|2259aR) 3168(3169bR|2617aR) 3169(3170bR|2282aR) 3170(3171bR|3748aR) 3171(3172bR|2927aR) 3172(3173bR|2582aR) 3173(3174bR|1950aR) 3174(3175bR|3460aR) 3175(3176bR|2912aR) 3176(3177bR|1928aR) 3177(3178bR|2457aR) 3178(3179bR|2777aR) 3179(3180bR|3690aR) 3180(3181bR|3460aR) 3181(3182bR|3625aR) 3182(3183bR|2105aR) 3183(3184bR|3738aR) 3184(3185bR|2820aR) 3185(3186bR|3838aR) 3186(3187bR|3420aR) 3187(3188bR|2999aR) 3188(3189bR|3089aR) 3189(3190bR|2253aR) 3190(3191bR|2859aR) 3191(3192bR|2250aR) 3192(3193bR|2105aR) 3193(3194bR|2330aR) 3194(3195bR|2859aR) 3195(3196bR|3674aR) 3196(3197bR|3044aR) 3197(3198bR|3562aR) 3198(3199bR|2779aR) 3199(3200bR|3690aR) 3200(3201bR|3420aR) 3201(3202bR|3565aR) 3202(3203bR|2065aR) 3203(3204bR|2003aR) 3204(3205bR|2052aR) 3205(3206bR|2976aR) 3206(3207bR|1927aR) 3207(3208bR|1911aR) 3208(3209bR|3553aR) 3209(3210bR|3680aR) 3210(3211bR|2404aR) 3211(3212bR|3256aR) 3212(3213bR|3405aR) 3213(3214bR|1993aR) 3214(3215bR|2777aR) 3215(3216bR|3678aR) 3216(3217bR|3452aR) 3217(3218bR|3625aR) 3218(3219bR|2105aR) 3219(3220bR|3434aR) 3220(3221bR|3589aR) 3221(3222bR|2167aR) 3222(3223bR|2917aR) 3223(3224bR|3273aR) 3224(3225bR|2105aR) 3225(3226bR|3357aR) 3226(3227bR|2867aR) 3227(3228bR|1909aR) 3228(3229bR|2107aR) 3229(3230bR|3760aR) 3230(3231bR|2070aR) 3231(3232bR|1962aR) 3232(3233bR|3612aR) 3233(3234bR|1970aR) 3234(3235bR|2107aR) 3235(3236bR|3674aR) 3236(3237bR|3428aR) 3237(3238bR|2976aR) 3238(3239bR|2056aR) 3239(3240bR|2413aR) 3240(3241bR|2825aR) 3241(3242bR|3487aR) 3242(3243bR|2067aR) 3243(3244bR|2329aR) 3244(3245bR|2779aR) 3245(3246bR|3690aR) 3246(3247bR|3460aR) 3247(3248bR|3561aR) 3248(3249bR|2023aR) 3249(3250bR|2253aR) 3250(3251bR|2820aR) 3251(3252bR|2912aR) 3252(3253bR|3207aR) 3253(3254bR|1910aR) 3254(3255bR|2639aR) 3255(3256bR|2912aR) 3256(3257bR|1929aR) 3257(3258bR|1968aR) 3258(3259bR|2584aR) 3259(3260bR|2881aR) 3260(3261bR|2081aR) 3261(3262bR|3674aR) 3262(3263bR|3428aR) 3263(3264bR|2976aR) 3264(3265bR|3321aR) 3265(3266bR|2330aR) 3266(3267bR|2919aR) 3267(3268bR|3257aR) 3268(3269bR|2859aR) 3269(3270bR|3674aR) 3270(3271bR|2952aR) 3271(3272bR|2485aR) 3272(3273bR|2617aR) 3273(3274bR|2369aR) 3274(3275bR|2535aR) 3275(3276bR|3441aR) 3276(3277bR|2619aR) 3277(3278bR|3744aR) 3278(3279bR|1895aR) 3279(3280bR|1973aR) 3280(3281bR|2449aR) 3281(3282bR|3283aR) 3282(3283bR|2617aR) 3283(3284bR|2346aR) 3284(3285bR|3777aR) 3285(3286bR|1993aR) 3286(3287bR|2657aR) 3287(3288bR|3278aR) 3288(3289bR|2896aR) 3289(3290bR|3546aR) 3290(3291bR|2951aR) 3291(3292bR|1909aR) 3292(3293bR|2065aR) 3293(3294bR|3028aR) 3294(3295bR|2396aR) 3295(3296bR|2934aR) 3296(3297bR|3591aR) 3297(3298bR|3838aR) 3298(3299bR|2691aR) 3299(3300bR|3674aR) 3300(3301bR|3790aR) 3301(3302bR|2298aR) 3302(3303bR|2053aR) 3303(3304bR|2237aR) 3304(3305bR|2857aR) 3305(3306bR|2223aR) 3306(3307bR|2085aR) 3307(3308bR|3306aR) 3308(3309bR|3079aR) 3309(3310bR|2003aR) 3310(3311bR|2593aR) 3311(3312bR|3511aR) 3312(3313bR|3452aR) 3313(3314bR|2933aR) 3314(3315bR|2115aR) 3315(3316bR|3905aR) 3316(3317bR|2021aR) 3317(3318bR|3446aR) 3318(3319bR|2684aR) 3319(3320bR|2229aR) 3320(3321bR|2125aR) 3321(3322bR|2206aR) 3322(3323bR|2823aR) 3323(3324bR|3509aR) 3324(3325bR|2652aR) 3325(3326bR|2162aR) 3326(3327bR|2617aR) 3327(3328bR|3833aR) 3328(3329bR|2405aR) 3329(3330bR|2233aR) 3330(3331bR|2684aR) 3331(3332bR|2229aR) 3332(3333bR|2125aR) 3333(3334bR|1978aR) 3334(3335bR|3045aR) 3335(3336bR|3273aR) 3336(3337bR|2859aR) 3337(3338bR|2143aR) 3338(3339bR|2085aR) 3339(3340bR|3306aR) 3340(3341bR|3079aR) 3341(3342bR|2004aR) 3342(3343bR|2023aR) 3343(3344bR|1978aR) 3344(3345bR|3589aR) 3345(3346bR|2375aR) 3346(3347bR|3213aR) 3347(3348bR|2234aR) 3348(3349bR|3094aR) 3349(3350bR|2238aR) 3350(3351bR|3151aR) 3351(3352bR|2249aR) 3352(3353bR|2684aR) 3353(3354bR|3256aR) 3354(3355bR|3205aR) 3355(3356bR|3383aR) 3356(3357bR|3463aR) 3357(3358bR|1977aR) 3358(3359bR|2820aR) 3359(3360bR|3863aR) 3360(3361bR|3077aR) 3361(3362bR|2288aR) 3362(3363bR|3405aR) 3363(3364bR|3312aR) 3364(3365bR|2865aR) 3365(3366bR|3354aR) 3366(3367bR|3559aR) 3367(3368bR|3287aR) 3368(3369bR|3077aR) 3369(3370bR|2288aR) 3370(3371bR|3408aR) 3371(3372bR|2547aR) 3372(3373bR|2019aR) 3373(3374bR|3761aR) 3374(3375bR|2125aR) 3375(3376bR|2154aR) 3376(3377bR|2964aR) 3377(3378bR|2233aR) 3378(3379bR|2660aR) 3379(3380bR|2167aR) 3380(3381bR|3081aR) 3381(3382bR|2288aR) 3382(3383bR|3289aR) 3383(3384bR|3679aR) 3384(3385bR|2451aR) 3385(3386bR|3411aR) 3386(3387bR|1916aR) 3387(3388bR|2272aR) 3388(3389bR|2073aR) 3389(3390bR|3439aR) 3390(3391bR|2454aR) 3391(3392bR|1897aR) 3392(3393bR|2823aR) 3393(3394bR|2266aR) 3394(3395bR|2949aR) 3395(3396bR|3027aR) 3396(3397bR|1924aR) 3397(3398bR|1999aR) 3398(3399bR|2585aR) 3399(3400bR|3439aR) 3400(3401bR|2454aR) 3401(3402bR|2912aR) 3402(3403bR|1929aR) 3403(3404bR|3759aR) 3404(3405bR|2128aR) 3405(3406bR|2547aR) 3406(3407bR|2395aR) 3407(3408bR|3442aR) 3408(3409bR|2105aR) 3409(3410bR|3370aR) 3410(3411bR|2817aR) 3411(3412bR|3758aR) 3412(3413bR|3739aR) 3413(3414bR|3488aR) 3414(3415bR|2440aR) 3415(3416bR|2486aR) 3416(3417bR|2059aR) 3417(3418bR|2370aR) 3418(3419bR|1939aR) 3419(3420bR|3021aR) 3420(3421bR|2639aR) 3421(3422bR|3232aR) 3422(3423bR|1929aR) 3423(3424bR|1974aR) 3424(3425bR|2646aR) 3425(3426bR|1886aR) 3426(3427bR|2617aR) 3427(3428bR|3306aR) 3428(3429bR|3079aR) 3429(3430bR|1997aR) 3430(3431bR|2627aR) 3431(3432bR|2350aR) 3432(3433bR|2859aR) 3433(3434bR|3742aR) 3434(3435bR|3432aR) 3435(3436bR|2486aR) 3436(3437bR|2441aR) 3437(3438bR|3393aR) 3438(3439bR|2567aR) 3439(3440bR|3254aR) 3440(3441bR|2067aR) 3441(3442bR|3028aR) 3442(3443bR|2564aR) 3443(3444bR|2934aR) 3444(3445bR|1939aR) 3445(3446bR|3411aR) 3446(3447bR|1923aR) 3447(3448bR|2293aR) 3448(3449bR|2059aR) 3449(3450bR|2286aR) 3450(3451bR|3452aR) 3451(3452bR|3184aR) 3452(3453bR|2535aR) 3453(3454bR|3255aR) 3454(3455bR|3033aR) 3455(3456bR|2250aR) 3456(3457bR|2108aR) 3457(3458bR|2162aR) 3458(3459bR|2617aR) 3459(3460bR|3438aR) 3460(3461bR|3036aR) 3461(3462bR|2992aR) 3462(3463bR|2439aR) 3463(3464bR|3190aR) 3464(3465bR|2582aR) 3465(3466bR|2253aR) 3466(3467bR|2619aR) 3467(3468bR|2370aR) 3468(3469bR|2817aR) 3469(3470bR|3694aR) 3470(3471bR|3739aR) 3471(3472bR|1968aR) 3472(3473bR|2567aR) 3473(3474bR|2998aR) 3474(3475bR|2454aR) 3475(3476bR|2250aR) 3476(3477bR|2115aR) 3477(3478bR|3394aR) 3478(3479bR|2524aR) 3479(3480bR|2333aR) 3480(3481bR|2684aR) 3481(3482bR|2229aR) 3482(3483bR|2107aR) 3483(3484bR|3743aR) 3484(3485bR|2053aR) 3485(3486bR|2934aR) 3486(3487bR|2593aR) 3487(3488bR|3370aR) 3488(3489bR|2817aR) 3489(3490bR|3758aR) 3490(3491bR|3748aR) 3491(3492bR|3184aR) 3492(3493bR|2582aR) 3493(3494bR|3230aR) 3494(3495bR|3431aR) 3495(3496bR|2934aR) 3496(3497bR|2646aR) 3497(3498bR|1961aR) 3498(3499bR|2593aR) 3499(3500bR|3296aR) 3500(3501bR|2684aR) 3501(3502bR|2229aR) 3502(3503bR|2107aR) 3503(3504bR|3744aR) 3504(3505bR|2439aR) 3505(3506bR|3190aR) 3506(3507bR|2569aR) 3507(3508bR|1911aR) 3508(3509bR|3091aR) 3509(3510bR|2253aR) 3510(3511bR|2867aR) 3511(3512bR|2997aR) 3512(3513bR|2113aR) 3513(3514bR|2350aR) 3514(3515bR|3881aR) 3515(3516bR|2986aR) 3516(3517bR|3044aR) 3517(3518bR|2353aR) 3518(3519bR|2821aR) 3519(3520bR|1978aR) 3520(3521bR|2693aR) 3521(3522bR|3006aR) 3522(3523bR|2780aR) 3523(3524bR|2998aR) 3524(3525bR|2115aR) 3525(3526bR|3418aR) 3526(3527bR|2916aR) 3527(3528bR|1951aR) 3528(3529bR|2053aR) 3529(3530bR|2930aR) 3530(3531bR|2443aR) 3531(3532bR|3273aR) 3532(3533bR|2113aR) 3533(3534bR|2346aR) 3534(3535bR|2817aR) 3535(3536bR|3758aR) 3536(3537bR|2905aR) 3537(3538bR|3418aR) 3538(3539bR|2859aR) 3539(3540bR|3742aR) 3540(3541bR|3548aR) 3541(3542bR|3629aR) 3542(3543bR|2617aR) 3543(3544bR|3738aR) 3544(3545bR|3592aR) 3545(3546bR|3667aR) 3546(3547bR|2401aR) 3547(3548bR|3744aR) 3548(3549bR|2440aR) 3549(3550bR|2397aR) 3550(3551bR|2617aR) 3551(3552bR|3370aR) 3552(3553bR|3079aR) 3553(3554bR|1994aR) 3554(3555bR|2684aR) 3555(3556bR|1968aR) 3556(3557bR|2439aR) 3557(3558bR|3190aR) 3558(3559bR|2582aR) 3559(3560bR|2253aR) 3560(3561bR|2587aR) 3561(3562bR|3394aR) 3562(3563bR|2779aR) 3563(3564bR|3742aR) 3564(3565bR|3779aR) 3565(3566bR|2249aR) 3566(3567bR|2663aR) 3567(3568bR|3258aR) 3568(3569bR|3739aR) 3569(3570bR|3738aR) 3570(3571bR|3556aR) 3571(3572bR|3651aR) 3572(3573bR|2721aR) 3573(3574bR|2225aR) 3574(3575bR|2595aR) 3575(3576bR|3422aR) 3576(3577bR|3748aR) 3577(3578bR|1903aR) 3578(3579bR|2070aR) 3579(3580bR|2234aR) 3580(3581bR|2108aR) 3581(3582bR|3562aR) 3582(3583bR|2948aR) 3583(3584bR|1951aR) 3584(3585bR|1928aR) 3585(3586bR|2654aR) 3586(3587bR|2780aR) 3587(3588bR|2158aR) 3588(3589bR|3592aR) 3589(3590bR|3567aR) 3590(3591bR|1927aR) 3591(3592bR|3838aR) 3592(3593bR|2940aR) 3593(3594bR|1975aR) 3594(3595bR|3067aR) 3595(3596bR|3438aR) 3596(3597bR|2916aR) 3597(3598bR|1952aR) 3598(3599bR|3291aR) 3599(3600bR|3354aR) 3600(3601bR|3035aR) 3601(3602bR|3760aR) 3602(3603bR|3748aR) 3603(3604bR|1903aR) 3604(3605bR|2125aR) 3605(3606bR|3293aR) 3606(3607bR|2660aR) 3607(3608bR|3561aR) 3608(3609bR|1929aR) 3609(3610bR|3257aR) 3610(3611bR|2587aR) 3611(3612bR|2370aR) 3612(3613bR|2652aR) 3613(3614bR|1968aR) 3614(3615bR|3195aR) 3615(3616bR|2282aR) 3616(3617bR|3077aR) 3617(3618bR|1975aR) 3618(3619bR|3081aR) 3619(3620bR|2237aR) 3620(3621bR|2859aR) 3621(3622bR|3434aR) 3622(3623bR|3478aR) 3623(3624bR|2161aR) 3624(3625bR|2569aR) 3625(3626bR|3001aR) 3626(3627bR|2639aR) 3627(3628bR|1968aR) 3628(3629bR|1929aR) 3629(3630bR|2934aR) 3630(3631bR|2646aR) 3631(3632bR|2233aR) 3632(3633bR|2600aR) 3633(3634bR|2481aR) 3634(3635bR|2443aR) 3635(3636bR|2370aR) 3636(3637bR|2437aR) 3637(3638bR|2225aR) 3638(3639bR|2057aR) 3639(3640bR|3028aR) 3640(3641bR|1924aR) 3641(3642bR|2929aR) 3642(3643bR|2023aR) 3643(3644bR|3287aR) 3644(3645bR|3081aR) 3645(3646bR|2234aR) 3646(3647bR|3748aR) 3647(3648bR|1951aR) 3648(3649bR|2454aR) 3649(3650bR|2137aR) 3650(3651bR|2659aR) 3651(3652bR|3422aR) 3652(3653bR|2940aR) 3653(3654bR|3568aR) 3654(3655bR|2587aR) 3655(3656bR|3394aR) 3656(3657bR|2436aR) 3657(3658bR|3257aR) 3658(3659bR|2780aR) 3659(3660bR|3190aR) 3660(3661bR|2619aR) 3661(3662bR|3422aR) 3662(3663bR|2940aR) 3663(3664bR|1968aR) 3664(3665bR|3203aR) 3665(3666bR|3262aR) 3666(3667bR|2788aR) 3667(3668bR|1903aR) 3668(3669bR|2021aR) 3669(3670bR|2231aR) 3670(3671bR|3035aR) 3671(3672bR|3278aR) 3672(3673bR|2919aR) 3673(3674bR|1997aR) 3674(3675bR|2820aR) 3675(3676bR|2993aR) 3676(3677bR|2057aR) 3677(3678bR|3002aR) 3678(3679bR|2593aR) 3679(3680bR|2282aR) 3680(3681bR|2857aR) 3681(3682bR|3186aR) 3682(3683bR|2789aR) 3683(3684bR|2237aR) 3684(3685bR|2780aR) 3685(3686bR|1903aR) 3686(3687bR|2021aR) 3687(3688bR|2226aR) 3688(3689bR|1942aR) 3689(3690bR|1965aR) 3690(3691bR|2105aR) 3691(3692bR|3695aR) 3692(3693bR|1925aR) 3693(3694bR|2161aR) 3694(3695bR|2582aR) 3695(3696bR|2222aR) 3696(3697bR|2692aR) 3697(3698bR|2929aR) 3698(3699bR|2057aR) 3699(3700bR|3002aR) 3700(3701bR|2637aR) 3701(3702bR|1981aR) 3702(3703bR|2651aR) 3703(3704bR|3434aR) 3704(3705bR|2948aR) 3705(3706bR|3609aR) 3706(3707bR|2617aR) 3707(3708bR|3418aR) 3708(3709bR|3589aR) 3709(3710bR|1911aR) 3710(3711bR|3081aR) 3711(3712bR|2238aR) 3712(3713bR|2859aR) 3713(3714bR|1978aR) 3714(3715bR|2073aR) 3715(3716bR|3397aR) 3716(3717bR|2023aR) 3717(3718bR|1981aR) 3718(3719bR|2067aR) 3719(3720bR|3411aR) 3720(3721bR|2057aR) 3721(3722bR|1977aR) 3722(3723bR|2640aR) 3723(3724bR|3546aR) 3724(3725bR|2949aR) 3725(3726bR|1969aR) 3726(3727bR|2441aR) 3727(3728bR|3284aR) 3728(3729bR|1892aR) 3729(3730bR|2934aR) 3730(3731bR|3591aR) 3731(3732bR|3838aR) 3732(3733bR|2649aR) 3733(3734bR|3422aR) 3734(3735bR|3790aR) 3735(3736bR|2302aR) 3736(3737bR|3144aR) 3737(3738bR|2478aR) 3738(3739bR|2710aR) 3739(3740bR|3583aR) 3740(3741bR|3041aR) 3741(3742bR|3446aR) 3742(3743bR|2652aR) 3743(3744bR|3566aR) 3744(3745bR|2441aR) 3745(3746bR|3535aR) 3746(3747bR|2620aR) 3747(3748bR|1905aR) 3748(3749bR|2593aR) 3749(3750bR|3838aR) 3750(3751bR|2917aR) 3751(3752bR|3394aR) 3752(3753bR|2660aR) 3753(3754bR|3027aR) 3754(3755bR|1889aR) 3755(3756bR|3487aR) 3756(3757bR|1929aR) 3757(3758bR|3397aR) 3758(3759bR|2449aR) 3759(3760bR|2329aR) 3760(3761bR|2820aR) 3761(3762bR|3833aR) 3762(3763bR|2396aR) 3763(3764bR|2271aR) 3764(3765bR|2585aR) 3765(3766bR|3439aR) 3766(3767bR|2443aR) 3767(3768bR|3354aR) 3768(3769bR|3737aR) 3769(3770bR|2912aR) 3770(3771bR|2663aR) 3771(3772bR|2266aR) 3772(3773bR|2949aR) 3773(3774bR|3006aR) 3774(3775bR|2780aR) 3775(3776bR|3567aR) 3776(3777bR|1931aR) 3777(3778bR|3838aR) 3778(3779bR|2951aR) 3779(3780bR|2370aR) 3780(3781bR|2081aR) 3781(3782bR|3439aR) 3782(3783bR|1942aR) 3783(3784bR|2976aR) 3784(3785bR|1929aR) 3785(3786bR|1998aR) 3786(3787bR|2893aR) 3787(3788bR|3278aR) 3788(3789bR|2920aR) 3789(3790bR|2882aR) 3790(3791bR|2652aR) 3791(3792bR|2259aR) 3792(3793bR|2587aR) 3793(3794bR|3703aR) 3794(3795bR|3045aR) 3795(3796bR|3001aR) 3796(3797bR|2660aR) 3797(3798bR|3613aR) 3798(3799bR|2780aR) 3799(3800bR|2160aR) 3800(3801bR|2584aR) 3801(3802bR|2881aR) 3802(3803bR|2076aR) 3803(3804bR|1905aR) 3804(3805bR|2593aR) 3805(3806bR|3838aR) 3806(3807bR|2533aR) 3807(3808bR|2294aR) 3808(3809bR|3089aR) 3809(3810bR|3423aR) 3810(3811bR|2059aR) 3811(3812bR|2373aR) 3812(3813bR|1937aR) 3813(3814bR|2272aR) 3814(3815bR|2569aR) 3815(3816bR|3838aR) 3816(3817bR|2917aR) 3817(3818bR|3838aR) 3818(3819bR|1929aR) 3819(3820bR|3290aR) 3820(3821bR|3748aR) 3821(3822bR|1910aR) 3822(3823bR|3747aR) 3823(3824bR|1981aR) 3824(3825bR|2652aR) 3825(3826bR|3562aR) 3826(3827bR|2917aR) 3827(3828bR|2237aR) 3828(3829bR|2865aR) 3829(3830bR|1993aR) 3830(3831bR|2819aR) 3831(3832bR|3760aR) 3832(3833bR|3798aR) 3833(3834bR|2230aR) 3834(3835bR|2652aR) 3835(3836bR|3651aR) 3836(3837bR|2715aR) 3837(3838bR|2289aR) 3838(3839bR|2695aR) 3839(3840bR|3290aR) 3840(3841bR|2952aR) 3841(3842bR|2410aR) 3842(3843bR|3612aR) 3843(3844bR|3022aR) 3844(3845bR|2917aR) 3845(3846bR|2260aR) 3846(3847bR|3238aR) 3847(3848bR|2202aR) 3848(3849bR|2710aR) 3849(3850bR|1967aR) 3850(3851bR|2870aR) 3851(3852bR|2498aR) 3852(3853bR|2019aR) 3853(3854bR|3539aR) 3854(3855bR|2053aR) 3855(3856bR|3859aR) 3856(3857bR|3115aR) 3857(3858bR|2142aR) 3858(3859bR|3149aR) 3859(3860bR|2226aR) 3860(3861bR|3790aR) 3861(3862bR|2301aR) 3862(3863bR|2532aR) 3863(3864bR|2369aR) 3864(3865bR|2572aR) 3865(3866bR|3651aR) 3866(3867bR|2721aR) 3867(3868bR|2975aR) 3868(3869bR|2857aR) 3869(3870bR|3264aR) 3870(3871bR|3373aR) 3871(3872bR|3567aR) 3872(3873bR|2838aR) 3873(3874bR|2142aR) 3874(3875bR|3149aR) 3875(3876bR|2158aR) 3876(3877bR|2697aR) 3877(3878bR|3859aR) 3878(3879bR|3115aR) 3879(3880bR|2202aR) 3880(3881bR|3149aR) 3881(3882bR|2226aR) 3882(3883bR|3790aR) 3883(3884bR|2297aR) 3884(3885bR|2088aR) 3885(3886bR|2394aR) 3886(3887bR|2068aR) 3887(3888bR|2004aR) 3888(3889bR|1939aR) 3889(3890bR|3017aR) 3890(3891bR|2792aR) 3891(3892bR|2742aR) 3892(3893bR|2396aR) 3893(3894bR|2929aR) 3894(3895bR|2125aR) 3895(3896bR|2974aR) 3896(3897bR|2966aR) 3897(3898bR|3018aR) 3898(3899bR|2055aR) 3899(3900bR|3514aR) 3900(3901bR|3737aR) 3901(3902bR|1906aR) 3902(3903bR|3612aR) 3903(3904bR|3546aR) 3904(3905bR|2437aR) 3905(3906bR|3770aR) 3906(3907bR|3790aR) 3907(3908bR|2301aR) 3908(3909bR|2060aR) 3909(3910bR|3565aR) 3910(3911bR|2710aR) 3911(3912bR|3278aR) 3912(3913bR|2652aR) 3913(3914bR|2912aR) 3914(3915bR|3335aR) 3915(3916bR|3894aR) 3916(3917bR|2641aR) 3917(3918bR|2387aR) 3918(3919bR|2533aR) 3919(3920bR|3850aR) 3920(3921bR|3429aR) 3921(3922bR|2259aR) 3922(3923bR|2012aR) 3923(3924bR|2370aR) 3924(3925bR|2533aR) 3925(3926bR|3795aR) 3926(3927bR|1892aR) 3927(3928bR|2359aR) 3928(3929aR|3928aR) 3929(3930aL|4049aL) 3930(3931aL|ERROR-) 3931(3932aL|3931bL) 3932(3933aR|3932bL) 3933(3935aR|3934aR) 3934(3935bR|3934bR) 3935(3937aR|3936aR) 3936(3938bL|3936bR) 3937(3938aL|ERROR-) 3938(3939aL|3938bL) 3939(3940aR|3939bL) 3940(3942aR|3941aR) 3941(3942bR|3941bR) 3942(3944aR|3943aR) 3943(3945bL|3943bR) 3944(3945aL|ERROR-) 3945(3946aL|3945bL) 3946(3947aR|3946bL) 3947(3949aR|3948aR) 3948(3949bR|3948bR) 3949(3951aR|3950aR) 3950(3952bL|3950bR) 3951(3952aL|ERROR-) 3952(3953aL|3952bL) 3953(3954aR|3953bL) 3954(3956aR|3955aR) 3955(3956bR|3955bR) 3956(3958aR|3957aR) 3957(3959bL|3957bR) 3958(3959aL|ERROR-) 3959(3960aL|3959bL) 3960(3961aR|3960bL) 3961(3963aR|3962aR) 3962(3963bR|3962bR) 3963(3965aR|3964aR) 3964(3966bL|3964bR) 3965(3966aL|ERROR-) 3966(3967aL|3966bL) 3967(3968aR|3967bL) 3968(3970aR|3969aR) 3969(3970bR|3969bR) 3970(3972aR|3971aR) 3971(3973bL|3971bR) 3972(3973aL|ERROR-) 3973(3974aL|3973bL) 3974(3975aR|3974bL) 3975(3977aR|3976aR) 3976(3977bR|3976bR) 3977(3979aR|3978aR) 3978(3980bL|3978bR) 3979(3980aL|ERROR-) 3980(3981aL|3980bL) 3981(3982aR|3981bL) 3982(3984aR|3983aR) 3983(3984bR|3983bR) 3984(3986aR|3985aR) 3985(3987bL|3985bR) 3986(3987aL|ERROR-) 3987(3988aL|3987bL) 3988(3989aR|3988bL) 3989(3991aR|3990aR) 3990(3991bR|3990bR) 3991(3993aR|3992aR) 3992(3994bL|3992bR) 3993(3994aL|ERROR-) 3994(3995aL|3994bL) 3995(3996aR|3995bL) 3996(3998aR|3997aR) 3997(3998bR|3997bR) 3998(4000aR|3999aR) 3999(4001bL|3999bR) 4000(4001aL|ERROR-) 4001(4002aL|4001bL) 4002(4003aR|4002bL) 4003(4005aR|4004aR) 4004(4005bR|4004bR) 4005(4006aR|4007aR) 4006(4008aR|ERROR-) 4007(4009bR|4007bR) 4008(4009aL|ERROR-) 4009(4010aL|4009bR) 4010(4014aR|4011aL) 4011(4012aL|4011bL) 4012(4017aL|4012bL) 4013(4009aR|4013bR) 4014(4015bL|ERROR-) 4015(4016aL|ERROR-) 4016(0395aR|4016aL) 4017(4019bL|4018aL) 4018(4022bL|4021aL) 4019(4022aL|4022bL) 4020(4023aR|4023bR) 4021(4025bL|4024aL) 4022(4025aL|4025bL) 4023(4026aR|4026bR) 4024(4028bL|4027aL) 4025(4028aL|4028bL) 4026(4029aR|4029bR) 4027(4031bL|4030aL) 4028(4031aL|4031bL) 4029(4032aR|4032bR) 4030(4034bL|4033aL) 4031(4034aL|4034bL) 4032(4035aR|4035bR) 4033(4037bL|4036aL) 4034(4037aL|4037bL) 4035(4038aR|4038bR) 4036(4040bL|4039aL) 4037(4040aL|4040bL) 4038(4041aR|4041bR) 4039(4043bL|4042aL) 4040(4043aL|4043bL) 4041(4044aR|4044bR) 4042(4046bL|4045aL) 4043(4046aL|4046bL) 4044(4047aR|4047bR) 4045(4020bR|ERROR-) 4046(4020aR|4020bR) 4047(4048aR|4048bR) 4048(4013aR|ERROR-) 4049(0392aL|ERROR-) 4050(4051aR|ERROR-) 4051(4052aR|4052bR) 4052(4053bL|ERROR-) 4053(4054aL|4054bL) 4054(4055aR|4060bR) 4055(4056aL|4058bL) 4056(4057aL|4057bL) 4057(4054aL|4054bL) 4058(4059aL|4059bL) 4059(4054aL|4054bL) 4060(4061aL|4063bL) 4061(4062aL|4062bL) 4062(4065aL|4065bL) 4063(4064aL|4064bL) 4064(4054aL|4054bL) 4065(4066aR|4069bR) 4066(4099aR|4067bL) 4067(4068bL|4068bL) 4068(4065aL|4065bL) 4069(4137aR|4070bL) 4070(4071aL|4071aL) 4071(4054aL|4054bL) 4072(4073aR|4076bR) 4073(4072aR|4074aL) 4074(4075aR|4075aR) 4075(4081aR|4081bR) 4076(4077aL|4079aL) 4077(4078aR|4078aR) 4078(4108aR|4108bR) 4079(4080aR|4080aR) 4080(4099aR|4099bR) 4081(4082aR|4085bR) 4082(4083bL|4081bR) 4083(4084aR|4084aR) 4084(4072aR|4072bR) 4085(4086bL|4088bL) 4086(4087aR|4087aR) 4087(4090aR|4090bR) 4088(4089aR|4089aR) 4089(4099aR|4099bR) 4090(4091aR|4096bR) 4091(4092aL|4094aL) 4092(4093bR|4093bR) 4093(4072aR|4072bR) 4094(4095bR|4095bR) 4095(4081aR|4081bR) 4096(4090aR|4097aL) 4097(4098bR|4098bR) 4098(4099aR|4099bR) 4099(4100aR|4105bR) 4100(4101bL|4103bL) 4101(4102bR|4102bR) 4102(4072aR|4072bR) 4103(4104bR|4104bR) 4104(4081aR|4081bR) 4105(4106bL|4099bR) 4106(4107bR|4107bR) 4107(4090aR|4090bR) 4108(4109aR|4114bR) 4109(4110aL|4112aL) 4110(4111bL|4111bL) 4111(4117aL|4117bL) 4112(4113bR|4113bR) 4113(4081aR|4081bR) 4114(4090aR|4115aL) 4115(4116bR|4116bR) 4116(4099aR|4099bR) 4117(4118aR|4123bR) 4118(4119aL|4121bL) 4119(4120aL|4120bL) 4120(4117aL|4117bL) 4121(4122aL|4122bL) 4122(4117aL|4117bL) 4123(4124aL|4126bL) 4124(4125aL|4125bL) 4125(4128aL|4128bL) 4126(4127aL|4127bL) 4127(4117aL|4117bL) 4128(4129aR|4134bR) 4129(4130aL|4132bL) 4130(4131aR|4131bR) 4131(4065aL|4065bL) 4132(4133aR|4133bR) 4133(4065aL|4065bL) 4134(4137aR|4135bL) 4135(4136aL|4136bL) 4136(4117aL|4117bL) 4137(ERROR-|4138bR) 4138(4139aR|4139bR) 4139(4140aR|4141bR) 4140(4144aR|4139bR) 4141(4142aL|4139bR) 4142(4143aR|4143aR) 4143(4149aR|4149bR) 4144(4145aR|4146bR) 4145(4144aR|4139bR) 4146(4147aL|4139bR) 4147(4148aL|4148bL) 4148(4152aL|4152bL) 4149(4150aR|4151bR) 4150(4168aR|4149bR) 4151(4149aR|4149bR) 4152(4153aR|4158bR) 4153(4154aL|4156bL) 4154(4155aL|4155bL) 4155(4152aL|4152bL) 4156(4157aL|4157bL) 4157(4152aL|4152bL) 4158(4159aL|4161bL) 4159(4160aL|4160aL) 4160(4173aL|4173bL) 4161(4162aL|4162bL) 4162(4152aL|4152bL) 4163(4164aR|4167bR) 4164(4165aL|ERROR-) 4165(4166aR|4166bR) 4166(4054aL|4054bL) 4167(4184aR|ERROR-) 4168(4169aR|4170bR) 4169(4168aR|4149bR) 4170(4171aL|4149bR) 4171(4172aL|4172bL) 4172(4054aL|4054bL) 4173(4174aR|4179bR) 4174(4175aL|4177bL) 4175(4176aL|4176bL) 4176(4152aL|4152bL) 4177(4178aL|4178bL) 4178(4152aL|4152bL) 4179(4180aL|4182bL) 4180(4181aL|4181aL) 4181(4163aL|4163bL) 4182(4183aL|4183bL) 4183(4152aL|4152bL) 4184(4185aR|4186bR) 4185(4198aR|4189bR) 4186(4187aL|ERROR-) 4187(4188aL|4188bL) 4188(4226aL|4226bL) 4189(4190aR|4191bR) 4190(4192aR|4192bR) 4191(4192aR|4192bR) 4192(4193aR|4194bR) 4193(4195aR|4195bR) 4194(4195aR|4195bR) 4195(4196aR|4197bR) 4196(4211aR|4211bR) 4197(4211aR|4211bR) 4198(4199aR|4200bR) 4199(4224aR|4201bR) 4200(ERROR-|4201bR) 4201(4202aR|4203bR) 4202(4204aR|4201bR) 4203(4201aR|4201bR) 4204(4205aR|4208bR) 4205(4184aR|4206bL) 4206(4207aR|4207bR) 4207(4211aL|4211bL) 4208(ERROR-|4209bL) 4209(4210aR|4210bR) 4210(4214aL|4214bL) 4211(4212aR|4213bR) 4212(4184aR|4211bR) 4213(4211aR|4211bR) 4214(4215aR|4216bR) 4215(4217aR|4214bR) 4216(4214aR|4214bR) 4217(4218aR|4221bR) 4218(4198aR|4219bL) 4219(4220aR|4220bR) 4220(4214aL|4214bL) 4221(ERROR-|4222bL) 4222(4223aR|4223bR) 4223(4214aL|4214bL) 4224(ERROR-|4225bR) 4225(ERROR-|4184aR) 4226(4227aR|4232bR) 4227(4228aL|4230bL) 4228(4229aL|4229bL) 4229(4226aL|4226bL) 4230(4231aL|4231bL) 4231(4226aL|4226bL) 4232(4233aL|4235bL) 4233(4234aL|4234bL) 4234(4237aL|4237bL) 4235(4236aL|4236bL) 4236(4226aL|4226bL) 4237(4238aR|4241bR) 4238(4271aR|4239bL) 4239(4240bL|4240bL) 4240(4237aL|4237bL) 4241(4302aR|4242bL) 4242(4243aL|4243aL) 4243(4226aL|4226bL) 4244(4245aR|4248bR) 4245(4244aR|4246aL) 4246(4247aR|4247aR) 4247(4253aR|4253bR) 4248(4249aL|4251aL) 4249(4250aR|4250aR) 4250(4280aR|4280bR) 4251(4252aR|4252aR) 4252(4271aR|4271bR) 4253(4254aR|4257bR) 4254(4255bL|4253bR) 4255(4256aR|4256aR) 4256(4244aR|4244bR) 4257(4258bL|4260bL) 4258(4259aR|4259aR) 4259(4262aR|4262bR) 4260(4261aR|4261aR) 4261(4271aR|4271bR) 4262(4263aR|4268bR) 4263(4264aL|4266aL) 4264(4265bR|4265bR) 4265(4244aR|4244bR) 4266(4267bR|4267bR) 4267(4253aR|4253bR) 4268(4262aR|4269aL) 4269(4270bR|4270bR) 4270(4271aR|4271bR) 4271(4272aR|4277bR) 4272(4273bL|4275bL) 4273(4274bR|4274bR) 4274(4244aR|4244bR) 4275(4276bR|4276bR) 4276(4253aR|4253bR) 4277(4278bL|4271bR) 4278(4279bR|4279bR) 4279(4262aR|4262bR) 4280(4281bL|ERROR-) 4281(4282aL|4282bL) 4282(4283aR|4288bR) 4283(4284aL|4286bL) 4284(4285aL|4285bL) 4285(4282aL|4282bL) 4286(4287aL|4287bL) 4287(4282aL|4282bL) 4288(4289aL|4291bL) 4289(4290aL|4290bL) 4290(4293aL|4293bL) 4291(4292aL|4292bL) 4292(4282aL|4282bL) 4293(4294aR|4299bR) 4294(4295aL|4297bL) 4295(4296aR|4296bR) 4296(4237aL|4237bL) 4297(4298aR|4298bR) 4298(4237aL|4237bL) 4299(4302aR|4300bL) 4300(4301aR|4301bR) 4301(4282aL|4282bL) 4302(ERROR-|4303bR) 4303(4304aR|4304bR) 4304(4305aR|4306bR) 4305(4309aR|4304bR) 4306(4307aL|4304bR) 4307(4308aR|4308aR) 4308(4320aR|4320bR) 4309(4310aR|4315bR) 4310(4311aL|4313bL) 4311(4312aR|4312bR) 4312(4304aL|4304bL) 4313(4314aR|4314bR) 4314(4304aL|4304bL) 4315(4316aL|4318bL) 4316(4317aL|4317bL) 4317(4328aL|4328bL) 4318(4319aR|4319bR) 4319(4304aL|4304bL) 4320(4321aR|4322bR) 4321(4323aR|4320bR) 4322(4320aR|4320bR) 4323(4324aR|4325bR) 4324(4323aR|4320bR) 4325(4326aL|4320bR) 4326(4327aL|4327bL) 4327(4226aL|4226bL) 4328(4329aR|4334bR) 4329(4330aL|4332bL) 4330(4331aL|4331bL) 4331(4328aL|4328bL) 4332(4333aL|4333bL) 4333(4328aL|4328bL) 4334(4335aL|4337bL) 4335(4336aL|4336bL) 4336(4369aL|4369bL) 4337(4338aL|4338bL) 4338(4328aL|4328bL) 4339(4340aR|4341bR) 4340(4378aR|4378bR) 4341(4378aR|4378bR) 4342(4343bR|ERROR-) 4343(4344aR|4344bR) 4344(4345aR|4346bR) 4345(4344aR|4344bR) 4346(4347aR|4344bR) 4347(4348bR|ERROR-) 4348(4349aR|4349bR) 4349(4350bL|ERROR-) 4350(4351aL|4351bL) 4351(4352aR|4355bR) 4352(4353aL|ERROR-) 4353(4354aR|4354bR) 4354(4358aL|4358bL) 4355(4356aL|ERROR-) 4356(4357aL|4357bL) 4357(4351aL|4351bL) 4358(4359aR|4364bR) 4359(4360aL|4362bL) 4360(4361aL|4361bL) 4361(4358aL|4358bL) 4362(4363aL|4363bL) 4363(4358aL|4358bL) 4364(4365aL|4367bL) 4365(4366aL|4366bL) 4366(4381aL|4381bL) 4367(4368aL|4368bL) 4368(4358aL|4358bL) 4369(4370aR|4375bR) 4370(4371aL|4373bL) 4371(4372aL|4372bL) 4372(4328aL|4328bL) 4373(4374aL|4374bL) 4374(4328aL|4328bL) 4375(4339aR|4376bL) 4376(4377aL|4377bL) 4377(4328aL|4328bL) 4378(4379aR|4380bR) 4379(4342aR|4342bR) 4380(4342aR|4342bR) 4381(4382aR|4387bR) 4382(4383aL|4385bL) 4383(4384aL|4384bL) 4384(4381aL|4381bL) 4385(4386aR|4386bR) 4386(4392aL|4392bL) 4387(4388aL|4390bL) 4388(4389aR|4389aR) 4389(4381aL|4381bL) 4390(4391aR|4391bR) 4391(4392aL|4392bL) 4392(4393aR|4398bR) 4393(4394aL|4396bL) 4394(4395aL|4395bL) 4395(4405aL|4405bL) 4396(4397aL|4397bL) 4397(4392aL|4392bL) 4398(4399aL|4401bL) 4399(4400aL|4400bL) 4400(4392aL|4392bL) 4401(4402aL|4402bL) 4402(4392aL|4392bL) 4403(4404bR|ERROR-) 4404(4414aR|4414bR) 4405(4406aR|4409bR) 4406(4403aR|4407bL) 4407(4408aL|4408bL) 4408(4392aL|4392bL) 4409(4410aL|4412bL) 4410(4411aL|4411bL) 4411(4392aL|4392bL) 4412(4413aL|4413bL) 4413(4392aL|4392bL) 4414(4415aR|4420bR) 4415(4416aL|4418aL) 4416(4417aL|4417bL) 4417(4427aL|4427bL) 4418(4419bL|4419bL) 4419(4423aL|4423bL) 4420(ERROR-|4421aL) 4421(4422bL|4422bL) 4422(4425aL|4425bL) 4423(ERROR-|4424aR) 4424(4459bR|ERROR-) 4425(ERROR-|4426bR) 4426(4461bR|ERROR-) 4427(4428aR|4431bR) 4428(4436aR|4429bL) 4429(4430aL|4430bL) 4430(4427aL|4427bL) 4431(4432aL|4434bL) 4432(4433aL|4433bL) 4433(4427aL|4427bL) 4434(4435aL|4435bL) 4435(4427aL|4427bL) 4436(4437aR|4440bR) 4437(ERROR-|4438aL) 4438(4439bR|4439bR) 4439(4445aR|4445bR) 4440(4441aL|4443aL) 4441(4442aR|4442bR) 4442(4463aL|4463bL) 4443(4444bR|4444bR) 4444(4452aR|4452bR) 4445(4446aR|4447bR) 4446(ERROR-|4445bR) 4447(4448bL|4450bL) 4448(4449aR|4449aR) 4449(4463aL|4463bL) 4450(4451aR|4451aR) 4451(4452aR|4452bR) 4452(4453aR|4456bR) 4453(ERROR-|4454bL) 4454(4455bR|4455bR) 4455(4445aR|4445bR) 4456(4457bL|4452bR) 4457(4458bR|4458bR) 4458(4463aL|4463bL) 4459(ERROR-|4460bR) 4460(4466aR|4466bR) 4461(ERROR-|4462bR) 4462(4469aR|4469bR) 4463(4464aR|4465bR) 4464(4463aR|4463bR) 4465(4472aR|4463bR) 4466(4467aR|4468bR) 4467(4466aR|4466bR) 4468(4481aR|4466bR) 4469(4470aR|4471bR) 4470(4469aR|4469bR) 4471(4492aR|4469bR) 4472(4473aR|4478bR) 4473(4474aL|4476bL) 4474(4475aL|4475bL) 4475(4646aL|4646bL) 4476(4477aR|4477bR) 4477(4532aL|4532bL) 4478(ERROR-|4479bL) 4479(4480aR|4480bR) 4480(4532aL|4532bL) 4481(4482aR|4487bR) 4482(4483aL|4485aL) 4483(4484aR|4484bR) 4484(4532aL|4532bL) 4485(4486bL|4486bL) 4486(4490aL|4490bL) 4487(ERROR-|4488bL) 4488(4489aR|4489bR) 4489(4532aL|4532bL) 4490(ERROR-|4491aR) 4491(4503bL|ERROR-) 4492(4493aR|4498bR) 4493(4494aL|4496bL) 4494(4495aR|4495bR) 4495(4532aL|4532bL) 4496(4497aR|4497bR) 4497(4532aL|4532bL) 4498(ERROR-|4499aL) 4499(4500bL|4500bL) 4500(4501aL|4501bL) 4501(ERROR-|4502bR) 4502(4503bL|ERROR-) 4503(4504aR|4509bR) 4504(4505aL|4507bL) 4505(4506aL|4506bL) 4506(4503aL|4503bL) 4507(4508aL|4508bL) 4508(4503aL|4503bL) 4509(4414aR|4510bL) 4510(4511aL|4511bL) 4511(4503aL|4503bL) 4512(4513aR|ERROR-) 4513(4517aR|4514bR) 4514(4515aR|4516bR) 4515(4556aR|4556bR) 4516(4556aR|4556bR) 4517(4518aR|4521bR) 4518(4519aL|4522bR) 4519(4520bL|4520bL) 4520(4545aL|4545bL) 4521(ERROR-|4522bR) 4522(4523aR|4524bR) 4523(4525aR|4522bR) 4524(4522aR|4522bR) 4525(4526aR|4529bR) 4526(4512aR|4527bL) 4527(4528aR|4528bR) 4528(4532aL|4532bL) 4529(ERROR-|4530bL) 4530(4531aR|4531bR) 4531(4535aL|4535bL) 4532(4533aR|4534bR) 4533(4512aR|4532bR) 4534(4532aR|4532bR) 4535(4536aR|4537bR) 4536(4538aR|4535bR) 4537(4535aR|4535bR) 4538(4539aR|4542bR) 4539(4517aR|4540bL) 4540(4541aR|4541bR) 4541(4535aL|4535bL) 4542(ERROR-|4543bL) 4543(4544aR|4544bR) 4544(4535aL|4535bL) 4545(4546aR|4551bR) 4546(4547aL|4549bL) 4547(4548aL|4548bL) 4548(4545aL|4545bL) 4549(4550aL|4550bL) 4550(4545aL|4545bL) 4551(4552aL|4554bL) 4552(4553aR|4553bR) 4553(4562aL|4562bL) 4554(4555aL|4555bL) 4555(4545aL|4545bL) 4556(4557aR|4558bR) 4557(4559aR|4559bR) 4558(4559aR|4559bR) 4559(4560aR|4561bR) 4560(4532aR|4532bR) 4561(4532aR|4532bR) 4562(4563aR|4566bR) 4563(4571aR|4564bL) 4564(4565aL|4565bL) 4565(4562aL|4562bL) 4566(4567aL|4569bL) 4567(4568aL|4568bL) 4568(4562aL|4562bL) 4569(4570aL|4570bL) 4570(4562aL|4562bL) 4571(4572aR|4575bR) 4572(ERROR-|4573aL) 4573(4574aR|4574aR) 4574(4580aR|4580bR) 4575(4576aL|4578aL) 4576(4577aR|4577aR) 4577(4594aL|4594bL) 4578(4579aR|4579aR) 4579(4587aR|4587bR) 4580(4581aR|4582bR) 4581(ERROR-|4580bR) 4582(4583bL|4585bL) 4583(4584aR|4584aR) 4584(4594aL|4594bL) 4585(4586aR|4586aR) 4586(4587aR|4587bR) 4587(4588aR|4591bR) 4588(ERROR-|4589bL) 4589(4590bR|4590bR) 4590(4580aR|4580bR) 4591(4592bL|4587bR) 4592(4593bR|4593bR) 4593(4594aL|4594bL) 4594(4595aR|4600bR) 4595(4596aL|4598bL) 4596(4597aL|4597bL) 4597(4594aL|4594bL) 4598(4599aL|4599bL) 4599(4594aL|4594bL) 4600(4601aL|4603bL) 4601(4602aR|4602bR) 4602(4605aL|4605bL) 4603(4604aL|4604bL) 4604(4594aL|4594bL) 4605(4606aR|4609bR) 4606(4614aR|4607bL) 4607(4608aL|4608bL) 4608(4605aL|4605bL) 4609(4610aL|4612bL) 4610(4611aL|4611bL) 4611(4605aL|4605bL) 4612(4613aL|4613bL) 4613(4605aL|4605bL) 4614(4615aR|4618bR) 4615(ERROR-|4616aL) 4616(4617bR|4617bR) 4617(4623aR|4623bR) 4618(4619aL|4621aL) 4619(4620aR|4620bR) 4620(4637aL|4637bL) 4621(4622bR|4622bR) 4622(4630aR|4630bR) 4623(4624aR|4625bR) 4624(ERROR-|4623bR) 4625(4626bL|4628bL) 4626(4627aR|4627aR) 4627(4637aL|4637bL) 4628(4629aR|4629aR) 4629(4630aR|4630bR) 4630(4631aR|4634bR) 4631(ERROR-|4632bL) 4632(4633bR|4633bR) 4633(4623aR|4623bR) 4634(4635bL|4630bR) 4635(4636bR|4636bR) 4636(4637aL|4637bL) 4637(4638aR|4643bR) 4638(4639aL|4641bL) 4639(4640aL|4640bL) 4640(4637aL|4637bL) 4641(4642aL|4642bL) 4642(4637aL|4637bL) 4643(4414aR|4644bL) 4644(4645aL|4645bL) 4645(4637aL|4637bL) 4646(4647aR|4650bR) 4647(4655aR|4648bL) 4648(4649aL|4649bL) 4649(4646aL|4646bL) 4650(4651aL|4653bL) 4651(4652aL|4652bL) 4652(4646aL|4646bL) 4653(4654aL|4654bL) 4654(4646aL|4646bL) 4655(4656aR|4659bR) 4656(ERROR-|4657aL) 4657(4658bR|4658bR) 4658(4664aR|4664bR) 4659(4660aL|4662aL) 4660(4661aR|4661bR) 4661(4678aL|4678bL) 4662(4663bR|4663bR) 4663(4671aR|4671bR) 4664(4665aR|4666bR) 4665(ERROR-|4664bR) 4666(4667bL|4669bL) 4667(4668aR|4668aR) 4668(4678aL|4678bL) 4669(4670aR|4670aR) 4670(4671aR|4671bR) 4671(4672aR|4675bR) 4672(ERROR-|4673bL) 4673(4674bR|4674bR) 4674(4664aR|4664bR) 4675(4676bL|4671bR) 4676(4677bR|4677bR) 4677(4678aL|4678bL) 4678(4679aR|4684bR) 4679(4680aL|4682bL) 4680(4681aL|4681bL) 4681(4678aL|4678bL) 4682(4683aL|4683bL) 4683(4678aL|4678bL) 4684(4685aL|4687bL) 4685(4686aL|4686bL) 4686(4689aL|4689bL) 4687(4688aL|4688bL) 4688(4678aL|4678bL) 4689(4690aR|4695bR) 4690(4691aL|4693bL) 4691(4692aL|4692bL) 4692(4689aL|4689bL) 4693(4694aL|4694bL) 4694(4689aL|4689bL) 4695(4698aR|4696bL) 4696(4697aL|4697bL) 4697(4689aL|4689bL) 4698(4699aR|4702bR) 4699(4707aR|4700bL) 4700(4701aL|4701bL) 4701(4698aL|4698bL) 4702(4703aL|4705bL) 4703(4704aL|4704bL) 4704(4698aL|4698bL) 4705(4706aL|4706bL) 4706(4698aL|4698bL) 4707(4708aR|4711bR) 4708(ERROR-|4709aL) 4709(4710aR|4710aR) 4710(4716aR|4716bR) 4711(4712aL|4714aL) 4712(4713aR|4713aR) 4713(4730aL|4730bL) 4714(4715aR|4715aR) 4715(4723aR|4723bR) 4716(4717aR|4718bR) 4717(ERROR-|4716bR) 4718(4719bL|4721bL) 4719(4720aR|4720aR) 4720(4730aL|4730bL) 4721(4722aR|4722aR) 4722(4723aR|4723bR) 4723(4724aR|4727bR) 4724(ERROR-|4725bL) 4725(4726bR|4726bR) 4726(4716aR|4716bR) 4727(4728bL|4723bR) 4728(4729bR|4729bR) 4729(4730aL|4730bL) 4730(4731aR|4736bR) 4731(4732aL|4734bL) 4732(4733aL|4733bL) 4733(4780aL|4780bL) 4734(4735aL|4735bL) 4735(4730aL|4730bL) 4736(4737aL|4739bL) 4737(4738aL|4738bL) 4738(4730aL|4730bL) 4739(4740aL|4740bL) 4740(4730aL|4730bL) 4741(4742aR|4747bR) 4742(4743aL|4745bL) 4743(4744aL|4744bL) 4744(4741aL|4741bL) 4745(4746aR|4746bR) 4746(4750aL|4750bL) 4747(ERROR-|4748bL) 4748(4749aR|4749bR) 4749(4750aL|4750bL) 4750(4751aR|4756bR) 4751(4752aL|4754bL) 4752(4753bR|4753bR) 4753(4761aR|4761bR) 4754(4755aL|4755bL) 4755(4750aL|4750bL) 4756(4757aL|4759bL) 4757(4758aL|4758bL) 4758(4750aL|4750bL) 4759(4760aL|4760bL) 4760(4750aL|4750bL) 4761(4762aR|4763bR) 4762(4761aR|4761bR) 4763(4764aL|4761bR) 4764(4765aR|4765aR) 4765(4766aR|4766bR) 4766(4767aR|4768bR) 4767(4769aR|4766bR) 4768(4766aR|4766bR) 4769(4770bL|ERROR-) 4770(4771aL|4771bL) 4771(4772aR|4777bR) 4772(4773aL|4775bL) 4773(4774aL|4774bL) 4774(4771aL|4771bL) 4775(4776aL|4776bL) 4776(4771aL|4771bL) 4777(4802aR|4778bL) 4778(4779aL|4779bL) 4779(4771aL|4771bL) 4780(4781aR|4786bR) 4781(4782aL|4784bL) 4782(4783aL|4783bL) 4783(4791aL|4791bL) 4784(4785aL|4785bL) 4785(4730aL|4730bL) 4786(4787aL|4789bL) 4787(4788aL|4788bL) 4788(4730aL|4730bL) 4789(4790aL|4790bL) 4790(4730aL|4730bL) 4791(4792aR|4797bR) 4792(4793aL|4795bL) 4793(4794aR|4794bR) 4794(4741aL|4741bL) 4795(4796aL|4796bL) 4796(4730aL|4730bL) 4797(4798aL|4800bL) 4798(4799aL|4799bL) 4799(4730aL|4730bL) 4800(4801aL|4801bL) 4801(4730aL|4730bL) 4802(4803aR|4808bR) 4803(4804aL|4806aL) 4804(4805aL|4805bL) 4805(4815aL|4815bL) 4806(4807bL|4807bL) 4807(4811aL|4811bL) 4808(ERROR-|4809aL) 4809(4810bL|4810bL) 4810(4813aL|4813bL) 4811(ERROR-|4812aR) 4812(4847bR|ERROR-) 4813(ERROR-|4814bR) 4814(4849bR|ERROR-) 4815(4816aR|4819bR) 4816(4824aR|4817bL) 4817(4818aL|4818bL) 4818(4815aL|4815bL) 4819(4820aL|4822bL) 4820(4821aL|4821bL) 4821(4815aL|4815bL) 4822(4823aL|4823bL) 4823(4815aL|4815bL) 4824(4825aR|4828bR) 4825(ERROR-|4826aL) 4826(4827bR|4827bR) 4827(4833aR|4833bR) 4828(4829aL|4831aL) 4829(4830aR|4830bR) 4830(4851aL|4851bL) 4831(4832bR|4832bR) 4832(4840aR|4840bR) 4833(4834aR|4835bR) 4834(ERROR-|4833bR) 4835(4836bL|4838bL) 4836(4837aR|4837aR) 4837(4851aL|4851bL) 4838(4839aR|4839aR) 4839(4840aR|4840bR) 4840(4841aR|4844bR) 4841(ERROR-|4842bL) 4842(4843bR|4843bR) 4843(4833aR|4833bR) 4844(4845bL|4840bR) 4845(4846bR|4846bR) 4846(4851aL|4851bL) 4847(ERROR-|4848bR) 4848(4854aR|4854bR) 4849(ERROR-|4850bR) 4850(4857aR|4857bR) 4851(4852aR|4853bR) 4852(4851aR|4851bR) 4853(4860aR|4851bR) 4854(4855aR|4856bR) 4855(4854aR|4854bR) 4856(4869aR|4854bR) 4857(4858aR|4859bR) 4858(4857aR|4857bR) 4859(4880aR|4857bR) 4860(4861aR|4866bR) 4861(4862aL|4864bL) 4862(4863aL|4863bL) 4863(5033aL|5033bL) 4864(4865aR|4865bR) 4865(4900aL|4900bL) 4866(ERROR-|4867bL) 4867(4868aR|4868bR) 4868(4900aL|4900bL) 4869(4870aR|4875bR) 4870(4871aL|4873aL) 4871(4872aR|4872bR) 4872(4900aL|4900bL) 4873(4874bL|4874bL) 4874(4878aL|4878bL) 4875(ERROR-|4876bL) 4876(4877aR|4877bR) 4877(4900aL|4900bL) 4878(ERROR-|4879aR) 4879(4891bL|ERROR-) 4880(4881aR|4886bR) 4881(4882aL|4884bL) 4882(4883aR|4883bR) 4883(4900aL|4900bL) 4884(4885aR|4885bR) 4885(4900aL|4900bL) 4886(ERROR-|4887aL) 4887(4888bL|4888bL) 4888(4889aL|4889bL) 4889(ERROR-|4890bR) 4890(4891bL|ERROR-) 4891(4892aR|4897bR) 4892(4893aL|4895bL) 4893(4894aL|4894bL) 4894(4891aL|4891bL) 4895(4896aL|4896bL) 4896(4891aL|4891bL) 4897(4802aR|4898bL) 4898(4899aL|4899bL) 4899(4891aL|4891bL) 4900(4901aR|4902bR) 4901(4910aR|4900bR) 4902(4900aR|4900bR) 4903(4904aR|ERROR-) 4904(4905aL|4907bR) 4905(4906bL|4906bL) 4906(4932aL|4932bL) 4907(4908aR|4909bR) 4908(4943aR|4943bR) 4909(4943aR|4943bR) 4910(4911aR|4914bR) 4911(4903aR|4912bL) 4912(4913aR|4913bR) 4913(4917aL|4917bL) 4914(ERROR-|4915bL) 4915(4916aR|4916bR) 4916(4920aL|4920bL) 4917(4918aR|4919bR) 4918(4903aR|4917bR) 4919(4917aR|4917bR) 4920(4921aR|4922bR) 4921(4923aR|4920bR) 4922(4920aR|4920bR) 4923(4924aR|4929bR) 4924(4925aL|4927bL) 4925(4926bL|4926bL) 4926(4932aL|4932bL) 4927(4928aR|4928bR) 4928(4920aL|4920bL) 4929(ERROR-|4930bL) 4930(4931aR|4931bR) 4931(4920aL|4920bL) 4932(4933aR|4938bR) 4933(4934aL|4936bL) 4934(4935aL|4935bL) 4935(4932aL|4932bL) 4936(4937aL|4937bL) 4937(4932aL|4932bL) 4938(4939aL|4941bL) 4939(4940aR|4940bR) 4940(4949aL|4949bL) 4941(4942aL|4942bL) 4942(4932aL|4932bL) 4943(4944aR|4945bR) 4944(4946aR|4946bR) 4945(4946aR|4946bR) 4946(4947aR|4948bR) 4947(4917aR|4917bR) 4948(4917aR|4917bR) 4949(4950aR|4953bR) 4950(4958aR|4951bL) 4951(4952aL|4952bL) 4952(4949aL|4949bL) 4953(4954aL|4956bL) 4954(4955aL|4955bL) 4955(4949aL|4949bL) 4956(4957aL|4957bL) 4957(4949aL|4949bL) 4958(4959aR|4962bR) 4959(ERROR-|4960aL) 4960(4961aR|4961aR) 4961(4967aR|4967bR) 4962(4963aL|4965aL) 4963(4964aR|4964aR) 4964(4981aL|4981bL) 4965(4966aR|4966aR) 4966(4974aR|4974bR) 4967(4968aR|4969bR) 4968(ERROR-|4967bR) 4969(4970bL|4972bL) 4970(4971aR|4971aR) 4971(4981aL|4981bL) 4972(4973aR|4973aR) 4973(4974aR|4974bR) 4974(4975aR|4978bR) 4975(ERROR-|4976bL) 4976(4977bR|4977bR) 4977(4967aR|4967bR) 4978(4979bL|4974bR) 4979(4980bR|4980bR) 4980(4981aL|4981bL) 4981(4982aR|4987bR) 4982(4983aL|4985bL) 4983(4984aL|4984bL) 4984(4981aL|4981bL) 4985(4986aL|4986bL) 4986(4981aL|4981bL) 4987(4988aL|4990bL) 4988(4989aR|4989bR) 4989(4992aL|4992bL) 4990(4991aL|4991bL) 4991(4981aL|4981bL) 4992(4993aR|4996bR) 4993(5001aR|4994bL) 4994(4995aL|4995bL) 4995(4992aL|4992bL) 4996(4997aL|4999bL) 4997(4998aL|4998bL) 4998(4992aL|4992bL) 4999(5000aL|5000bL) 5000(4992aL|4992bL) 5001(5002aR|5005bR) 5002(ERROR-|5003aL) 5003(5004bR|5004bR) 5004(5010aR|5010bR) 5005(5006aL|5008aL) 5006(5007aR|5007bR) 5007(5024aL|5024bL) 5008(5009bR|5009bR) 5009(5017aR|5017bR) 5010(5011aR|5012bR) 5011(ERROR-|5010bR) 5012(5013bL|5015bL) 5013(5014aR|5014aR) 5014(5024aL|5024bL) 5015(5016aR|5016aR) 5016(5017aR|5017bR) 5017(5018aR|5021bR) 5018(ERROR-|5019bL) 5019(5020bR|5020bR) 5020(5010aR|5010bR) 5021(5022bL|5017bR) 5022(5023bR|5023bR) 5023(5024aL|5024bL) 5024(5025aR|5030bR) 5025(5026aL|5028bL) 5026(5027aL|5027bL) 5027(5024aL|5024bL) 5028(5029aL|5029bL) 5029(5024aL|5024bL) 5030(4802aR|5031bL) 5031(5032aL|5032bL) 5032(5024aL|5024bL) 5033(5034aR|5037bR) 5034(5042aR|5035bL) 5035(5036aL|5036bL) 5036(5033aL|5033bL) 5037(5038aL|5040bL) 5038(5039aL|5039bL) 5039(5033aL|5033bL) 5040(5041aL|5041bL) 5041(5033aL|5033bL) 5042(5043aR|5046bR) 5043(ERROR-|5044aL) 5044(5045bR|5045bR) 5045(5051aR|5051bR) 5046(5047aL|5049aL) 5047(5048aR|5048bR) 5048(5065aL|5065bL) 5049(5050bR|5050bR) 5050(5058aR|5058bR) 5051(5052aR|5053bR) 5052(ERROR-|5051bR) 5053(5054bL|5056bL) 5054(5055aR|5055aR) 5055(5065aL|5065bL) 5056(5057aR|5057aR) 5057(5058aR|5058bR) 5058(5059aR|5062bR) 5059(ERROR-|5060bL) 5060(5061bR|5061bR) 5061(5051aR|5051bR) 5062(5063bL|5058bR) 5063(5064bR|5064bR) 5064(5065aL|5065bL) 5065(5066aR|5071bR) 5066(5067aL|5069bL) 5067(5068aL|5068bL) 5068(5065aL|5065bL) 5069(5070aL|5070bL) 5070(5065aL|5065bL) 5071(5072aL|5074bL) 5072(5073aL|5073bL) 5073(5076aL|5076bL) 5074(5075aL|5075bL) 5075(5065aL|5065bL) 5076(5077aR|5082bR) 5077(5078aL|5080bL) 5078(5079aL|5079bL) 5079(5076aL|5076bL) 5080(5081aL|5081bL) 5081(5076aL|5076bL) 5082(5085aR|5083bL) 5083(5084aL|5084bL) 5084(5076aL|5076bL) 5085(5086aR|5089bR) 5086(5094aR|5087bL) 5087(5088aL|5088bL) 5088(5085aL|5085bL) 5089(5090aL|5092bL) 5090(5091aL|5091bL) 5091(5085aL|5085bL) 5092(5093aL|5093bL) 5093(5085aL|5085bL) 5094(5095aR|5098bR) 5095(ERROR-|5096aL) 5096(5097aR|5097aR) 5097(5103aR|5103bR) 5098(5099aL|5101aL) 5099(5100aR|5100aR) 5100(5117aL|5117bL) 5101(5102aR|5102aR) 5102(5110aR|5110bR) 5103(5104aR|5105bR) 5104(ERROR-|5103bR) 5105(5106bL|5108bL) 5106(5107aR|5107aR) 5107(5117aL|5117bL) 5108(5109aR|5109aR) 5109(5110aR|5110bR) 5110(5111aR|5114bR) 5111(ERROR-|5112bL) 5112(5113bR|5113bR) 5113(5103aR|5103bR) 5114(5115bL|5110bR) 5115(5116bR|5116bR) 5116(5117aL|5117bL) 5117(5118aR|5119bR) 5118(5120aR|5120bR) 5119(5120aR|5120bR) 5120(5121aR|5122bR) 5121(5123aR|5120bR) 5122(5120aR|5120bR) 5123(5124bR|ERROR-) 5124(5125aL|5125bL) 5125(5126aR|5127bR) 5126(5125aR|5125bR) 5127(5125aR|5128bL) 5128(5129aR|5129bR) 5129(5130aL|5130bL) 5130(5131aR|5132bR) 5131(5165aR|5130bR) 5132(5130aR|5130bR) 5133(5134aR|ERROR-) 5134(5200bR|ERROR-) 5135(5136aR|5141bR) 5136(5137aL|5139bL) 5137(5138aL|5138bL) 5138(5135aL|5135bL) 5139(5140aL|5140bL) 5140(5135aL|5135bL) 5141(5144aR|5142bL) 5142(5143aL|5143bL) 5143(5135aL|5135bL) 5144(5145aR|ERROR-) 5145(5146aL|5148aL) 5146(5147aR|5147bR) 5147(5150aL|5150bL) 5148(5149aR|5149aR) 5149(5200aL|5200bL) 5150(5151aR|5152bR) 5151(5150aR|5150bR) 5152(5153aL|5150bR) 5153(5154aR|5154aR) 5154(5155aR|5155bR) 5155(5156aR|5157bR) 5156(5158aR|5155bR) 5157(5155aR|5155bR) 5158(5159aR|5164bR) 5159(5160aL|5162aL) 5160(5161aR|5161bR) 5161(7699aL|7699bL) 5162(5163bL|5163bL) 5163(5238aL|5238bL) 5164(ERROR-|6887bR) 5165(5166aR|5167bR) 5166(5168aR|5130bR) 5167(5130aR|5130bR) 5168(5169aR|5172bR) 5169(5170aL|5130bR) 5170(5171bR|5171bR) 5171(5133aR|5133bR) 5172(5130aR|5130bR) 5173(5174aR|5177bR) 5174(5175bL|5173bR) 5175(5176aR|5176aR) 5176(5200aR|5200bR) 5177(5178bL|5180bL) 5178(5179aR|5179aR) 5179(5182aR|5182bR) 5180(5181aR|5181aR) 5181(5191aR|5191bR) 5182(5183aR|5188bR) 5183(5184aL|5186aL) 5184(5185bR|5185bR) 5185(5200aR|5200bR) 5186(5187bR|5187bR) 5187(5173aR|5173bR) 5188(5209aR|5189aL) 5189(5190bR|5190bR) 5190(5191aR|5191bR) 5191(5192aR|5197bR) 5192(5193bL|5195bL) 5193(5194bR|5194bR) 5194(5200aR|5200bR) 5195(5196bR|5196bR) 5196(5173aR|5173bR) 5197(5198bL|5191bR) 5198(5199bR|5199bR) 5199(5182aR|5182bR) 5200(5201aR|5204bR) 5201(5200aR|5202aL) 5202(5203aR|5203aR) 5203(5173aR|5173bR) 5204(5205aL|5207aL) 5205(5206aR|5206aR) 5206(5182aR|5182bR) 5207(5208aR|5208aR) 5208(5191aR|5191bR) 5209(5210aR|5215bR) 5210(5211aL|5213aL) 5211(5212bR|5212bR) 5212(5200aR|5200bR) 5213(5214bR|5214bR) 5214(5173aR|5173bR) 5215(5218aR|5216aL) 5216(5217bR|5217bR) 5217(5191aR|5191bR) 5218(5219bR|ERROR-) 5219(5220aL|5220bL) 5220(5221aR|5224bR) 5221(5222aL|ERROR-) 5222(5223aR|5223bR) 5223(5227aL|5227bL) 5224(5225aL|ERROR-) 5225(5226aL|5226bL) 5226(5220aL|5220bL) 5227(5228aR|5233bR) 5228(5229aL|5231bL) 5229(5230aL|5230bL) 5230(5227aL|5227bL) 5231(5232aL|5232bL) 5232(5227aL|5227bL) 5233(5234aL|5236bL) 5234(5235aL|5235bL) 5235(5135aL|5135bL) 5236(5237aL|5237bL) 5237(5227aL|5227bL) 5238(5239aR|5244bR) 5239(5240aL|5242bL) 5240(5241aL|5241bL) 5241(5238aL|5238bL) 5242(5243aL|5243bL) 5243(5238aL|5238bL) 5244(5245aL|5247bL) 5245(5246aR|5246aR) 5246(5249aL|5249bL) 5247(5248aL|5248bL) 5248(5238aL|5238bL) 5249(5250aR|5251bR) 5250(5249aR|5249bR) 5251(5252aR|5249bR) 5252(5253aR|5258bR) 5253(5254aL|5256aL) 5254(5255aL|5255bL) 5255(5292aL|5292bL) 5256(5257bL|5257bL) 5257(5261aL|5261bL) 5258(ERROR-|5259aL) 5259(5260bL|5260bL) 5260(5263aL|5263bL) 5261(ERROR-|5262aR) 5262(5267bL|ERROR-) 5263(ERROR-|5264bR) 5264(5278bL|ERROR-) 5265(5266bR|ERROR-) 5266(5289aR|5289bR) 5267(5268aR|5273bR) 5268(5269aL|5271bL) 5269(5270aL|5270bL) 5270(5267aL|5267bL) 5271(5272aL|5272bL) 5272(5267aL|5267bL) 5273(5274bL|5276bL) 5274(5275aR|5275aR) 5275(5265aR|5265bR) 5276(5277aL|5277bL) 5277(5267aL|5267bL) 5278(5279aR|5284bR) 5279(5280aL|5282bL) 5280(5281aL|5281bL) 5281(5278aL|5278bL) 5282(5283aL|5283bL) 5283(5278aL|5278bL) 5284(5285bL|5287bL) 5285(5286bR|5286bR) 5286(5265aR|5265bR) 5287(5288aL|5288bL) 5288(5278aL|5278bL) 5289(5290aR|5291bR) 5290(5289aR|5289bR) 5291(5252aR|5289bR) 5292(5293aR|5296bR) 5293(5301aR|5294bL) 5294(5295aL|5295bL) 5295(5292aL|5292bL) 5296(5297aL|5299bL) 5297(5298aL|5298bL) 5298(5292aL|5292bL) 5299(5300aL|5300bL) 5300(5292aL|5292bL) 5301(5302aR|5303bR) 5302(ERROR-|5308bR) 5303(5304bL|5306bL) 5304(5305aR|5305aR) 5305(5322aL|5322bL) 5306(5307aR|5307aR) 5307(5315aR|5315bR) 5308(5309aR|5310bR) 5309(ERROR-|5308bR) 5310(5311bL|5313bL) 5311(5312aR|5312aR) 5312(5322aL|5322bL) 5313(5314aR|5314aR) 5314(5315aR|5315bR) 5315(5316aR|5319bR) 5316(ERROR-|5317bL) 5317(5318bR|5318bR) 5318(5308aR|5308bR) 5319(5320bL|5315bR) 5320(5321bR|5321bR) 5321(5322aL|5322bL) 5322(5323aR|5326bR) 5323(5324aL|5322bR) 5324(5325bL|5325bL) 5325(5327aL|5327bL) 5326(5322aR|5322bR) 5327(5328aR|5333bR) 5328(5329aL|5331bL) 5329(5330aL|5330bL) 5330(5327aL|5327bL) 5331(5332aL|5332bL) 5332(5327aL|5327bL) 5333(5336aR|5334bL) 5334(5335aL|5335bL) 5335(5327aL|5327bL) 5336(5337aR|5338bR) 5337(5336aR|5336bR) 5338(5336aR|5339bL) 5339(5340aR|5340bR) 5340(5341aL|5341bL) 5341(5342aR|5343bR) 5342(5370aR|5341bR) 5343(5341aR|5341bR) 5344(5345aL|ERROR-) 5345(5346aL|5346bL) 5346(5347aR|5352bR) 5347(5348aL|5350bL) 5348(5349aL|5349bL) 5349(5378aL|5378bL) 5350(5351aL|5351bL) 5351(5346aL|5346bL) 5352(5353aL|5355bL) 5353(5354aL|5354bL) 5354(5346aL|5346bL) 5355(5356aL|5356bL) 5356(5346aL|5346bL) 5357(5358bL|ERROR-) 5358(5359aL|5359bL) 5359(5360aR|5365bR) 5360(5361aL|5363bL) 5361(5362aL|5362bL) 5362(5359aL|5359bL) 5363(5364aL|5364bL) 5364(5359aL|5359bL) 5365(5366aL|5368bL) 5366(5367aL|5367bL) 5367(5387aL|5387bL) 5368(5369aL|5369bL) 5369(5359aL|5359bL) 5370(5371aR|5372bR) 5371(5373aR|5341bR) 5372(5341aR|5341bR) 5373(5374aR|5377bR) 5374(5375aL|5341bR) 5375(5376aL|5376bL) 5376(5344aL|5344bL) 5377(5341aR|5341bR) 5378(5379aR|5382bR) 5379(5357aR|5380bL) 5380(5381aL|5381bL) 5381(5346aL|5346bL) 5382(5383aL|5385bL) 5383(5384aL|5384bL) 5384(5346aL|5346bL) 5385(5386aL|5386bL) 5386(5346aL|5346bL) 5387(5388aR|5391bR) 5388(5394aR|5389bL) 5389(5390bR|5390bR) 5390(5401aL|5401bL) 5391(ERROR-|5392bL) 5392(5393aL|5393aL) 5393(5387aL|5387bL) 5394(5395aR|5398bR) 5395(ERROR-|5396aL) 5396(5397aR|5397aR) 5397(5401aR|5401bR) 5398(5399aL|ERROR-) 5399(5400aR|5400aR) 5400(5425aR|5425bR) 5401(5402aR|5403bR) 5402(5401aR|5401bR) 5403(5404aR|5401bR) 5404(5405aR|5406bR) 5405(5404aR|5404bR) 5406(5407aL|5404bR) 5407(5408aR|5408aR) 5408(5409aR|5409bR) 5409(5410aR|5413bR) 5410(5411aL|5409bR) 5411(5412bL|5412bL) 5412(5414aL|5414bL) 5413(5409aR|5409bR) 5414(5415aR|5420bR) 5415(5416aL|5418bL) 5416(5417aL|5417bL) 5417(5414aL|5414bL) 5418(5419aL|5419bL) 5419(5414aL|5414bL) 5420(5421aL|5423bL) 5421(5422aL|5422bL) 5422(5387aL|5387bL) 5423(5424aL|5424bL) 5424(5414aL|5414bL) 5425(5426aR|5431bR) 5426(5427aL|5429bL) 5427(5428aL|5428bL) 5428(5425aL|5425bL) 5429(5430aL|5430bL) 5430(5425aL|5425bL) 5431(5432aL|5434bL) 5432(5433aL|5433bL) 5433(5488aL|5488bL) 5434(5435aL|5435bL) 5435(5425aL|5425bL) 5436(5437aR|5442bR) 5437(5438aL|5440bL) 5438(5439aL|5439bL) 5439(5436aL|5436bL) 5440(5441aR|5441bR) 5441(5447aL|5447bL) 5442(5443aL|5445bL) 5443(5444aL|5444bL) 5444(5436aL|5436bL) 5445(5446aR|5446bR) 5446(5447aL|5447bL) 5447(5448aR|5453bR) 5448(5449aL|5451bL) 5449(5450aR|5450bR) 5450(5458aL|5458bL) 5451(5452aL|5452bL) 5452(5447aL|5447bL) 5453(5454aL|5456bL) 5454(5455aL|5455bL) 5455(5447aL|5447bL) 5456(5457aL|5457bL) 5457(5447aL|5447bL) 5458(5459aR|5464bR) 5459(5460aL|5462bL) 5460(5461aL|5461bL) 5461(5458aL|5458bL) 5462(5463aR|5463bR) 5463(5469aL|5469bL) 5464(5465aL|5467bL) 5465(5466aL|5466bL) 5466(5458aL|5458bL) 5467(5468aR|5468bR) 5468(5469aL|5469bL) 5469(5470aR|5475bR) 5470(5471aL|5473bL) 5471(5472bR|5472bR) 5472(5480aL|5480bL) 5473(5474aL|5474bL) 5474(5469aL|5469bL) 5475(5476aL|5478bL) 5476(5477aL|5477bL) 5477(5469aL|5469bL) 5478(5479aL|5479bL) 5479(5469aL|5469bL) 5480(5481aR|5482bR) 5481(5480aR|5480bR) 5482(5499aR|5480bR) 5483(5484aR|5485bR) 5484(5483aR|5483bR) 5485(5486aL|5483bR) 5486(5487aR|5487aR) 5487(5502aR|5502bR) 5488(5489aR|5494bR) 5489(5490aL|5492bL) 5490(5491aL|5491bL) 5491(5425aL|5425bL) 5492(5493aL|5493bL) 5493(5425aL|5425bL) 5494(5495aL|5497bL) 5495(5496aL|5496bL) 5496(5436aL|5436bL) 5497(5498aL|5498bL) 5498(5425aL|5425bL) 5499(5500aR|5501bR) 5500(5480aR|5480bR) 5501(5483aR|5480bR) 5502(5503aR|5508bR) 5503(5504aL|5506aL) 5504(5505aL|5505bL) 5505(5511aL|5511bL) 5506(5507bL|5507bL) 5507(5611aL|5611bL) 5508(ERROR-|5509aL) 5509(5510bL|5510bL) 5510(5675aL|5675bL) 5511(5512aR|5517bR) 5512(5513aL|5515bL) 5513(5514bL|5514bL) 5514(5522aL|5522bL) 5515(5516aL|5516bL) 5516(5511aL|5511bL) 5517(5518aL|5520bL) 5518(5519aL|5519bL) 5519(5511aL|5511bL) 5520(5521aL|5521bL) 5521(5511aL|5511bL) 5522(5523aR|5528bR) 5523(5524aL|5526bL) 5524(5525aL|5525bL) 5525(5522aL|5522bL) 5526(5527aL|5527bL) 5527(5522aL|5522bL) 5528(5529aL|5531bL) 5529(5530aL|5530bL) 5530(5600aL|5600bL) 5531(5532aL|5532bL) 5532(5522aL|5522bL) 5533(5534aR|5539bR) 5534(5535aL|5537bL) 5535(5536aL|5536bL) 5536(5533aL|5533bL) 5537(5538aR|5538bR) 5538(5553aL|5553bL) 5539(5540aL|5542bL) 5540(5541aL|5541bL) 5541(5544aL|5544bL) 5542(5543aR|5543bR) 5543(5553aL|5553bL) 5544(5545aR|5550bR) 5545(5546aL|5548bL) 5546(5547aL|5547bL) 5547(5544aL|5544bL) 5548(5549aR|5549bR) 5549(5564aL|5564bL) 5550(ERROR-|5551bL) 5551(5552aR|5552bR) 5552(5564aL|5564bL) 5553(5554aR|5559bR) 5554(5555aL|5557bL) 5555(5556aR|5556bR) 5556(5575aL|5575bL) 5557(5558aL|5558bL) 5558(5553aL|5553bL) 5559(5560aL|5562bL) 5560(5561aL|5561bL) 5561(5564aL|5564bL) 5562(5563aL|5563bL) 5563(5553aL|5553bL) 5564(5565aR|5570bR) 5565(5566aL|5568bL) 5566(5567aR|5567bR) 5567(5582aL|5582bL) 5568(5569aL|5569bL) 5569(5564aL|5564bL) 5570(5571aL|5573bL) 5571(5572aL|5572bL) 5572(5564aL|5564bL) 5573(5574aL|5574bL) 5574(5564aL|5564bL) 5575(5576aR|5579bR) 5576(5577aL|ERROR-) 5577(5578aL|5578bL) 5578(5575aL|5575bL) 5579(5580aL|ERROR-) 5580(5581aL|5581bL) 5581(5582aL|5582bL) 5582(5583aR|5588bR) 5583(5584aL|5586bL) 5584(5585aL|5585bL) 5585(5582aL|5582bL) 5586(5587aR|5587bR) 5587(5591aL|5591bL) 5588(5739aR|5589bL) 5589(5590aR|5590bR) 5590(5591aL|5591bL) 5591(5592aR|5597bR) 5592(5593aL|5595bL) 5593(5594aR|5594bR) 5594(5533aL|5533bL) 5595(5596aL|5596bL) 5596(5591aL|5591bL) 5597(5739aR|5598bL) 5598(5599aL|5599bL) 5599(5591aL|5591bL) 5600(5601aR|5606bR) 5601(5602aL|5604bL) 5602(5603aL|5603bL) 5603(5522aL|5522bL) 5604(5605aL|5605bL) 5605(5522aL|5522bL) 5606(5607aL|5609bL) 5607(5608aL|5608bL) 5608(5533aL|5533bL) 5609(5610aL|5610bL) 5610(5522aL|5522bL) 5611(5612aR|5617bR) 5612(5613aL|5615bL) 5613(5614aL|5614bL) 5614(5611aL|5611bL) 5615(5616aL|5616bL) 5616(5611aL|5611bL) 5617(5618aL|5620bL) 5618(5619aL|5619bL) 5619(5664aL|5664bL) 5620(5621aL|5621bL) 5621(5611aL|5611bL) 5622(5623aR|5628bR) 5623(5624aL|5626bL) 5624(5625aL|5625bL) 5625(5622aL|5622bL) 5626(5627aR|5627bR) 5627(5633aL|5633bL) 5628(5629aL|5631bL) 5629(5630aL|5630bL) 5630(5622aL|5622bL) 5631(5632aR|5632bR) 5632(5633aL|5633bL) 5633(5634aR|5639bR) 5634(5635aL|5637bL) 5635(5636aR|5636bR) 5636(5644aL|5644bL) 5637(5638aL|5638bL) 5638(5633aL|5633bL) 5639(5640aL|5642bL) 5640(5641aL|5641bL) 5641(5633aL|5633bL) 5642(5643aL|5643bL) 5643(5633aL|5633bL) 5644(5645aR|5650bR) 5645(5646aL|5648bL) 5646(5647aL|5647bL) 5647(5644aL|5644bL) 5648(5649aR|5649bR) 5649(5655aL|5655bL) 5650(5651aL|5653bL) 5651(5652aL|5652bL) 5652(5644aL|5644bL) 5653(5654aR|5654bR) 5654(5655aL|5655bL) 5655(5656aR|5661bR) 5656(5657aL|5659bL) 5657(5658aR|5658bR) 5658(5622aL|5622bL) 5659(5660aL|5660bL) 5660(5655aL|5655bL) 5661(5748aR|5662bL) 5662(5663aL|5663bL) 5663(5655aL|5655bL) 5664(5665aR|5670bR) 5665(5666aL|5668bL) 5666(5667aL|5667bL) 5667(5611aL|5611bL) 5668(5669aL|5669bL) 5669(5611aL|5611bL) 5670(5671aL|5673bL) 5671(5672aL|5672bL) 5672(5622aL|5622bL) 5673(5674aL|5674bL) 5674(5611aL|5611bL) 5675(5676aR|5681bR) 5676(5677aL|5679bL) 5677(5678aL|5678bL) 5678(5675aL|5675bL) 5679(5680aL|5680bL) 5680(5675aL|5675bL) 5681(5682aL|5684bL) 5682(5683aL|5683bL) 5683(5728aL|5728bL) 5684(5685aL|5685bL) 5685(5675aL|5675bL) 5686(5687aR|5692bR) 5687(5688aL|5690bL) 5688(5689aL|5689bL) 5689(5686aL|5686bL) 5690(5691aR|5691bR) 5691(5697aL|5697bL) 5692(5693aL|5695bL) 5693(5694aL|5694bL) 5694(5686aL|5686bL) 5695(5696aR|5696bR) 5696(5697aL|5697bL) 5697(5698aR|5703bR) 5698(5699aL|5701bL) 5699(5700aR|5700bR) 5700(5708aL|5708bL) 5701(5702aL|5702bL) 5702(5697aL|5697bL) 5703(5704aL|5706bL) 5704(5705aL|5705bL) 5705(5697aL|5697bL) 5706(5707aL|5707bL) 5707(5697aL|5697bL) 5708(5709aR|5714bR) 5709(5710aL|5712bL) 5710(5711aL|5711bL) 5711(5708aL|5708bL) 5712(5713aR|5713bR) 5713(5719aL|5719bL) 5714(5715aL|5717bL) 5715(5716aL|5716bL) 5716(5708aL|5708bL) 5717(5718aR|5718bR) 5718(5719aL|5719bL) 5719(5720aR|5725bR) 5720(5721aL|5723bL) 5721(5722aR|5722bR) 5722(5686aL|5686bL) 5723(5724aL|5724bL) 5724(5719aL|5719bL) 5725(5759aR|5726bL) 5726(5727aL|5727bL) 5727(5719aL|5719bL) 5728(5729aR|5734bR) 5729(5730aL|5732bL) 5730(5731aL|5731bL) 5731(5675aL|5675bL) 5732(5733aL|5733bL) 5733(5675aL|5675bL) 5734(5735aL|5737bL) 5735(5736aL|5736bL) 5736(5686aL|5686bL) 5737(5738aL|5738bL) 5738(5675aL|5675bL) 5739(5740aR|5745bR) 5740(5741aL|5743bL) 5741(5742aL|5742bL) 5742(6003aL|6003bL) 5743(5744aR|5744bR) 5744(5897aL|5897bL) 5745(ERROR-|5746bL) 5746(5747aR|5747bR) 5747(5897aL|5897bL) 5748(5749aR|5754bR) 5749(5750aL|5752aL) 5750(5751aR|5751bR) 5751(5809aL|5809bL) 5752(5753bL|5753bL) 5753(5757aL|5757bL) 5754(ERROR-|5755bL) 5755(5756aR|5756bR) 5756(5809aL|5809bL) 5757(ERROR-|5758aR) 5758(5770bL|ERROR-) 5759(5760aR|5765bR) 5760(5761aL|5763bL) 5761(5762aR|5762bR) 5762(5820aL|5820bL) 5763(5764aR|5764bR) 5764(5820aL|5820bL) 5765(ERROR-|5766aL) 5766(5767bL|5767bL) 5767(5768aL|5768bL) 5768(ERROR-|5769bR) 5769(5773bL|ERROR-) 5770(5771aR|5772bR) 5771(5770aR|5770bR) 5772(5776aR|5770bR) 5773(5774aR|5775bR) 5774(5773aR|5773bR) 5775(5787aR|5773bR) 5776(5777aR|5778bR) 5777(5776aR|5776bR) 5778(5784aR|5776bR) 5779(5780aR|5781bR) 5780(5779aR|5779bR) 5781(5782bL|5779bR) 5782(5783aR|5783aR) 5783(5502aR|5502bR) 5784(5785aR|5786bR) 5785(5776aR|5776bR) 5786(5779aR|5776bR) 5787(5788aR|5789bR) 5788(5787aR|5787bR) 5789(5795aR|5787bR) 5790(5791aR|5792bR) 5791(5790aR|5790bR) 5792(5793bL|5790bR) 5793(5794bR|5794bR) 5794(5502aR|5502bR) 5795(5796aR|5797bR) 5796(5787aR|5787bR) 5797(5790aR|5787bR) 5798(5799aR|5804bR) 5799(5800aL|5802bL) 5800(5801bL|5801bL) 5801(5831aL|5831bL) 5802(5803aL|5803bL) 5803(5798aL|5798bL) 5804(5805aL|5807bL) 5805(5806aL|5806bL) 5806(5798aL|5798bL) 5807(5808aL|5808bL) 5808(5798aL|5798bL) 5809(5810aR|5811bR) 5810(5809aR|5809bR) 5811(5817aR|5809bR) 5812(5813aR|5814bR) 5813(5812aR|5812bR) 5814(5815bL|5812bR) 5815(5816aR|5816aR) 5816(5798aL|5798bL) 5817(5818aR|5819bR) 5818(5809aR|5809bR) 5819(5812aR|5809bR) 5820(5821aR|5822bR) 5821(5820aR|5820bR) 5822(5828aR|5820bR) 5823(5824aR|5825bR) 5824(5823aR|5823bR) 5825(5826bL|5823bR) 5826(5827bR|5827bR) 5827(5798aL|5798bL) 5828(5829aR|5830bR) 5829(5820aR|5820bR) 5830(5823aR|5820bR) 5831(5832aR|5837bR) 5832(5833aL|5835bL) 5833(5834aL|5834bL) 5834(5831aL|5831bL) 5835(5836aL|5836bL) 5836(5831aL|5831bL) 5837(5838aL|5840bL) 5838(5839aL|5839bL) 5839(5886aL|5886bL) 5840(5841aL|5841bL) 5841(5831aL|5831bL) 5842(5843aR|5848bR) 5843(5844aL|5846bL) 5844(5845aL|5845bL) 5845(5842aL|5842bL) 5846(5847aR|5847bR) 5847(5853aL|5853bL) 5848(5849aL|5851bL) 5849(5850aL|5850bL) 5850(5842aL|5842bL) 5851(5852aR|5852bR) 5852(5853aL|5853bL) 5853(5854aR|5859bR) 5854(5855aL|5857bL) 5855(5856aR|5856bR) 5856(5864aL|5864bL) 5857(5858aL|5858bL) 5858(5853aL|5853bL) 5859(5860aL|5862bL) 5860(5861aL|5861bL) 5861(5853aL|5853bL) 5862(5863aL|5863bL) 5863(5853aL|5853bL) 5864(5865aR|5870bR) 5865(5866aL|5868bL) 5866(5867aL|5867bL) 5867(5864aL|5864bL) 5868(5869aR|5869bR) 5869(5875aL|5875bL) 5870(5871aL|5873bL) 5871(5872aL|5872bL) 5872(5864aL|5864bL) 5873(5874aR|5874bR) 5874(5875aL|5875bL) 5875(5876aR|5881bR) 5876(5877aL|5879bL) 5877(5878aR|5878bR) 5878(5842aL|5842bL) 5879(5880aL|5880bL) 5880(5875aL|5875bL) 5881(5882aL|5884bL) 5882(5883aR|5883bR) 5883(5897aL|5897bL) 5884(5885aL|5885bL) 5885(5875aL|5875bL) 5886(5887aR|5892bR) 5887(5888aL|5890bL) 5888(5889aL|5889bL) 5889(5831aL|5831bL) 5890(5891aL|5891bL) 5891(5831aL|5831bL) 5892(5893aL|5895bL) 5893(5894aL|5894bL) 5894(5842aL|5842bL) 5895(5896aL|5896bL) 5896(5831aL|5831bL) 5897(5898aR|5903bR) 5898(5899aL|5901bL) 5899(5900aR|5900bR) 5900(5908aL|5908bL) 5901(5902aL|5902bL) 5902(5897aL|5897bL) 5903(5904aL|5906bL) 5904(5905aL|5905bL) 5905(5897aL|5897bL) 5906(5907aL|5907bL) 5907(5897aL|5897bL) 5908(5909aR|5914bR) 5909(5910aL|5912bL) 5910(5911aL|5911bL) 5911(5908aL|5908bL) 5912(5913aR|5913bR) 5913(5919aL|5919bL) 5914(5915aL|5917bL) 5915(5916aL|5916bL) 5916(5908aL|5908bL) 5917(5918aR|5918bR) 5918(5919aL|5919bL) 5919(5920aR|5925bR) 5920(5921aL|5923bL) 5921(5922aR|5922bR) 5922(5930aL|5930bL) 5923(5924aL|5924bL) 5924(5919aL|5919bL) 5925(5926aL|5928bL) 5926(5927aL|5927bL) 5927(5919aL|5919bL) 5928(5929aL|5929bL) 5929(5919aL|5919bL) 5930(5931aR|5936bR) 5931(5932aL|5934bL) 5932(5933aL|5933bL) 5933(5930aL|5930bL) 5934(5935aR|5935bR) 5935(5941aL|5941bL) 5936(5937aL|5939bL) 5937(5938aL|5938bL) 5938(5930aL|5930bL) 5939(5940aR|5940bR) 5940(5941aL|5941bL) 5941(5942aR|5947bR) 5942(5943aL|5945bL) 5943(5944bR|5944bR) 5944(5952aR|5952bR) 5945(5946aL|5946bL) 5946(5941aL|5941bL) 5947(5948aL|5950bL) 5948(5949aL|5949bL) 5949(5941aL|5941bL) 5950(5951aL|5951bL) 5951(5941aL|5941bL) 5952(5953aR|5954bR) 5953(5952aR|5952bR) 5954(5955aR|5952bR) 5955(5956aR|5957bR) 5956(5955aR|5955bR) 5957(5958aL|5955bR) 5958(5959aR|5959bR) 5959(5960aL|5960bL) 5960(5961aR|5964bR) 5961(5969aR|5962bL) 5962(5963aL|5963bL) 5963(5960aL|5960bL) 5964(5965aL|5967bL) 5965(5966aL|5966bL) 5966(5960aL|5960bL) 5967(5968aL|5968bL) 5968(5960aL|5960bL) 5969(5970aR|5973bR) 5970(ERROR-|5971aL) 5971(5972aR|5972aR) 5972(5978aR|5978bR) 5973(5974aL|5976aL) 5974(5975aR|5975aR) 5975(5992aL|5992bL) 5976(5977aR|5977aR) 5977(5985aR|5985bR) 5978(5979aR|5980bR) 5979(ERROR-|5978bR) 5980(5981bL|5983bL) 5981(5982aR|5982aR) 5982(5992aL|5992bL) 5983(5984aR|5984aR) 5984(5985aR|5985bR) 5985(5986aR|5989bR) 5986(ERROR-|5987bL) 5987(5988bR|5988bR) 5988(5978aR|5978bR) 5989(5990bL|5985bR) 5990(5991bR|5991bR) 5991(5992aL|5992bL) 5992(5993aR|5994bR) 5993(5992aR|5992bR) 5994(6000aR|5992bR) 5995(5996aR|5997bR) 5996(5995aR|5995bR) 5997(5998aL|5995bR) 5998(5999aR|5999aR) 5999(5502aR|5502bR) 6000(6001aR|6002bR) 6001(5992aR|5992bR) 6002(5995aR|5992bR) 6003(6004aR|6007bR) 6004(6012aR|6005bL) 6005(6006aL|6006bL) 6006(6003aL|6003bL) 6007(6008aL|6010bL) 6008(6009aL|6009bL) 6009(6003aL|6003bL) 6010(6011aL|6011bL) 6011(6003aL|6003bL) 6012(6013aR|6016bR) 6013(ERROR-|6014aL) 6014(6015bR|6015bR) 6015(6021aR|6021bR) 6016(6017aL|6019aL) 6017(6018aR|6018bR) 6018(6035aL|6035bL) 6019(6020bR|6020bR) 6020(6028aR|6028bR) 6021(6022aR|6023bR) 6022(ERROR-|6021bR) 6023(6024bL|6026bL) 6024(6025aR|6025aR) 6025(6035aL|6035bL) 6026(6027aR|6027aR) 6027(6028aR|6028bR) 6028(6029aR|6032bR) 6029(ERROR-|6030bL) 6030(6031bR|6031bR) 6031(6021aR|6021bR) 6032(6033bL|6028bR) 6033(6034bR|6034bR) 6034(6035aL|6035bL) 6035(6036aR|6037bR) 6036(6035aR|6035bR) 6037(6038aR|6035bR) 6038(6039aR|6044bR) 6039(6040aL|6042bL) 6040(6041aR|6041bR) 6041(6047aL|6047bL) 6042(6043aR|6043bR) 6043(6058aL|6058bL) 6044(ERROR-|6045bL) 6045(6046aR|6046bR) 6046(6069aL|6069bL) 6047(6048aR|6049bR) 6048(6047aR|6047bR) 6049(6055aR|6047bR) 6050(6051aR|6052bR) 6051(6050aR|6050bR) 6052(6053aL|6050bR) 6053(6054aR|6054aR) 6054(6080aL|6080bL) 6055(6056aR|6057bR) 6056(6047aR|6047bR) 6057(6050aR|6047bR) 6058(6059aR|6060bR) 6059(6058aR|6058bR) 6060(6066aR|6058bR) 6061(6062aR|6063bR) 6062(6061aR|6061bR) 6063(6064aL|6061bR) 6064(6065aR|6065aR) 6065(6085aL|6085bL) 6066(6067aR|6068bR) 6067(6058aR|6058bR) 6068(6061aR|6058bR) 6069(6070aR|6071bR) 6070(6069aR|6069bR) 6071(6077aR|6069bR) 6072(6073aR|6074bR) 6073(6072aR|6072bR) 6074(6075aL|6072bR) 6075(6076aR|6076aR) 6076(6090aL|6090bL) 6077(6078aR|6079bR) 6078(6069aR|6069bR) 6079(6072aR|6069bR) 6080(6081aR|6082bR) 6081(6080aR|6080bR) 6082(6083aL|6080bR) 6083(6084aR|6084aR) 6084(6095aR|6095bR) 6085(6086aR|6087bR) 6086(6085aR|6085bR) 6087(6088aL|6085bR) 6088(6089aR|6089aR) 6089(6118aR|6118bR) 6090(6091aR|6092bR) 6091(6090aR|6090bR) 6092(6093aL|6090bR) 6093(6094aR|6094aR) 6094(6141aR|6141bR) 6095(6096bR|ERROR-) 6096(ERROR-|6097aR) 6097(6098aR|6101bR) 6098(6164aR|6099bL) 6099(6100aL|6100bL) 6100(6104aL|6104bL) 6101(ERROR-|6102bL) 6102(6103aL|6103bL) 6103(6104aL|6104bL) 6104(ERROR-|6105aR) 6105(6106bR|ERROR-) 6106(6107aR|6108bR) 6107(6112aR|6112bR) 6108(6112aR|6112bR) 6109(6110aR|6111bR) 6110(6095aR|6109bR) 6111(6109aR|6109bR) 6112(6113aR|6114bR) 6113(6115aR|6115bR) 6114(6115aR|6115bR) 6115(6116aR|6117bR) 6116(6109aR|6109bR) 6117(6109aR|6109bR) 6118(6119bR|ERROR-) 6119(ERROR-|6120aR) 6120(6121aR|6124bR) 6121(6122aL|6171bR) 6122(6123aL|6123bL) 6123(6127aL|6127bL) 6124(ERROR-|6125bL) 6125(6126aL|6126bL) 6126(6127aL|6127bL) 6127(ERROR-|6128aR) 6128(6129bR|ERROR-) 6129(6130aR|6131bR) 6130(6135aR|6135bR) 6131(6135aR|6135bR) 6132(6133aR|6134bR) 6133(6118aR|6132bR) 6134(6132aR|6132bR) 6135(6136aR|6137bR) 6136(6138aR|6138bR) 6137(6138aR|6138bR) 6138(6139aR|6140bR) 6139(6132aR|6132bR) 6140(6132aR|6132bR) 6141(6142bR|ERROR-) 6142(ERROR-|6143aR) 6143(6144aR|6149bR) 6144(6145aL|6147bL) 6145(6146aL|6146bL) 6146(6150aL|6150bL) 6147(6148aL|6148bL) 6148(6150aL|6150bL) 6149(ERROR-|6178bR) 6150(ERROR-|6151aR) 6151(6152bR|ERROR-) 6152(6153aR|6154bR) 6153(6158aR|6158bR) 6154(6158aR|6158bR) 6155(6156aR|6157bR) 6156(6141aR|6155bR) 6157(6155aR|6155bR) 6158(6159aR|6160bR) 6159(6161aR|6161bR) 6160(6161aR|6161bR) 6161(6162aR|6163bR) 6162(6155aR|6155bR) 6163(6155aR|6155bR) 6164(6165aR|6168bR) 6165(6207aR|6166bL) 6166(6167aR|6167bR) 6167(6196aL|6196bL) 6168(ERROR-|6169bL) 6169(6170aR|6170bR) 6170(6185aL|6185bL) 6171(6172aR|6175bR) 6172(6173aL|6207bR) 6173(6174aR|6174bR) 6174(6185aL|6185bL) 6175(ERROR-|6176bL) 6176(6177aR|6177bR) 6177(6196aL|6196bL) 6178(6179aR|6184bR) 6179(6180aL|6182bL) 6180(6181aR|6181bR) 6181(6196aL|6196bL) 6182(6183aR|6183bR) 6183(6185aL|6185bL) 6184(ERROR-|6207bR) 6185(6186aR|6191bR) 6186(6187aL|6189bL) 6187(6188aL|6188bL) 6188(6185aL|6185bL) 6189(6190aL|6190bL) 6190(6185aL|6185bL) 6191(6192aL|6194bL) 6192(6193aL|6193bL) 6193(6294aL|6294bL) 6194(6195aL|6195bL) 6195(6185aL|6185bL) 6196(6197aR|6202bR) 6197(6198aL|6200bL) 6198(6199aL|6199bL) 6199(6196aL|6196bL) 6200(6201aL|6201bL) 6201(6196aL|6196bL) 6202(6203aL|6205bL) 6203(6204aL|6204bL) 6204(6358aL|6358bL) 6205(6206aL|6206bL) 6206(6196aL|6196bL) 6207(6208aR|6213bR) 6208(6209aL|6211bL) 6209(6210aR|6210bR) 6210(6494aL|6494bL) 6211(6212aR|6212bR) 6212(6560aL|6560bL) 6213(ERROR-|6214bL) 6214(6215aR|6215bR) 6215(6626aL|6626bL) 6216(6217aR|6222bR) 6217(6218bL|6220bL) 6218(6219aR|6219aR) 6219(6234aL|6234bL) 6220(6221bR|6221bR) 6221(6234aL|6234bL) 6222(ERROR-|6223aL) 6223(6224aR|6224aR) 6224(6234aL|6234bL) 6225(6226aR|6231bR) 6226(6227bL|6229aL) 6227(6228bR|6228bR) 6228(6234aL|6234bL) 6229(6230aR|6230aR) 6230(6234aL|6234bL) 6231(ERROR-|6232bL) 6232(6233aR|6233aR) 6233(6234aL|6234bL) 6234(6235aR|6236bR) 6235(6455aR|6234bR) 6236(6234aR|6234bR) 6237(6238aR|6239bR) 6238(6491aR|6491bR) 6239(6491aR|6491bR) 6240(6241aR|6246bR) 6241(6242aL|6244aL) 6242(6243bR|6243bR) 6243(6258aR|6258bR) 6244(6245bR|6245bR) 6245(6260aR|6260bR) 6246(ERROR-|6247aL) 6247(6248bR|6248bR) 6248(6262aR|6262bR) 6249(6250aR|6255bR) 6250(6251aL|6253aL) 6251(6252bL|6252bL) 6252(6258aL|6258bL) 6253(6254bL|6254bL) 6254(6260aL|6260bL) 6255(ERROR-|6256aL) 6256(6257bL|6257bL) 6257(6262aL|6262bL) 6258(ERROR-|6259aR) 6259(6264aL|6264bL) 6260(ERROR-|6261aR) 6261(6264bL|ERROR-) 6262(ERROR-|6263bR) 6263(6264bL|ERROR-) 6264(6265aR|6266bR) 6265(6264aR|6264bR) 6266(6700aR|6264bR) 6267(6268aR|6271bR) 6268(6267aR|6269bL) 6269(6270aR|6270bR) 6270(6274aL|6274bL) 6271(ERROR-|6272bL) 6272(6273aR|6273bR) 6273(6274aL|6274bL) 6274(6275aR|6278bR) 6275(6276aL|6274bR) 6276(6277bR|6277bR) 6277(6279aR|6279bR) 6278(6274aR|6274bR) 6279(6280aR|6281bR) 6280(6279aR|6279bR) 6281(6282aR|6279bR) 6282(6283aR|6284bR) 6283(6703aR|6703bR) 6284(6703aR|6703bR) 6285(6286aR|6291bR) 6286(6287aL|6289bL) 6287(6288aR|6288bR) 6288(6709aL|6709bL) 6289(6290aR|6290bR) 6290(6745aL|6745bL) 6291(ERROR-|6292bL) 6292(6293aR|6293bR) 6293(6745aL|6745bL) 6294(6295aR|6300bR) 6295(6296aL|6298bL) 6296(6297aL|6297bL) 6297(6294aL|6294bL) 6298(6299aL|6299bL) 6299(6294aL|6294bL) 6300(6301aL|6303bL) 6301(6302aL|6302bL) 6302(6347aL|6347bL) 6303(6304aL|6304bL) 6304(6294aL|6294bL) 6305(6306aR|6311bR) 6306(6307aL|6309bL) 6307(6308aL|6308bL) 6308(6305aL|6305bL) 6309(6310aR|6310bR) 6310(6316aL|6316bL) 6311(6312aL|6314bL) 6312(6313aL|6313bL) 6313(6305aL|6305bL) 6314(6315aR|6315bR) 6315(6316aL|6316bL) 6316(6317aR|6322bR) 6317(6318aL|6320bL) 6318(6319aR|6319bR) 6319(6327aL|6327bL) 6320(6321aL|6321bL) 6321(6316aL|6316bL) 6322(6323aL|6325bL) 6323(6324aL|6324bL) 6324(6316aL|6316bL) 6325(6326aL|6326bL) 6326(6316aL|6316bL) 6327(6328aR|6333bR) 6328(6329aL|6331bL) 6329(6330aL|6330bL) 6330(6327aL|6327bL) 6331(6332aR|6332bR) 6332(6338aL|6338bL) 6333(6334aL|6336bL) 6334(6335aL|6335bL) 6335(6327aL|6327bL) 6336(6337aR|6337bR) 6337(6338aL|6338bL) 6338(6339aR|6344bR) 6339(6340aL|6342bL) 6340(6341aR|6341bR) 6341(6305aL|6305bL) 6342(6343aL|6343bL) 6343(6338aL|6338bL) 6344(6422aR|6345bL) 6345(6346aL|6346bL) 6346(6338aL|6338bL) 6347(6348aR|6353bR) 6348(6349aL|6351bL) 6349(6350aL|6350bL) 6350(6294aL|6294bL) 6351(6352aL|6352bL) 6352(6294aL|6294bL) 6353(6354aL|6356bL) 6354(6355aL|6355bL) 6355(6305aL|6305bL) 6356(6357aL|6357bL) 6357(6294aL|6294bL) 6358(6359aR|6364bR) 6359(6360aL|6362bL) 6360(6361aL|6361bL) 6361(6358aL|6358bL) 6362(6363aL|6363bL) 6363(6358aL|6358bL) 6364(6365aL|6367bL) 6365(6366aL|6366bL) 6366(6411aL|6411bL) 6367(6368aL|6368bL) 6368(6358aL|6358bL) 6369(6370aR|6375bR) 6370(6371aL|6373bL) 6371(6372aL|6372bL) 6372(6369aL|6369bL) 6373(6374aR|6374bR) 6374(6380aL|6380bL) 6375(6376aL|6378bL) 6376(6377aL|6377bL) 6377(6369aL|6369bL) 6378(6379aR|6379bR) 6379(6380aL|6380bL) 6380(6381aR|6386bR) 6381(6382aL|6384bL) 6382(6383aR|6383bR) 6383(6391aL|6391bL) 6384(6385aL|6385bL) 6385(6380aL|6380bL) 6386(6387aL|6389bL) 6387(6388aL|6388bL) 6388(6380aL|6380bL) 6389(6390aL|6390bL) 6390(6380aL|6380bL) 6391(6392aR|6397bR) 6392(6393aL|6395bL) 6393(6394aL|6394bL) 6394(6391aL|6391bL) 6395(6396aR|6396bR) 6396(6402aL|6402bL) 6397(6398aL|6400bL) 6398(6399aL|6399bL) 6399(6391aL|6391bL) 6400(6401aR|6401bR) 6401(6402aL|6402bL) 6402(6403aR|6408bR) 6403(6404aL|6406bL) 6404(6405aR|6405bR) 6405(6369aL|6369bL) 6406(6407aL|6407bL) 6407(6402aL|6402bL) 6408(6425aR|6409bL) 6409(6410aL|6410bL) 6410(6402aL|6402bL) 6411(6412aR|6417bR) 6412(6413aL|6415bL) 6413(6414aL|6414bL) 6414(6358aL|6358bL) 6415(6416aL|6416bL) 6416(6358aL|6358bL) 6417(6418aL|6420bL) 6418(6419aL|6419bL) 6419(6369aL|6369bL) 6420(6421aL|6421bL) 6421(6358aL|6358bL) 6422(6423aR|6424bR) 6423(6422aR|6422bR) 6424(6225aR|6422bR) 6425(6426aR|6427bR) 6426(6425aR|6425bR) 6427(6216aR|6425bR) 6428(6429aR|6432bR) 6429(6430bL|6428bR) 6430(6431aR|6431aR) 6431(6455aR|6455bR) 6432(6433bL|6435bL) 6433(6434aR|6434aR) 6434(6437aR|6437bR) 6435(6436aR|6436aR) 6436(6446aR|6446bR) 6437(6438aR|6443bR) 6438(6439aL|6441aL) 6439(6440bR|6440bR) 6440(6455aR|6455bR) 6441(6442bR|6442bR) 6442(6428aR|6428bR) 6443(6464aR|6444aL) 6444(6445bR|6445bR) 6445(6446aR|6446bR) 6446(6447aR|6452bR) 6447(6448bL|6450bL) 6448(6449bR|6449bR) 6449(6455aR|6455bR) 6450(6451bR|6451bR) 6451(6428aR|6428bR) 6452(6453bL|6446bR) 6453(6454bR|6454bR) 6454(6437aR|6437bR) 6455(6456aR|6459bR) 6456(6455aR|6457aL) 6457(6458aR|6458aR) 6458(6428aR|6428bR) 6459(6460aL|6462aL) 6460(6461aR|6461aR) 6461(6437aR|6437bR) 6462(6463aR|6463aR) 6463(6446aR|6446bR) 6464(6465aR|6470bR) 6465(6466aL|6468aL) 6466(6467bR|6467bR) 6467(6455aR|6455bR) 6468(6469bR|6469bR) 6469(6428aR|6428bR) 6470(6473aR|6471aL) 6471(6472bR|6472bR) 6472(6446aR|6446bR) 6473(6474bR|ERROR-) 6474(6475aL|6475bL) 6475(6476aR|6479bR) 6476(6477aL|ERROR-) 6477(6478aR|6478bR) 6478(6482aL|6482bL) 6479(6480aL|ERROR-) 6480(6481aL|6481bL) 6481(6475aL|6475bL) 6482(6483aR|6488bR) 6483(6484aL|6486bL) 6484(6485aL|6485bL) 6485(6482aL|6482bL) 6486(6487aL|6487bL) 6487(6482aL|6482bL) 6488(6237aR|6489bL) 6489(6490aL|6490bL) 6490(6482aL|6482bL) 6491(6492aR|6493bR) 6492(6207aR|6207bR) 6493(6207aR|6207bR) 6494(6495aR|6500bR) 6495(6496aL|6498bL) 6496(6497aL|6497bL) 6497(6494aL|6494bL) 6498(6499aL|6499bL) 6499(6494aL|6494bL) 6500(6501aL|6503bL) 6501(6502aL|6502bL) 6502(6549aL|6549bL) 6503(6504aL|6504bL) 6504(6494aL|6494bL) 6505(6506aR|6511bR) 6506(6507aL|6509bL) 6507(6508aL|6508bL) 6508(6505aL|6505bL) 6509(6510aR|6510bR) 6510(6516aL|6516bL) 6511(6512aL|6514bL) 6512(6513aL|6513bL) 6513(6505aL|6505bL) 6514(6515aR|6515bR) 6515(6516aL|6516bL) 6516(6517aR|6522bR) 6517(6518aL|6520bL) 6518(6519aR|6519bR) 6519(6527aL|6527bL) 6520(6521aL|6521bL) 6521(6516aL|6516bL) 6522(6523aL|6525bL) 6523(6524aL|6524bL) 6524(6516aL|6516bL) 6525(6526aL|6526bL) 6526(6516aL|6516bL) 6527(6528aR|6533bR) 6528(6529aL|6531bL) 6529(6530aL|6530bL) 6530(6527aL|6527bL) 6531(6532aR|6532bR) 6532(6538aL|6538bL) 6533(6534aL|6536bL) 6534(6535aL|6535bL) 6535(6527aL|6527bL) 6536(6537aR|6537bR) 6537(6538aL|6538bL) 6538(6539aR|6544bR) 6539(6540aL|6542bL) 6540(6541aR|6541bR) 6541(6505aL|6505bL) 6542(6543aL|6543bL) 6543(6538aL|6538bL) 6544(6545aL|6547bL) 6545(6546aR|6546aR) 6546(6264aL|6264bL) 6547(6548aL|6548bL) 6548(6538aL|6538bL) 6549(6550aR|6555bR) 6550(6551aL|6553bL) 6551(6552aL|6552bL) 6552(6494aL|6494bL) 6553(6554aL|6554bL) 6554(6494aL|6494bL) 6555(6556aL|6558bL) 6556(6557aL|6557bL) 6557(6505aL|6505bL) 6558(6559aL|6559bL) 6559(6494aL|6494bL) 6560(6561aR|6566bR) 6561(6562aL|6564bL) 6562(6563aL|6563bL) 6563(6560aL|6560bL) 6564(6565aL|6565bL) 6565(6560aL|6560bL) 6566(6567aL|6569bL) 6567(6568aL|6568bL) 6568(6615aL|6615bL) 6569(6570aL|6570bL) 6570(6560aL|6560bL) 6571(6572aR|6577bR) 6572(6573aL|6575bL) 6573(6574aL|6574bL) 6574(6571aL|6571bL) 6575(6576aR|6576bR) 6576(6582aL|6582bL) 6577(6578aL|6580bL) 6578(6579aL|6579bL) 6579(6571aL|6571bL) 6580(6581aR|6581bR) 6581(6582aL|6582bL) 6582(6583aR|6588bR) 6583(6584aL|6586bL) 6584(6585aR|6585bR) 6585(6593aL|6593bL) 6586(6587aL|6587bL) 6587(6582aL|6582bL) 6588(6589aL|6591bL) 6589(6590aL|6590bL) 6590(6582aL|6582bL) 6591(6592aL|6592bL) 6592(6582aL|6582bL) 6593(6594aR|6599bR) 6594(6595aL|6597bL) 6595(6596aL|6596bL) 6596(6593aL|6593bL) 6597(6598aR|6598bR) 6598(6604aL|6604bL) 6599(6600aL|6602bL) 6600(6601aL|6601bL) 6601(6593aL|6593bL) 6602(6603aR|6603bR) 6603(6604aL|6604bL) 6604(6605aR|6610bR) 6605(6606aL|6608bL) 6606(6607aR|6607bR) 6607(6571aL|6571bL) 6608(6609aL|6609bL) 6609(6604aL|6604bL) 6610(6611aL|6613bL) 6611(6612aR|6612aR) 6612(6692aR|6692bR) 6613(6614aL|6614bL) 6614(6604aL|6604bL) 6615(6616aR|6621bR) 6616(6617aL|6619bL) 6617(6618aL|6618bL) 6618(6560aL|6560bL) 6619(6620aL|6620bL) 6620(6560aL|6560bL) 6621(6622aL|6624bL) 6622(6623aL|6623bL) 6623(6571aL|6571bL) 6624(6625aL|6625bL) 6625(6560aL|6560bL) 6626(6627aR|6632bR) 6627(6628aL|6630bL) 6628(6629aL|6629bL) 6629(6626aL|6626bL) 6630(6631aL|6631bL) 6631(6626aL|6626bL) 6632(6633aL|6635bL) 6633(6634aL|6634bL) 6634(6681aL|6681bL) 6635(6636aL|6636bL) 6636(6626aL|6626bL) 6637(6638aR|6643bR) 6638(6639aL|6641bL) 6639(6640aL|6640bL) 6640(6637aL|6637bL) 6641(6642aR|6642bR) 6642(6648aL|6648bL) 6643(6644aL|6646bL) 6644(6645aL|6645bL) 6645(6637aL|6637bL) 6646(6647aR|6647bR) 6647(6648aL|6648bL) 6648(6649aR|6654bR) 6649(6650aL|6652bL) 6650(6651aR|6651bR) 6651(6659aL|6659bL) 6652(6653aL|6653bL) 6653(6648aL|6648bL) 6654(6655aL|6657bL) 6655(6656aL|6656bL) 6656(6648aL|6648bL) 6657(6658aL|6658bL) 6658(6648aL|6648bL) 6659(6660aR|6665bR) 6660(6661aL|6663bL) 6661(6662aL|6662bL) 6662(6659aL|6659bL) 6663(6664aR|6664bR) 6664(6670aL|6670bL) 6665(6666aL|6668bL) 6666(6667aL|6667bL) 6667(6659aL|6659bL) 6668(6669aR|6669bR) 6669(6670aL|6670bL) 6670(6671aR|6676bR) 6671(6672aL|6674bL) 6672(6673aR|6673bR) 6673(6637aL|6637bL) 6674(6675aL|6675bL) 6675(6670aL|6670bL) 6676(6677aL|6679bL) 6677(6678aR|6678aR) 6678(6697aR|6697bR) 6679(6680aL|6680bL) 6680(6670aL|6670bL) 6681(6682aR|6687bR) 6682(6683aL|6685bL) 6683(6684aL|6684bL) 6684(6626aL|6626bL) 6685(6686aL|6686bL) 6686(6626aL|6626bL) 6687(6688aL|6690bL) 6688(6689aL|6689bL) 6689(6637aL|6637bL) 6690(6691aL|6691bL) 6691(6626aL|6626bL) 6692(6693aR|6694bR) 6693(6692aR|6692bR) 6694(6695aL|6692bR) 6695(6696aL|6696bL) 6696(6240aL|6240bL) 6697(6698aR|6699bR) 6698(6697aR|6697bR) 6699(6249aR|6697bR) 6700(6701aR|6702bR) 6701(6264aR|6264bR) 6702(6267aR|6264bR) 6703(6704aR|6705bR) 6704(6706aR|6706bR) 6705(6706aR|6706bR) 6706(6707aR|6708bR) 6707(6285aR|6285bR) 6708(6285aR|6285bR) 6709(6710aR|6715bR) 6710(6711aL|6713bL) 6711(6712aL|6712bL) 6712(6709aL|6709bL) 6713(6714aL|6714bL) 6714(6709aL|6709bL) 6715(6716bL|6718bL) 6716(6717aL|6717aL) 6717(6720aL|6720bL) 6718(6719aL|6719bL) 6719(6709aL|6709bL) 6720(6721aR|6726bR) 6721(6722aL|6724bL) 6722(6723aL|6723bL) 6723(6720aL|6720bL) 6724(6725aL|6725bL) 6725(6720aL|6720bL) 6726(6727aL|6729bL) 6727(6728aL|6728bL) 6728(6731aL|6731bL) 6729(6730aL|6730bL) 6730(6720aL|6720bL) 6731(6732aR|6737bR) 6732(6733bL|6735bL) 6733(6734bR|6734bR) 6734(6740aL|6740bL) 6735(6736bL|6736bL) 6736(6731aL|6731bL) 6737(ERROR-|6738bL) 6738(6739aR|6739aR) 6739(6740aL|6740bL) 6740(6741aR|6742bR) 6741(6740aR|6740bR) 6742(6743aL|6740bR) 6743(6744aR|6744aR) 6744(7887aL|7887bL) 6745(6746aR|6751bR) 6746(6747aL|6749bL) 6747(6748aL|6748bL) 6748(6745aL|6745bL) 6749(6750aL|6750bL) 6750(6745aL|6745bL) 6751(6752aL|6754bL) 6752(6753aL|6753bL) 6753(6756aL|6756bL) 6754(6755aL|6755bL) 6755(6745aL|6745bL) 6756(6757aR|6762bR) 6757(6758aL|6760bL) 6758(6759aL|6759bL) 6759(6756aL|6756bL) 6760(6761aL|6761bL) 6761(6756aL|6756bL) 6762(6763aL|6765bL) 6763(6764aL|6764bL) 6764(6767aL|6767bL) 6765(6766aL|6766bL) 6766(6756aL|6756bL) 6767(6768aR|6773bR) 6768(6769aL|6771aL) 6769(6770aR|6770bR) 6770(6776aL|6776bL) 6771(6772aL|6772aL) 6772(6767aL|6767bL) 6773(ERROR-|6774aL) 6774(6775aL|6775aL) 6775(6767aL|6767bL) 6776(6777aR|6778bR) 6777(6776aR|6776bR) 6778(6779aR|6776bR) 6779(6780aR|6781bR) 6780(6779aR|6779bR) 6781(6782bL|6779bR) 6782(6783aR|6783aR) 6783(6784aR|6784bR) 6784(6785aR|6786bR) 6785(6881aR|6881bR) 6786(6881aR|6881bR) 6787(6788aR|6793bR) 6788(6789aL|6791aL) 6789(6790aR|6790bR) 6790(6870aL|6870bL) 6791(6792bL|6792bL) 6792(6796aL|6796bL) 6793(ERROR-|6794aL) 6794(6795bL|6795bL) 6795(6807aL|6807bL) 6796(6797aR|6802bR) 6797(6798aL|6800bL) 6798(6799aL|6799bL) 6799(6796aL|6796bL) 6800(6801aL|6801bL) 6801(6796aL|6796bL) 6802(6803aL|6805bL) 6803(6804aL|6804bL) 6804(6818aL|6818bL) 6805(6806aL|6806bL) 6806(6796aL|6796bL) 6807(6808aR|6813bR) 6808(6809aL|6811bL) 6809(6810aL|6810bL) 6810(6807aL|6807bL) 6811(6812aL|6812bL) 6812(6807aL|6807bL) 6813(6814aL|6816bL) 6814(6815aL|6815bL) 6815(6845aL|6845bL) 6816(6817aL|6817bL) 6817(6807aL|6807bL) 6818(6819aR|6824bR) 6819(6820bL|6822bL) 6820(6821aR|6821aR) 6821(6854aL|6854bL) 6822(6823aL|6823bL) 6823(6818aL|6818bL) 6824(ERROR-|6825bL) 6825(6826aL|6826aL) 6826(6827aL|6827bL) 6827(6828aR|6833bR) 6828(6829bL|6831bL) 6829(6830bR|6830bR) 6830(6854aL|6854bL) 6831(6832bL|6832bL) 6832(6818aL|6818bL) 6833(ERROR-|6834bL) 6834(6835aL|6835bL) 6835(6827aL|6827bL) 6836(6837aR|6842bR) 6837(6838bL|6840bL) 6838(6839aR|6839aR) 6839(6857aL|6857bL) 6840(6841aL|6841bL) 6841(6836aL|6836bL) 6842(ERROR-|6843bL) 6843(6844aL|6844aL) 6844(6845aL|6845bL) 6845(6846aR|6851bR) 6846(6847bL|6849bL) 6847(6848bR|6848bR) 6848(6857aL|6857bL) 6849(6850bL|6850bL) 6850(6836aL|6836bL) 6851(ERROR-|6852bL) 6852(6853aL|6853bL) 6853(6845aL|6845bL) 6854(6855aR|6856bR) 6855(6854aR|6854bR) 6856(6860aR|6854bR) 6857(6858aR|6859bR) 6858(6857aR|6857bR) 6859(6865aR|6857bR) 6860(6861aR|6862bR) 6861(6860aR|6860bR) 6862(6863bL|6860bR) 6863(6864aR|6864aR) 6864(6787aR|6787bR) 6865(6866aR|6867bR) 6866(6865aR|6865bR) 6867(6868bL|6865bR) 6868(6869bR|6869bR) 6869(6787aR|6787bR) 6870(6871aR|6876bR) 6871(6872aL|6874bL) 6872(6873aL|6873bL) 6873(6870aL|6870bL) 6874(6875aL|6875bL) 6875(6870aL|6870bL) 6876(6877aL|6879bL) 6877(6878aR|6878aR) 6878(7887aL|7887bL) 6879(6880aL|6880bL) 6880(6870aL|6870bL) 6881(6882aR|6883bR) 6882(6884aR|6884bR) 6883(6884aR|6884bR) 6884(6885aR|6886bR) 6885(6787aR|6787bR) 6886(6787aR|6787bR) 6887(6888aR|6891bR) 6888(7046aR|6889aL) 6889(6890bL|6890bL) 6890(6894aL|6894bL) 6891(ERROR-|6892aL) 6892(6893bL|6893bL) 6893(6905aL|6905bL) 6894(6895aR|6900bR) 6895(6896aL|6898bL) 6896(6897aL|6897bL) 6897(6894aL|6894bL) 6898(6899aL|6899bL) 6899(6894aL|6894bL) 6900(6901bL|6903bL) 6901(6902aR|6902aR) 6902(6925aR|6925bR) 6903(6904aL|6904bL) 6904(6894aL|6894bL) 6905(6906aR|6911bR) 6906(6907aL|6909bL) 6907(6908aL|6908bL) 6908(6905aL|6905bL) 6909(6910aL|6910bL) 6910(6905aL|6905bL) 6911(6912bL|6914bL) 6912(6913bR|6913bR) 6913(6979aR|6979bR) 6914(6915aL|6915bL) 6915(6905aL|6905bL) 6916(6917aR|6920bR) 6917(6918bL|6916bR) 6918(6919aR|6919aR) 6919(6943aR|6943bR) 6920(6921bL|6923bL) 6921(6922aR|6922aR) 6922(6925aR|6925bR) 6923(6924aR|6924aR) 6924(6934aR|6934bR) 6925(6926aR|6931bR) 6926(6927aL|6929aL) 6927(6928bR|6928bR) 6928(6943aR|6943bR) 6929(6930bR|6930bR) 6930(6916aR|6916bR) 6931(6952aR|6932aL) 6932(6933bR|6933bR) 6933(6934aR|6934bR) 6934(6935aR|6940bR) 6935(6936bL|6938bL) 6936(6937bR|6937bR) 6937(6943aR|6943bR) 6938(6939bR|6939bR) 6939(6916aR|6916bR) 6940(6941bL|6934bR) 6941(6942bR|6942bR) 6942(6925aR|6925bR) 6943(6944aR|6947bR) 6944(6943aR|6945aL) 6945(6946aR|6946aR) 6946(6916aR|6916bR) 6947(6948aL|6950aL) 6948(6949aR|6949aR) 6949(6925aR|6925bR) 6950(6951aR|6951aR) 6951(6934aR|6934bR) 6952(6953aR|6958bR) 6953(6954aL|6956aL) 6954(6955bR|6955bR) 6955(6943aR|6943bR) 6956(6957bR|6957bR) 6957(6916aR|6916bR) 6958(6961aR|6959aL) 6959(6960bR|6960bR) 6960(6934aR|6934bR) 6961(6962bR|ERROR-) 6962(6963aL|6963bL) 6963(6964aR|6967bR) 6964(6965aL|ERROR-) 6965(6966aR|6966bR) 6966(7024aL|7024bL) 6967(6968aL|ERROR-) 6968(6969aL|6969bL) 6969(6963aL|6963bL) 6970(6971aR|6974bR) 6971(6972bL|6970bR) 6972(6973aR|6973aR) 6973(6997aR|6997bR) 6974(6975bL|6977bL) 6975(6976aR|6976aR) 6976(6979aR|6979bR) 6977(6978aR|6978aR) 6978(6988aR|6988bR) 6979(6980aR|6985bR) 6980(6981aL|6983aL) 6981(6982bR|6982bR) 6982(6997aR|6997bR) 6983(6984bR|6984bR) 6984(6970aR|6970bR) 6985(7006aR|6986aL) 6986(6987bR|6987bR) 6987(6988aR|6988bR) 6988(6989aR|6994bR) 6989(6990bL|6992bL) 6990(6991bR|6991bR) 6991(6997aR|6997bR) 6992(6993bR|6993bR) 6993(6970aR|6970bR) 6994(6995bL|6988bR) 6995(6996bR|6996bR) 6996(6979aR|6979bR) 6997(6998aR|7001bR) 6998(6997aR|6999aL) 6999(7000aR|7000aR) 7000(6970aR|6970bR) 7001(7002aL|7004aL) 7002(7003aR|7003aR) 7003(6979aR|6979bR) 7004(7005aR|7005aR) 7005(6988aR|6988bR) 7006(7007aR|7012bR) 7007(7008aL|7010aL) 7008(7009bR|7009bR) 7009(6997aR|6997bR) 7010(7011bR|7011bR) 7011(6970aR|6970bR) 7012(7015aR|7013aL) 7013(7014bR|7014bR) 7014(6988aR|6988bR) 7015(7016bR|ERROR-) 7016(7017aL|7017bL) 7017(7018aR|7021bR) 7018(7019aL|ERROR-) 7019(7020aR|7020bR) 7020(7035aL|7035bL) 7021(7022aL|ERROR-) 7022(7023aL|7023bL) 7023(7017aL|7017bL) 7024(7025aR|7030bR) 7025(7026aL|7028bL) 7026(7027aL|7027bL) 7027(7024aL|7024bL) 7028(7029aL|7029bL) 7029(7024aL|7024bL) 7030(7031bL|7033bL) 7031(7032aR|7032aR) 7032(6887aR|6887bR) 7033(7034aL|7034bL) 7034(7024aL|7024bL) 7035(7036aR|7041bR) 7036(7037aL|7039bL) 7037(7038aL|7038bL) 7038(7035aL|7035bL) 7039(7040aL|7040bL) 7040(7035aL|7035bL) 7041(7042bL|7044bL) 7042(7043bR|7043bR) 7043(6887aR|6887bR) 7044(7045aL|7045bL) 7045(7035aL|7035bL) 7046(7047aR|7052bR) 7047(7048aL|7050aL) 7048(7049bL|7049bL) 7049(7119aL|7119bL) 7050(7051bL|7051bL) 7051(7055aL|7055bL) 7052(ERROR-|7053aL) 7053(7054bL|7054bL) 7054(7066aL|7066bL) 7055(7056aR|7061bR) 7056(7057aL|7059bL) 7057(7058aL|7058bL) 7058(7055aL|7055bL) 7059(7060aL|7060bL) 7060(7055aL|7055bL) 7061(7062aL|7064bL) 7062(7063aL|7063bL) 7063(7077aL|7077bL) 7064(7065aL|7065bL) 7065(7055aL|7055bL) 7066(7067aR|7072bR) 7067(7068aL|7070bL) 7068(7069aL|7069bL) 7069(7066aL|7066bL) 7070(7071aL|7071bL) 7071(7066aL|7066bL) 7072(7073aL|7075bL) 7073(7074aL|7074bL) 7074(7088aL|7088bL) 7075(7076aL|7076bL) 7076(7066aL|7066bL) 7077(7078aR|7083bR) 7078(7079aL|7081bL) 7079(7080aL|7080bL) 7080(7077aL|7077bL) 7081(7082aL|7082bL) 7082(7077aL|7077bL) 7083(7084bL|7086bL) 7084(7085aR|7085aR) 7085(7099aR|7099bR) 7086(7087aL|7087bL) 7087(7077aL|7077bL) 7088(7089aR|7094bR) 7089(7090aL|7092bL) 7090(7091aL|7091bL) 7091(7088aL|7088bL) 7092(7093aL|7093bL) 7093(7088aL|7088bL) 7094(7095bL|7097bL) 7095(7096bR|7096bR) 7096(7101aR|7101bR) 7097(7098aL|7098bL) 7098(7088aL|7088bL) 7099(7100bR|ERROR-) 7100(7103aR|7103bR) 7101(7102bR|ERROR-) 7102(7106aR|7106bR) 7103(7104aR|7105bR) 7104(7103aR|7103bR) 7105(7109aR|7103bR) 7106(7107aR|7108bR) 7107(7106aR|7106bR) 7108(7114aR|7106bR) 7109(7110aR|7111bR) 7110(7109aR|7109bR) 7111(7112bL|7109bR) 7112(7113aR|7113aR) 7113(7046aR|7046bR) 7114(7115aR|7116bR) 7115(7114aR|7114bR) 7116(7117bL|7114bR) 7117(7118bR|7118bR) 7118(7046aR|7046bR) 7119(7120aR|7125bR) 7120(7121aL|7123bL) 7121(7122aL|7122bL) 7122(7119aL|7119bL) 7123(7124aL|7124bL) 7124(7119aL|7119bL) 7125(7126aL|7128bL) 7126(7127aR|7127aR) 7127(7185aR|7185bR) 7128(7129aL|7129bL) 7129(7119aL|7119bL) 7130(7131aR|7136bR) 7131(7132aL|7134bL) 7132(7133aL|7133bL) 7133(7130aL|7130bL) 7134(7135aL|7135bL) 7135(7130aL|7130bL) 7136(7137aL|7139bL) 7137(7138aL|7138bL) 7138(7141aL|7141bL) 7139(7140aL|7140bL) 7140(7130aL|7130bL) 7141(7142aR|7147bR) 7142(7143aL|7145bL) 7143(7144aL|7144bL) 7144(7241aL|7241bL) 7145(7146aL|7146bL) 7146(7141aL|7141bL) 7147(7148aL|7150bL) 7148(7149aL|7149bL) 7149(7141aL|7141bL) 7150(7151aL|7151bL) 7151(7141aL|7141bL) 7152(7153aR|7158bR) 7153(7154aL|7156bL) 7154(7155aL|7155bL) 7155(7252aL|7252bL) 7156(7157aL|7157bL) 7157(7152aL|7152bL) 7158(7159aL|7161bL) 7159(7160aL|7160bL) 7160(7152aL|7152bL) 7161(7162aL|7162bL) 7162(7152aL|7152bL) 7163(7164bL|ERROR-) 7164(7165aL|7165bL) 7165(7166aR|7171bR) 7166(7167aL|7169bL) 7167(7168aL|7168bL) 7168(7165aL|7165bL) 7169(7170aL|7170bL) 7170(7165aL|7165bL) 7171(7172aL|7174bL) 7172(7173aL|7173bL) 7173(7261aL|7261bL) 7174(7175aL|7175bL) 7175(7165aL|7165bL) 7176(7177aR|7180bR) 7177(7178bL|7176bR) 7178(7179aR|7179aR) 7179(7203aR|7203bR) 7180(7181bL|7183bL) 7181(7182aR|7182aR) 7182(7185aR|7185bR) 7183(7184aR|7184aR) 7184(7194aR|7194bR) 7185(7186aR|7191bR) 7186(7187aL|7189aL) 7187(7188bR|7188bR) 7188(7203aR|7203bR) 7189(7190bR|7190bR) 7190(7176aR|7176bR) 7191(7212aR|7192aL) 7192(7193bR|7193bR) 7193(7194aR|7194bR) 7194(7195aR|7200bR) 7195(7196bL|7198bL) 7196(7197bR|7197bR) 7197(7203aR|7203bR) 7198(7199bR|7199bR) 7199(7176aR|7176bR) 7200(7201bL|7194bR) 7201(7202bR|7202bR) 7202(7185aR|7185bR) 7203(7204aR|7207bR) 7204(7203aR|7205aL) 7205(7206aR|7206aR) 7206(7176aR|7176bR) 7207(7208aL|7210aL) 7208(7209aR|7209aR) 7209(7185aR|7185bR) 7210(7211aR|7211aR) 7211(7194aR|7194bR) 7212(7213aR|7218bR) 7213(7214aL|7216aL) 7214(7215bR|7215bR) 7215(7203aR|7203bR) 7216(7217bR|7217bR) 7217(7176aR|7176bR) 7218(7221aR|7219aL) 7219(7220bR|7220bR) 7220(7194aR|7194bR) 7221(7222bR|ERROR-) 7222(7223aL|7223bL) 7223(7224aR|7227bR) 7224(7225aL|ERROR-) 7225(7226aR|7226bR) 7226(7230aL|7230bL) 7227(7228aL|ERROR-) 7228(7229aL|7229bL) 7229(7223aL|7223bL) 7230(7231aR|7236bR) 7231(7232aL|7234bL) 7232(7233aL|7233bL) 7233(7230aL|7230bL) 7234(7235aL|7235bL) 7235(7230aL|7230bL) 7236(7237aL|7239bL) 7237(7238aL|7238bL) 7238(7130aL|7130bL) 7239(7240aL|7240bL) 7240(7230aL|7230bL) 7241(7242aR|7247bR) 7242(7243aL|7245bL) 7243(7244aL|7244bL) 7244(7152aL|7152bL) 7245(7246aL|7246bL) 7246(7141aL|7141bL) 7247(7248aL|7250bL) 7248(7249aL|7249bL) 7249(7141aL|7141bL) 7250(7251aL|7251bL) 7251(7141aL|7141bL) 7252(7253aR|7256bR) 7253(7163aR|7254bL) 7254(7255aL|7255bL) 7255(7152aL|7152bL) 7256(7257aL|7259bL) 7257(7258aL|7258bL) 7258(7152aL|7152bL) 7259(7260aL|7260bL) 7260(7152aL|7152bL) 7261(7262aR|7265bR) 7262(7268aR|7263bL) 7263(7264bR|7264bR) 7264(7275aL|7275bL) 7265(ERROR-|7266bL) 7266(7267aL|7267aL) 7267(7261aL|7261bL) 7268(7269aR|7272bR) 7269(ERROR-|7270aL) 7270(7271aR|7271aR) 7271(7275aR|7275bR) 7272(7273aL|ERROR-) 7273(7274aR|7274aR) 7274(7299aR|7299bR) 7275(7276aR|7277bR) 7276(7275aR|7275bR) 7277(7278aR|7275bR) 7278(7279aR|7280bR) 7279(7278aR|7278bR) 7280(7281aL|7278bR) 7281(7282aR|7282aR) 7282(7283aR|7283bR) 7283(7284aR|7287bR) 7284(7285aL|7283bR) 7285(7286bL|7286bL) 7286(7288aL|7288bL) 7287(7283aR|7283bR) 7288(7289aR|7294bR) 7289(7290aL|7292bL) 7290(7291aL|7291bL) 7291(7288aL|7288bL) 7292(7293aL|7293bL) 7293(7288aL|7288bL) 7294(7295aL|7297bL) 7295(7296aL|7296bL) 7296(7261aL|7261bL) 7297(7298aL|7298bL) 7298(7288aL|7288bL) 7299(7300aR|7301bR) 7300(7299aR|7299bR) 7301(7302aL|7299bR) 7302(7303aR|7303aR) 7303(7304aR|7304bR) 7304(7305aR|7310bR) 7305(7306aL|7308aL) 7306(7307aR|7307bR) 7307(7497aL|7497bL) 7308(7309bR|7309bR) 7309(7313aR|7313bR) 7310(ERROR-|7311aL) 7311(7312bR|7312bR) 7312(7318aR|7318bR) 7313(7314aR|7315bR) 7314(7313aR|7313bR) 7315(7316bL|7313bR) 7316(7317aR|7317aR) 7317(7376aR|7376bR) 7318(7319aR|7320bR) 7319(7318aR|7318bR) 7320(7321bL|7318bR) 7321(7322bR|7322bR) 7322(7430aR|7430bR) 7323(7324aR|7329bR) 7324(7325aL|7327bL) 7325(7326aL|7326bL) 7326(7323aL|7323bL) 7327(7328aL|7328bL) 7328(7323aL|7323bL) 7329(7330aL|7332bL) 7330(7331aL|7331bL) 7331(7345aL|7345bL) 7332(7333aL|7333bL) 7333(7323aL|7323bL) 7334(7335aR|7340bR) 7335(7336aL|7338bL) 7336(7337aL|7337bL) 7337(7334aL|7334bL) 7338(7339aL|7339bL) 7339(7334aL|7334bL) 7340(7341aL|7343bL) 7341(7342aL|7342bL) 7342(7356aL|7356bL) 7343(7344aL|7344bL) 7344(7334aL|7334bL) 7345(7346aR|7351bR) 7346(7347aL|7349bL) 7347(7348aL|7348bL) 7348(7345aL|7345bL) 7349(7350aL|7350bL) 7350(7345aL|7345bL) 7351(7352bL|7354bL) 7352(7353aR|7353aR) 7353(7304aR|7304bR) 7354(7355aL|7355bL) 7355(7345aL|7345bL) 7356(7357aR|7362bR) 7357(7358aL|7360bL) 7358(7359aL|7359bL) 7359(7356aL|7356bL) 7360(7361aL|7361bL) 7361(7356aL|7356bL) 7362(7363bL|7365bL) 7363(7364bR|7364bR) 7364(7304aR|7304bR) 7365(7366aL|7366bL) 7366(7356aL|7356bL) 7367(7368aR|7371bR) 7368(7369bL|7367bR) 7369(7370aR|7370aR) 7370(7394aR|7394bR) 7371(7372bL|7374bL) 7372(7373aR|7373aR) 7373(7376aR|7376bR) 7374(7375aR|7375aR) 7375(7385aR|7385bR) 7376(7377aR|7382bR) 7377(7378aL|7380aL) 7378(7379bR|7379bR) 7379(7394aR|7394bR) 7380(7381bR|7381bR) 7381(7367aR|7367bR) 7382(7403aR|7383aL) 7383(7384bR|7384bR) 7384(7385aR|7385bR) 7385(7386aR|7391bR) 7386(7387bL|7389bL) 7387(7388bR|7388bR) 7388(7394aR|7394bR) 7389(7390bR|7390bR) 7390(7367aR|7367bR) 7391(7392bL|7385bR) 7392(7393bR|7393bR) 7393(7376aR|7376bR) 7394(7395aR|7398bR) 7395(7394aR|7396aL) 7396(7397aR|7397aR) 7397(7367aR|7367bR) 7398(7399aL|7401aL) 7399(7400aR|7400aR) 7400(7376aR|7376bR) 7401(7402aR|7402aR) 7402(7385aR|7385bR) 7403(7404aR|7409bR) 7404(7405aL|7407aL) 7405(7406bR|7406bR) 7406(7394aR|7394bR) 7407(7408bR|7408bR) 7408(7367aR|7367bR) 7409(7412aR|7410aL) 7410(7411bR|7411bR) 7411(7385aR|7385bR) 7412(7413bR|ERROR-) 7413(7414aL|7414bL) 7414(7415aR|7418bR) 7415(7416aL|ERROR-) 7416(7417aR|7417bR) 7417(7475aL|7475bL) 7418(7419aL|ERROR-) 7419(7420aL|7420bL) 7420(7414aL|7414bL) 7421(7422aR|7425bR) 7422(7423bL|7421bR) 7423(7424aR|7424aR) 7424(7448aR|7448bR) 7425(7426bL|7428bL) 7426(7427aR|7427aR) 7427(7430aR|7430bR) 7428(7429aR|7429aR) 7429(7439aR|7439bR) 7430(7431aR|7436bR) 7431(7432aL|7434aL) 7432(7433bR|7433bR) 7433(7448aR|7448bR) 7434(7435bR|7435bR) 7435(7421aR|7421bR) 7436(7457aR|7437aL) 7437(7438bR|7438bR) 7438(7439aR|7439bR) 7439(7440aR|7445bR) 7440(7441bL|7443bL) 7441(7442bR|7442bR) 7442(7448aR|7448bR) 7443(7444bR|7444bR) 7444(7421aR|7421bR) 7445(7446bL|7439bR) 7446(7447bR|7447bR) 7447(7430aR|7430bR) 7448(7449aR|7452bR) 7449(7448aR|7450aL) 7450(7451aR|7451aR) 7451(7421aR|7421bR) 7452(7453aL|7455aL) 7453(7454aR|7454aR) 7454(7430aR|7430bR) 7455(7456aR|7456aR) 7456(7439aR|7439bR) 7457(7458aR|7463bR) 7458(7459aL|7461aL) 7459(7460bR|7460bR) 7460(7448aR|7448bR) 7461(7462bR|7462bR) 7462(7421aR|7421bR) 7463(7466aR|7464aL) 7464(7465bR|7465bR) 7465(7439aR|7439bR) 7466(7467bR|ERROR-) 7467(7468aL|7468bL) 7468(7469aR|7472bR) 7469(7470aL|ERROR-) 7470(7471aR|7471bR) 7471(7486aL|7486bL) 7472(7473aL|ERROR-) 7473(7474aL|7474bL) 7474(7468aL|7468bL) 7475(7476aR|7481bR) 7476(7477aL|7479bL) 7477(7478aL|7478bL) 7478(7475aL|7475bL) 7479(7480aL|7480bL) 7480(7475aL|7475bL) 7481(7482aL|7484bL) 7482(7483aL|7483bL) 7483(7323aL|7323bL) 7484(7485aL|7485bL) 7485(7475aL|7475bL) 7486(7487aR|7492bR) 7487(7488aL|7490bL) 7488(7489aL|7489bL) 7489(7486aL|7486bL) 7490(7491aL|7491bL) 7491(7486aL|7486bL) 7492(7493aL|7495bL) 7493(7494aL|7494bL) 7494(7334aL|7334bL) 7495(7496aL|7496bL) 7496(7486aL|7486bL) 7497(7498aR|7503bR) 7498(7499aL|7501bL) 7499(7500aL|7500bL) 7500(7497aL|7497bL) 7501(7502aL|7502bL) 7502(7497aL|7497bL) 7503(7506aR|7504bL) 7504(7505aL|7505bL) 7505(7497aL|7497bL) 7506(7507aR|7510bR) 7507(7506aR|7508bL) 7508(7509aR|7509bR) 7509(7513aL|7513bL) 7510(ERROR-|7511bL) 7511(7512aR|7512bR) 7512(7513aL|7513bL) 7513(7514aR|7515bR) 7514(7516aR|7513bR) 7515(7513aR|7513bR) 7516(7517bR|ERROR-) 7517(7518aR|7518bR) 7518(7519aR|7520bR) 7519(7518aR|7518bR) 7520(7521aR|7518bR) 7521(7522aR|7523bR) 7522(7521aR|7521bR) 7523(7524aL|7521bR) 7524(7525aR|7525aR) 7525(7526aR|7526bR) 7526(7527aR|7532bR) 7527(7528aL|7530bL) 7528(7529aR|7529bR) 7529(7535aL|7535bL) 7530(7531aR|7531bR) 7531(7046aL|7046bL) 7532(ERROR-|7533bL) 7533(7534aR|7534bR) 7534(7046aL|7046bL) 7535(7536aR|7541bR) 7536(7537aL|7539bL) 7537(7538aL|7538bL) 7538(7535aL|7535bL) 7539(7540aL|7540bL) 7540(7535aL|7535bL) 7541(7542aL|7544bL) 7542(7543aL|7543bL) 7543(7546aL|7546bL) 7544(7545aL|7545bL) 7545(7535aL|7535bL) 7546(7547aR|7552bR) 7547(7548aL|7550bL) 7548(7549aL|7549bL) 7549(7546aL|7546bL) 7550(7551aL|7551bL) 7551(7546aL|7546bL) 7552(7553aL|7555bL) 7553(7554aR|7554aR) 7554(7557aL|7557bL) 7555(7556aL|7556bL) 7556(7546aL|7546bL) 7557(7558aR|7563bR) 7558(7559aL|7561bL) 7559(7560aL|7560bL) 7560(7557aL|7557bL) 7561(7562aR|7562bR) 7562(7566aL|7566bL) 7563(ERROR-|7564bL) 7564(7565aR|7565bR) 7565(7566aL|7566bL) 7566(7567aR|7572bR) 7567(7568aL|7570bL) 7568(7569bR|7569bR) 7569(7577aL|7577bL) 7570(7571aL|7571bL) 7571(7566aL|7566bL) 7572(7573aL|7575bL) 7573(7574aL|7574bL) 7574(7566aL|7566bL) 7575(7576aL|7576bL) 7576(7566aL|7566bL) 7577(7578aR|7583bR) 7578(7579aL|7581aL) 7579(7580aL|7580bL) 7580(7618aL|7618bL) 7581(7582bL|7582bL) 7582(7586aL|7586bL) 7583(7594aR|7584aL) 7584(7585bL|7585bL) 7585(7588aL|7588bL) 7586(ERROR-|7587aR) 7587(7590bR|ERROR-) 7588(ERROR-|7589bR) 7589(7592bR|ERROR-) 7590(ERROR-|7591bR) 7591(7599aR|7599bR) 7592(ERROR-|7593bR) 7593(7604aR|7604bR) 7594(7595aR|7596bR) 7595(7594aR|7594bR) 7596(7597aL|7594bR) 7597(7598aR|7598aR) 7598(7643aR|7643bR) 7599(7600aR|7601bR) 7600(7599aR|7599bR) 7601(7602bL|7599bR) 7602(7603aR|7603aR) 7603(7643aR|7643bR) 7604(7605aR|7606bR) 7605(7604aR|7604bR) 7606(7607bL|7604bR) 7607(7608bR|7608bR) 7608(7643aR|7643bR) 7609(7610aR|7615bR) 7610(7611aL|7613bL) 7611(7612aL|7612bL) 7612(7609aL|7609bL) 7613(7614aL|7614bL) 7614(7609aL|7609bL) 7615(7577aR|7616bL) 7616(7617aL|7617bL) 7617(7609aL|7609bL) 7618(ERROR-|7619bR) 7619(7620bL|ERROR-) 7620(7621aL|7621bL) 7621(7622aL|7622bL) 7622(7623aR|7628bR) 7623(7624aL|7626aL) 7624(7625aR|7625bR) 7625(7631aL|7631bL) 7626(7627aL|7627aL) 7627(7622aL|7622bL) 7628(ERROR-|7629aL) 7629(7630aL|7630aL) 7630(7622aL|7622bL) 7631(7632aR|7633bR) 7632(7631aR|7631bR) 7633(7631aR|7887bR) 7634(7635aR|7638bR) 7635(7636bL|7634bR) 7636(7637aR|7637aR) 7637(7661aR|7661bR) 7638(7639bL|7641bL) 7639(7640aR|7640aR) 7640(7643aR|7643bR) 7641(7642aR|7642aR) 7642(7652aR|7652bR) 7643(7644aR|7649bR) 7644(7645aL|7647aL) 7645(7646bR|7646bR) 7646(7661aR|7661bR) 7647(7648bR|7648bR) 7648(7634aR|7634bR) 7649(7670aR|7650aL) 7650(7651bR|7651bR) 7651(7652aR|7652bR) 7652(7653aR|7658bR) 7653(7654bL|7656bL) 7654(7655bR|7655bR) 7655(7661aR|7661bR) 7656(7657bR|7657bR) 7657(7634aR|7634bR) 7658(7659bL|7652bR) 7659(7660bR|7660bR) 7660(7643aR|7643bR) 7661(7662aR|7665bR) 7662(7661aR|7663aL) 7663(7664aR|7664aR) 7664(7634aR|7634bR) 7665(7666aL|7668aL) 7666(7667aR|7667aR) 7667(7643aR|7643bR) 7668(7669aR|7669aR) 7669(7652aR|7652bR) 7670(7671aR|7676bR) 7671(7672aL|7674aL) 7672(7673bR|7673bR) 7673(7661aR|7661bR) 7674(7675bR|7675bR) 7675(7634aR|7634bR) 7676(7679aR|7677aL) 7677(7678bR|7678bR) 7678(7652aR|7652bR) 7679(7680bR|ERROR-) 7680(7681aL|7681bL) 7681(7682aR|7685bR) 7682(7683aL|ERROR-) 7683(7684aR|7684bR) 7684(7688aL|7688bL) 7685(7686aL|ERROR-) 7686(7687aL|7687bL) 7687(7681aL|7681bL) 7688(7689aR|7694bR) 7689(7690aL|7692bL) 7690(7691aL|7691bL) 7691(7688aL|7688bL) 7692(7693aL|7693bL) 7693(7688aL|7688bL) 7694(7695aL|7697bL) 7695(7696aL|7696bL) 7696(7609aL|7609bL) 7697(7698aL|7698bL) 7698(7688aL|7688bL) 7699(7700aR|7705bR) 7700(7701aL|7703bL) 7701(7702aL|7702bL) 7702(7699aL|7699bL) 7703(7704aL|7704bL) 7704(7699aL|7699bL) 7705(7706aL|7708bL) 7706(7707aL|7707aL) 7707(7710aL|7710bL) 7708(7709aL|7709bL) 7709(7699aL|7699bL) 7710(7711aR|7716bR) 7711(7712aL|7714bL) 7712(7713aL|7713bL) 7713(7710aL|7710bL) 7714(7715aR|7715bR) 7715(7719aL|7719bL) 7716(ERROR-|7717bL) 7717(7718aR|7718bR) 7718(7719aL|7719bL) 7719(7720aR|7725bR) 7720(7721aL|7723bL) 7721(7722bL|7722bL) 7722(7730aL|7730bL) 7723(7724aL|7724bL) 7724(7719aL|7719bL) 7725(7726aL|7728bL) 7726(7727aL|7727bL) 7727(7719aL|7719bL) 7728(7729aL|7729bL) 7729(7719aL|7719bL) 7730(7731aR|7736bR) 7731(7732aL|7734bL) 7732(7733aL|7733bL) 7733(7730aL|7730bL) 7734(7735aL|7735bL) 7735(7730aL|7730bL) 7736(7737aL|7739bL) 7737(7738aL|7738aL) 7738(7741aL|7741bL) 7739(7740aL|7740bL) 7740(7730aL|7730bL) 7741(7742bL|ERROR-) 7742(7743aL|7743bL) 7743(7744aR|7749bR) 7744(7745aL|7747aL) 7745(7746aR|7746bR) 7746(7752aL|7752bL) 7747(7748aL|7748aL) 7748(7743aL|7743bL) 7749(ERROR-|7750aL) 7750(7751aL|7751aL) 7751(7743aL|7743bL) 7752(7753aR|7754bR) 7753(7752aR|7752bR) 7754(7755aR|7752bR) 7755(7756aR|7757bR) 7756(7755aR|7755bR) 7757(7758aL|7755bR) 7758(7759aR|7759aR) 7759(7760aR|7760bR) 7760(7761aR|7766bR) 7761(7762aL|7764aL) 7762(7763aR|7763bR) 7763(7843aL|7843bL) 7764(7765bL|7765bL) 7765(7769aL|7769bL) 7766(ERROR-|7767aL) 7767(7768bL|7768bL) 7768(7780aL|7780bL) 7769(7770aR|7775bR) 7770(7771aL|7773bL) 7771(7772aL|7772bL) 7772(7769aL|7769bL) 7773(7774aL|7774bL) 7774(7769aL|7769bL) 7775(7776aL|7778bL) 7776(7777aL|7777bL) 7777(7791aL|7791bL) 7778(7779aL|7779bL) 7779(7769aL|7769bL) 7780(7781aR|7786bR) 7781(7782aL|7784bL) 7782(7783aL|7783bL) 7783(7780aL|7780bL) 7784(7785aL|7785bL) 7785(7780aL|7780bL) 7786(7787aL|7789bL) 7787(7788aL|7788bL) 7788(7818aL|7818bL) 7789(7790aL|7790bL) 7790(7780aL|7780bL) 7791(7792aR|7797bR) 7792(7793bL|7795bL) 7793(7794aR|7794aR) 7794(7827aL|7827bL) 7795(7796aL|7796bL) 7796(7791aL|7791bL) 7797(ERROR-|7798bL) 7798(7799aL|7799aL) 7799(7800aL|7800bL) 7800(7801aR|7806bR) 7801(7802bL|7804bL) 7802(7803bR|7803bR) 7803(7827aL|7827bL) 7804(7805bL|7805bL) 7805(7791aL|7791bL) 7806(ERROR-|7807bL) 7807(7808aL|7808bL) 7808(7800aL|7800bL) 7809(7810aR|7815bR) 7810(7811bL|7813bL) 7811(7812aR|7812aR) 7812(7830aL|7830bL) 7813(7814aL|7814bL) 7814(7809aL|7809bL) 7815(ERROR-|7816bL) 7816(7817aL|7817aL) 7817(7818aL|7818bL) 7818(7819aR|7824bR) 7819(7820bL|7822bL) 7820(7821bR|7821bR) 7821(7830aL|7830bL) 7822(7823bL|7823bL) 7823(7809aL|7809bL) 7824(ERROR-|7825bL) 7825(7826aL|7826bL) 7826(7818aL|7818bL) 7827(7828aR|7829bR) 7828(7827aR|7827bR) 7829(7833aR|7827bR) 7830(7831aR|7832bR) 7831(7830aR|7830bR) 7832(7838aR|7830bR) 7833(7834aR|7835bR) 7834(7833aR|7833bR) 7835(7836bL|7833bR) 7836(7837aR|7837aR) 7837(7760aR|7760bR) 7838(7839aR|7840bR) 7839(7838aR|7838bR) 7840(7841bL|7838bR) 7841(7842bR|7842bR) 7842(7760aR|7760bR) 7843(7844aR|7849bR) 7844(7845aL|7847aL) 7845(7846aL|7846bL) 7846(7852aL|7852bL) 7847(7848aL|7848aL) 7848(7843aL|7843bL) 7849(ERROR-|7850aL) 7850(7851aL|7851aL) 7851(7843aL|7843bL) 7852(7853aR|7858bR) 7853(7854aL|7856aL) 7854(7855aL|7855bL) 7855(7861aL|7861bL) 7856(7857aL|7857aL) 7857(7843aL|7843bL) 7858(ERROR-|7859aL) 7859(7860aL|7860aL) 7860(7843aL|7843bL) 7861(7862aR|7863bR) 7862(HALT--|7864bR) 7863(ERROR-|7864bR) 7864(7865aR|7870bR) 7865(7866aL|7868bL) 7866(7867aL|7867bL) 7867(7864aL|7864bL) 7868(7869aL|7869bL) 7869(7864aL|7864bL) 7870(7871aL|7873bL) 7871(7872aL|7872aL) 7872(7875aL|7875bL) 7873(7874aL|7874bL) 7874(7864aL|7864bL) 7875(7876aR|7881bR) 7876(7877bL|7879bL) 7877(7878bR|7878bR) 7878(7884aL|7884bL) 7879(7880bL|7880bL) 7880(7875aL|7875bL) 7881(ERROR-|7882bL) 7882(7883aR|7883aR) 7883(7884aL|7884bL) 7884(7885aR|7886bR) 7885(7887aR|7884bR) 7886(7884aR|7884bR) 7887(7888aR|7889bR) 7888(7887aR|ERROR-) 7889(ERROR-|7890bL) 7890(7891aR|7891bR) 7891(7892aL|7892bL) 7892(7893aR|7894bR) 7893(7895aR|7892bR) 7894(7892aR|7892bR) 7895(7896aR|7897bR) 7896(7898aR|7892bR) 7897(7892aR|7892bR) 7898(7899aR|7902bR) 7899(7900aL|7892bR) 7900(7901aR|7901bR) 7901(7903aL|7903bL) 7902(7892aR|7892bR) 7903(7904aR|7905bR) 7904(7903aR|ERROR-) 7905(ERROR-|7906bL) 7906(7907aL|7907bL) 7907(7908aL|7908bL) 7908(7909bL|ERROR-) 7909(4381aL|4381bL) \par}

\end{appendices}

\end{document}